\documentclass[a4paper,12pt]{book}
\usepackage[utf8]{inputenc}
\usepackage{amsmath, amsthm}
\usepackage{slashed}
\usepackage[mathscr]{euscript}
\usepackage{graphicx}
\usepackage{subfigure}
\usepackage{enumerate}
\usepackage{fancyhdr}
\usepackage[explicit]{titlesec}
\usepackage{emptypage}
\usepackage{setspace}
\usepackage[T1]{fontenc}
\usepackage{txfonts}
\usepackage[margin=3.0cm]{geometry}
\usepackage{amssymb}
 \usepackage{braket}
 \usepackage[colorlinks, citecolor = magenta, linkcolor = blue, urlcolor =  blue, anchorcolor = blue, menucolor = black]{hyperref}
 \usepackage{color}
 \usepackage{xcolor}
 \definecolor{armygreen}{rgb}{0.29, 0.33, 0.13}
 \definecolor{awesome}{rgb}{1.0, 0.13, 0.32}
 \definecolor{byzantine}{rgb}{0.74, 0.2, 0.64}
 \usepackage{dsfont}
 \usepackage{caption}
 \usepackage{bbm}
\usepackage{times}
 \usepackage{newtxtext,newtxmath}
 \usepackage{enumitem}
\usepackage{cite}
\usepackage[toc,page]{appendix}
\usepackage[nottoc,notlot,notlof]{tocbibind}
\usepackage{lipsum}
\usepackage{comment}
\DeclareMathAlphabet{\mathcal}{OMS}{cmsy}{m}{n}

\begin{document}
\begin{center}
{\Large {\Huge A Phenomenological Study of WIMP Models }}
\vskip 0.70cm
{\bf {\em By}} 
\vskip -0.2cm
{\bf {\large Shivam Gola}}
\vskip 0.0cm
{\bf {\large PHYS10201704001}}
\vskip 0.5cm
{\bf {\large The Institute of Mathematical Sciences, Chennai}}
\vskip 2.6cm
{\bf {\em {\large A thesis submitted to the
\vskip 0.05cm
Board of Studies in Physical Sciences
\vskip 0.05cm
In partial fulfilment of requirements
\vskip 0.05cm
For the Degree of 
}}}
\vskip 0.05cm
{\bf {\large DOCTOR OF PHILOSOPHY}}
\vskip 0.1cm
{\bf {\em of}}
\vskip 0.1cm
{\bf {\large HOMI BHABHA NATIONAL INSTITUTE}}
\vfill
\includegraphics[height=3.5cm, width=3.5cm]{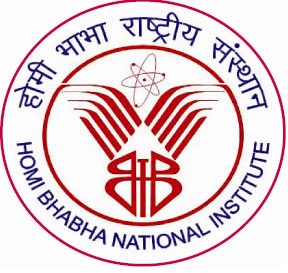}
\vfill
{\bf {\large 09-02-2024}}
\vfill
\end{center}

\newpage
\cleardoublepage
\begin{figure}[hbt]
\begin{center}
\includegraphics[width=16cm,height=24cm]{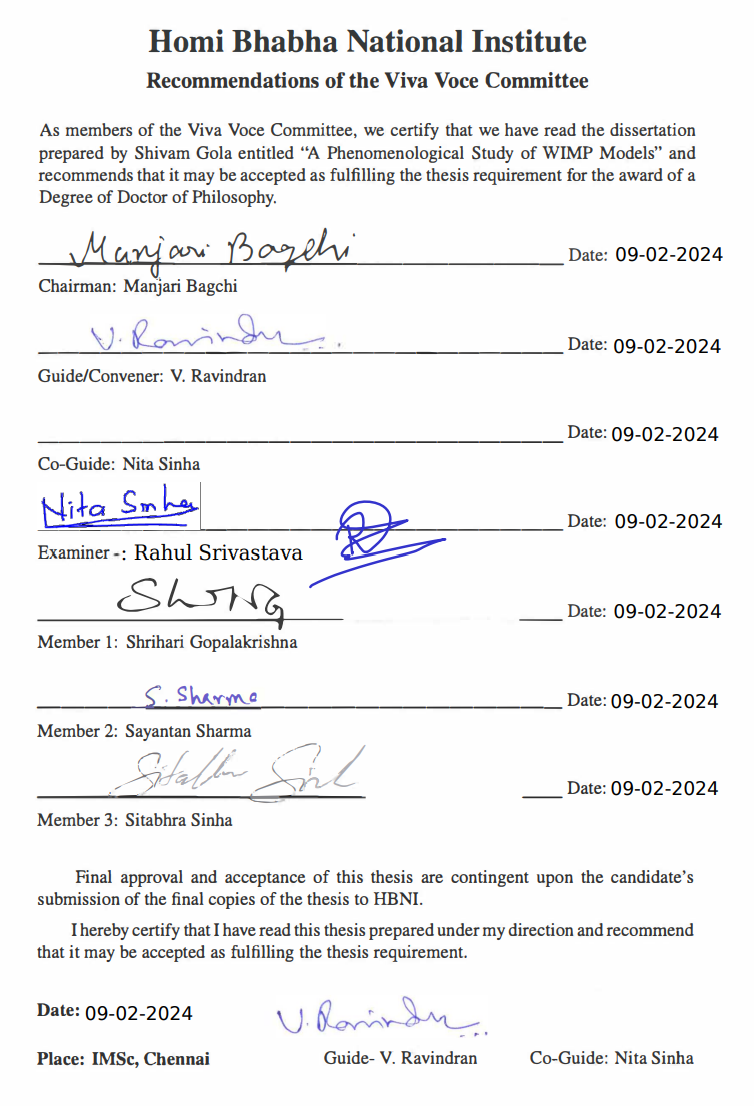}
\end{center}
\end{figure} 
\newpage
\cleardoublepage
\centerline{{\bf {\large STATEMENT BY AUTHOR}}}
\vskip 1.00cm
\doublespacing
This dissertation has been submitted in partial fulfillment of the requirements for an
advanced degree at Homi Bhabha National Institute (HBNI) and is deposited in the
Library to be made available to borrowers under the rules of the HBNI.
\vskip 0.6cm
Brief quotations from this dissertation are allowable without special permission.
provided that accurate acknowledgment of the source is made. Requests for permission
for extended quotation from or reproduction of this manuscript in whole or in part
may be granted by the Competent Authority of HBNI when in his or her judgment the
proposed use of the material is in the interests of scholarship. In all other instances,
however, permission must be obtained from the author.

\vskip 3.0cm

%$~$\hspace{11.0cm}-------------

$~$\hspace{10.2cm} Shivam Gola
\newpage
\cleardoublepage
~
\vskip 1.2cm
\centerline{{\bf{\large{DECLARATION}}}}
\vskip 1.2cm
I, hereby declare that the investigation presented in the thesis has been carried out by
me. The work is original and has not been submitted earlier as a whole or in part for a
degree/diploma at this or any other Institution / University.
\vskip 3.0cm
%
%\rightline{------------- \hspace{1.5cm}}
%
\rightline{Shivam Gola \hspace{0.9cm}}
\newpage
\cleardoublepage
%\pagenumbering{roman}
\rule{15cm}{0.3mm}
\centerline{\bf{\large List of Publications arising from the thesis}}
\rule{15cm}{0.3mm}
\begin{enumerate}
\item {\bf ``ALP-portal majorana dark matter''} \\
Shivam Gola, Sanjoy Mandal and Nita Sinha \\
~\href{https://doi.org/10.1142/S0217751X22501317}{Int.J.Mod.Phys.A 37 (2022) 22, 2250131}
\item {\bf ``Two-component scalar and fermionic dark matter candidates in a generic $\mathbf{U(1)_X}$ model''} \\
Arindam Das, Shivam Gola, Sanjoy Mandal and Nita Sinha \\
\href{https://doi.org/10.1016/j.physletb.2022.137117}{Phys.Lett.B 829 (2022) 137117}
\item {\bf ``Pseudo scalar dark matter in a generic U(1)$_X$ model''} \\
Shivam Gola \\
~\href{https://www.sciencedirect.com/science/article/pii/S0370269323003167?via\%3Dihub}{Phys.Lett.B 842 (2023) 137982}
\end{enumerate}
\vskip 7cm
\rightline{Shivam Gola \hspace{0.9cm}}

\newpage
\cleardoublepage
\rule{15cm}{0.3mm}
\centerline{\bf {\large Seminars presented}}
\rule{15cm}{0.3mm}
\begin{enumerate}
\item \textbf{A pseudo scalar dark matter in a generic U(1)$_X$ model} talk given at \href{https://indico.ictp.it/event/10182}{{\bf Summer School on Particle Physics~(smr 3848)}}, held at ICTP-Trieste, 23rd June, 2023.
\item {\bf ``Two-component scalar and fermionic dark matter candidates in a generic $\mathbf{U(1)_X}$ model''}, talk given at \href{https://indico.cern.ch/event/1146097/}{{\bf XXV DAE-BRNS High Energy Physics (HEP) Symposium 2022}}, held at IISER-Mohali, 12th Dec 2022.
 \item {\bf ``Two-component scalar and fermionic dark matter candidates in a generic $\mathbf{U(1)_X}$ model''}, poster presented at \href{https://2022.kashiwa-darkmatter-symposia.org/}{{\bf KASHIWA Dark Matter Symposium 2022~~(Online)}}, held at University of Tokyo, Japan, 1st December, 2022.
\item {\bf ``Two-component scalar and fermionic dark matter candidates in a generic $\mathbf{U(1)_X}$ model''}, talk given at
\href{http://www.ichep2018.org/}{\bf The Future is Illuminating~(Online),} at NCTS, Taiwan, 29th June, 2022
\item {\bf ``ALP-portal majorana dark matter''}, at
\href{http://www.ichep2018.org/}{\bf The 30th International Symposium on Lepton Photon Interactions at High Energies~(Online)}, poster presented at University of Manchester, 10th Jan, 2022.

\end{enumerate}
\vskip 7cm
\rightline{Shivam Gola \hspace{0.9cm}}
\newpage
\cleardoublepage
\rule{15cm}{0.3mm}
\centerline{\bf {\large Conferences and workshops attended}}
\rule{15cm}{0.3mm}
\begin{enumerate}
\item \href{https://indico.ictp.it/event/10182}{{\bf Summer School on Particle Physics~(smr 3848)}}, held at ICTP-Trieste, 23rd June, 2023.
\item \href{https://indico.cern.ch/event/1146097/}{{\bf XXV DAE-BRNS High Energy Physics (HEP) Symposium 2022}}, held at IISER-Mohali, 12th - 16th Dec 2022.
 \item \href{https://2022.kashiwa-darkmatter-symposia.org/}{{\bf KASHIWA Dark Matter Symposium 2022~(Online)}}, held at University of Tokyo, Japan, from 29th Nov. - 2nd Dec., 2022.
 \item \href{ https://indico.cern.ch/event/1119695/page/24895-overview}{{\bf FIPs 2022 workshop on Feebly-Interacting Particles~(Online)}}, held at CERN, Europe, from 17th - 21st October, 2022.
 \item \href{https://www.ictp-saifr.org/bsmp2022/}{{\bf Joint ICTP-SAIFR/MITP Summer School on Particle Physics beyond the Standard Model}}, held at ICTP - SAIFER, Sao Paulo, Brazil, from 12th - 23rd September, 2022.
 \item \href{https://www.ictp-saifr.org/ndfp2022/}{{\bf Program on New Directions in Particle Physics}}, held at ICTP - SAIFER, Sao Paulo, Brazil, from 5th - 23rd September, 2022. 
 \item \href{https://phys.ncts.ntu.edu.tw/act/actnews/The-64547473/home/introduction}{{\bf The Future is Illuminating~(Online)}}, held at NCTS, Taiwan, from 28th - 30th  June, 2022.
 \item \href{https://indico.cern.ch/event/949705/}{{\bf The 30th International Symposium on Lepton Photon Interactions at High Energies~(Online)}}, held at University of Manchester, from 10th - 14th January, 2021.
\item \href{https://www.icts.res.in/program/LTPDM2020}{{\bf Less Travelled Path of Dark Matter: Axions and Primordial Black Holes~(Online)}}, held at ICTS-TIFR, from November 9th–13th, 2020.
 \item \href{https://sites.google.com/view/serbmainschool2019/}{{\bf XXXIII SERB Main School 2019 in THEP at SGTB Khalsa College}}, held at the University of Delhi, from December 7th–26th, 2019.
\item \href{https://indico.cern.ch/event/829653/}{{\bf Madgraph School 2019}}, held at IMSc-Chennai, from November 18th–22nd, 2019.
\item \href{https://people.iith.ac.in/anomalies19/}{{\bf Anomalies 2019 workshop}}, held at IIT-Hyderabad, from July 18th–20th, 2019.
\item \href{https://web.iitm.ac.in/dae2018/index.php}{{\bf XXIIIth DAE-BRNS High Energy Physics Symposium 2018}}, held at IIT-Madras, from 8th–12th December 2018.
\end{enumerate}
\vskip 3cm
\rightline{Shivam Gola \hspace{0.9cm}}

\newpage
\cleardoublepage
%
%\centerline{{\bf {\large DEDICATIONS (OPTIONAL)}}}
%

\begin{center}
\vspace{5cm}
\centerline{\em \Large{ {\fontsize{30}{40}\selectfont Dedicated  } }}
\end{center}    
\begin{center}
\vspace{0cm}
\centerline{\em \Large{ {\fontsize{30}{40}\selectfont  to } }}
\end{center}    
\begin{center}
\centerline{\em \Large{ {\fontsize{30}{40}\selectfont  My Parents, Siblings \& Teachers } }}
\end{center}

\newpage
\cleardoublepage
~
\vskip 0.5cm
\centerline{{\bf{\large ACKNOWLEDGEMENTS}}}
\vskip 0 cm
%
%\centerline{$<$One page Write up, Times New Roman, font 12, Double
%Spacing.$>$}
%
I would like to express my sincere gratitude to Prof. Nita Sinha and Prof. V. Ravindran for being my advisors and supporting me through the highs and lows of my PhD life. I am grateful to them for giving me the freedom to learn the subject the way I wanted.

I am thankful to my doctoral committee members, Dr. Manjari Bagchi, Prof. Shrihari Gopalakrishna, Dr. Sayantan Sharma, and Dr.Sitabhra Sinha, for their suggestions and guidance.

I am thankful to Prof. M.V.N. Murthy, Prof. D. Indumathi, Prof. Bala Sathiyapalan, Prof. Rahul Sinha, and other faculties at IMSc for teaching me beautiful courses.

I am thankful to IMSc for supporting my visits to conferences and seminars within and outside India.

I am thankful to my senior, Dr. Sanjoy Mondal, for helping me learn a completely             new field and for supporting me in my bad times. I am thankful to Dr. Arindam Das and Dr. Disha Bhatia for allowing me the opportunity to work with them and learn from their experience and insight.

I am thankful to the particle physics group members Dr. Suprabh Prakash, Dr. K N Vishnudath, Pritam Sen, Ria Sen, Surabhi Tiwari, T. Ravi, and Aparna Sankar for their support and encouragement.

I am blessed to have many good friends here. It is impossible to list them all in this short PhD acknowledgment. Spending time, chatting, and going for dinner at various places with Vinay, Sahil, Farhina, Neelam, Amit, Nilakshi, Pavitra, Chandrani, Prabhat, Ravi Shankar, and others was the best part of my IMSc life.

Finally, I would not have reached this far if there had not been the constant support and faith my parents and siblings have shown in me. I am thankful to them for their love.

\newpage
\cleardoublepage
\pagestyle{plain}
\pagenumbering{roman}
\tableofcontents
 %\addcontentsline{toc}{chapter}{\listfigurename}
\listoffigures
\cleardoublepage
%\addcontentsline{toc}{chapter}{\listtablename}
\listoftables
\cleardoublepage

%%%%%%%%%%%%%%%%%%%%%%%%%%%%%%%%%%%%%%%%%%%%%%%%%%%%%%%
% \cleardoublepage
% \pagestyle{plain}
% \pagestyle{fancy}
% \renewcommand{\chaptermark}[1]{ \markboth{#1}{} }
% %\renewcommand{\sectionmark}[1]{ \markright{#1}{} }
% 
% \fancyhf{}
% \fancyhead[LE,RO]{\thepage}
% \fancyhead[RE]{\textit{ \nouppercase{\leftmark}} }
% \fancyhead[LO]{\textit{ \nouppercase{\rightmark}} }
% 
% \fancypagestyle{plain}{ %
%   \fancyhf{} % remove everything
%   \renewcommand{\headrulewidth}{0pt} % remove lines as well
%   \renewcommand{\footrulewidth}{0pt}
% }

%%%%%%%%%%%%%%  Page style %%%%%%%%%%%%%%%%%%%%
% \definecolor{armygreen}{rgb}{0.29, 0.33, 0.13}

\definecolor{headercolor}{gray}{0.55} % chapter numbers will be semi transparent .5 .55 .6 .0
%%%
%%% fancy header options
%%%
% 
% % redefine the plain style
% \fancypagestyle{plain}{%
% \fancyhf{} % clear all header and footer fields 
% % \fancyfoot[C]{\fontfamily{ppl}\selectfont \small\bfseries \thepage} % except the center 
% \renewcommand{\headrulewidth}{0.3pt} 
% \renewcommand{\footrulewidth}{0pt}}
%
% define the style fancy
\pagestyle{fancy}
\renewcommand{\chaptermark}[1]{\markboth{#1}{}}
\renewcommand{\sectionmark}[1]{\markright{\thesection\ \ #1}}
\fancyhf{}
\fancyhead[LE,RO]{\fontfamily{ppl}\selectfont \small\bfseries\thepage}
\fancyhead[LO]{\fontfamily{ppl}\selectfont \small\bfseries\nouppercase{\color{headercolor}\rightmark}}
\fancyhead[RE]{\fontfamily{ppl}\selectfont \small\bfseries\nouppercase{\color{headercolor}\leftmark}}
\renewcommand{\headrulewidth}{0.3pt}
\renewcommand{\footrulewidth}{0pt}
\addtolength{\headheight}{0.4pt} % leave space for the line
\renewcommand{\headrule}{{\color{headercolor} \hrule width\headwidth height\headrulewidth \vskip-\headrulewidth}}

% removes headings from empty pages
% \usepackage{emptypage}

\definecolor{halfgray}{gray}{0.55} % chapter numbers will be semi transparent .5 .55 .6 .0
\newfont{\chapNumFont}{eurb10 scaled 7000}
\newfont{\chapTitFont}{pplr9d}
\titleformat{\section}[hang]{\bfseries\Large}{\fontfamily{ppl}\selectfont \thesection}{15pt}{\fontfamily{ppl}\selectfont #1}
\titleformat{\subsection}[hang]{\bfseries\large}{\fontfamily{ppl}\selectfont \thesubsection}{15pt}{\fontfamily{ppl}\selectfont #1}
\titleformat{\subsubsection}[hang]{\bfseries}{\fontfamily{ppl}\selectfont \thesubsubsection}{15pt}{\fontfamily{ppl}\selectfont #1}
\titleformat{\chapter}[block]%
{\Huge}{\raggedleft{\color{halfgray}\chapNumFont\thechapter}}{20pt}%
{\raggedright{\fontfamily{ppl}\selectfont #1}}

 \setcounter{page}{0}
 \pagenumbering{arabic}

%=============================================================================================================
%main thesis body
%============================================================================================================
\chapter{Introduction to thesis}\label{chap_introduction}
Several observations ranging from star clusters to the cosmological scale are in favour of the existence of an unknown non-luminous matter, or dark matter~(DM)~\cite{Garrett_2011,Profumo:2009tb,Lisanti:2016jxe,Kolb:1990vq,Bartelmann:1999yn,Clowe:2003tk,Harvey:2015hha,Hinshaw:2012aka,Ade:2015xua}. Data from WMAP~\cite{Hinshaw:2012aka}, PLANCK~\cite{Aghanim:2018eyx}, which probe anisotropies in cosmic microwave background radiation~(CMBR), emphasize that the total mass-energy of the universe contains 4.9\% ordinary matter~(Baryonic matter), 26.8\% dark matter, and 68.3\% of an unknown form of energy called dark energy. However, only gravitational effects of DM have been observed to date; hence, the particle nature of DM is largely unknown.

This thesis focuses on the various WIMP models of DM. We divide this thesis into many chapters. In the first chapter, we discuss the basics of DM. The chapter is organized as follows: In section~\ref{history}, we discuss the strong evidence that historically supports this unknown matter. Section~\ref{facts}, we collect some of the known facts and features about DM. Section~\ref{dm_candi} briefly discusses several popular dark matter candidates. In the last section~\ref{detection}, we discuss three main detection techniques for DM.
\section{History and Evidence}
\label{history}
In 1884, Lord Kelvin estimated the number of dark bodies in the Milky Way from the observed velocity dispersion of the stars orbiting around the center of the galaxy. By using these measurements, he estimated the mass of the galaxy, which he determined to be different from the mass of visible stars. He thus concluded that many of our stars, perhaps a great majority of them, maybe dark bodies~\cite{011929196}.
 
In 1906, Henri Poincare, in "The Milky Way and Theory of Gases," used "dark matter" or “matière obscure” in French in discussing Kelvin's work~\cite{Garrett_2011}.
 
In 1930, J.H. Oort studied the Doppler shift of the stars moving near the galactic plane of the Milky Way, and from that, he calculated the velocities of stars. He found that stars are moving sufficiently fast that they can escape the gravitational pull of the luminous mass of the galaxy, calculated by the mass (M)/luminosity (L) ratio method, which implies that there must be an invisible gravity source that holds stars in the galactic plane~\cite{1932BAN.....6..249O}.
 
In 1933, F. Zwicky studied the Coma cluster, which is about 99 Mpc away from Earth. He calculated the velocity dispersion of the galaxies using Doppler shift data and applying the Virial theorem, and he found that the total mass of the cluster is just 2\% of the one calculated by the M/L ratio method, leading to the conclusion of invisible mass, which he called "Dunkle Materie" or dark matter~\cite{1937ApJ....86..217Z}.

\begin{figure}[t]
\centering
\includegraphics[height=10.5cm,width=12cm]{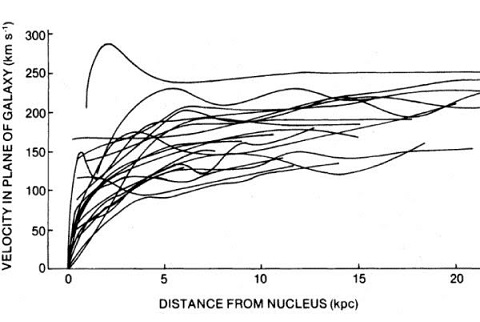}
\caption{Rotation curves of several galaxies as observed by Rubin. The figure has been taken from~\cite{Sofue_1999}.}
\label{rubin}
\end{figure}
 
In 1960–1970, Vera Rubin, with others~\cite{22e63506-7b44-3c21-8701-7cb28a0f350f}, studied the rotation curves of sixty isolated galaxies, from which she estimated the rotation velocities of the region of a galaxy using Doppler shift and found that it is independent of distance, as shown in figure~\ref{rubin}. Assuming Newtonian gravity and circular orbits, the velocity of the star is given by
\begin{align}
\text{v}(r) = \sqrt{GM/r}
\end{align}

where G is Newton's constant. Therefore, if the mass distribution is uniform, then $ \text{v}(r) \ \propto \ \frac{1}{\sqrt{r}}$; however, the observed star velocity seems to be independent of distance for large r. The following observation can be explained if $M(r) \ \propto \ r $, or density follows $\rho(r) \  \ \propto \  \frac{M(r)}{r^3}  \ \sim 1/r^2 $ distribution.
 
The phenomenon of gravitational lensing, described by General Relativity (GR), states that whenever light from a distant source passes near a massive object, it forms a ring of light called ``Einstein ring" around the same object. In 1979, D. Walsh was the first to study two distant objects separated by only 5.6 arcseconds with very similar redshifts. The explanation for his discovery is that there was only one object there, and the other is the image of it formed by an unseen dark matter cloud~\cite{1979Natur.279..381W}.

\begin{figure}[htbp]
\centering
\includegraphics[width=14cm,height=7.5cm]{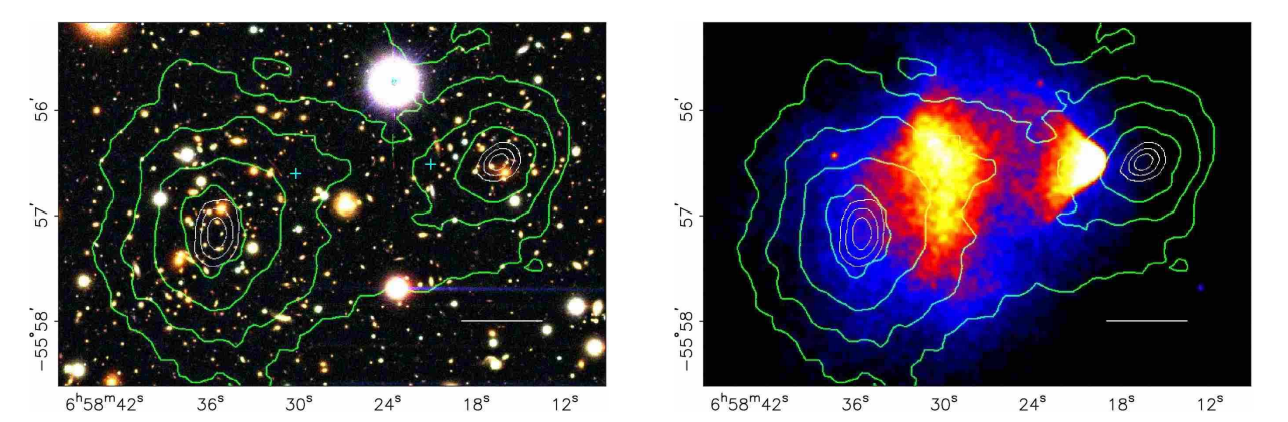}
\caption{A representation of the collision of two galaxy clusters is shown. The red region is due to the observation of electromagnetic radiation by the Chandra X-ray observatory.  The blue region is where the most mass is concentrated, as mapped by gravitational lensing. The figure has been taken from~\cite{Clowe_2006}.}
\label{bullet}
\end{figure}
  
In 2004, from Chandra's observations of the bullet cluster~(1E0657–56), Markevitch and Clowe found the most compelling reason for the presence of the invisible matter. A bullet cluster is a system of two colliding galactic clusters. During the collision, the galaxies, which are separated by large distances within the two clusters, pass right through without interacting. However, the majority of the cluster's baryonic mass exists in the hot gas in between galaxies. Collision dislocates the stellar gasses from their respective galaxies and heats the gases, resulting in a huge amount of X-ray radiation emission. However, the location of baryonic mass (stellar gases) seen by an X-ray telescope is different from the one found by weak gravitational lensing, i.e., the region of space where most masses are located, as shown in figure~\ref{bullet}. If an unseen mass is electromagnetically neutral, then during a collision, it won't be dragged and will remain in its separate clusters, which explains this puzzle~\cite{Clowe_2006}.
\begin{figure}[t]
\centering
\includegraphics[width=14cm,height=9cm]{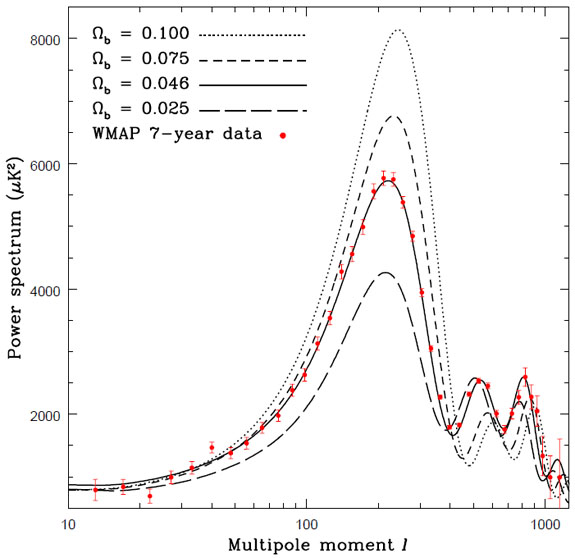}
\caption{The CMB power spectrum for various values of $\Omega_b$ and $\Omega_{\text{DM}}$ while keeping $\Omega_{\text{Total}}=1$. The figure has been taken from~\cite{Garrett_2011}. }
\label{power}
\end{figure} 
In 1964, two American radio astronomers, Arno Penzias, and Robert Wilson, accidentally discovered the earliest radiation (uniform radio waves) or so-called cosmic microwave background radiation~(CMBR) in the universe. Later in 1989, the COBE satellite confirmed this extreme uniform relic radiation at 2.73 K, with tiny spatial fluctuations in temperature. In 2001, WMAP and later Planck confirmed the prediction of COBE, measured the temperature fluctuation with greater accuracy, and found that it was of the order of $\mu$K. This whole observation was done via the power spectrum method, which is shown in figure~\ref{power}. It has been found that these temperature fluctuations in the early universe can be explained only if one accounts for the additional non-baryonic matter, which is nearly five times ordinary matter~\cite{1965ApJ...142..419P}.

\section[Facts on dark Matter]{Facts on dark matter}
\label{facts}
The following facts about dark matter are known:
\begin{itemize}
\item It is a non-baryonic matter, i.e., having no baryon quantum number; otherwise, it can interact with baryons (quarks, protons, neutrons, etc.), and therefore it could have been detected in colliders or other experimental searches~\cite{Garrett_2011,Bertone_2018}. MACHOs are also baryonic DM candidates, which have been ruled out by astronomical observations~(see \ref{non_thermal_DM}).
\item DM does not interact with light; therefore, it should not have any electric charge. However, millicharged particles were proposed as DM in the literature~\cite {Arg_elles_2021}.
\item A particle candidate for DM must be stable or have a lifetime greater than the age of the universe~\cite{Profumo:2019ujg}.
\item A large fraction of DM is non-relativistic or cold; however, some fractions could be warm or hot~\cite{Lisanti:2016jxe}.
\item The total DM abundance~(relic density) today is $\Omega_{\text{DM}} h^2 = 0.12\pm 0.001$~\cite{Planck:2018vyg}.
\end{itemize}
\section[Candidates to dark matter]{Candidates to dark matter}
\label{dm_candi}
In this section, we briefly mention some of the popular DM candidates.
\subsubsection{Looking into the Standard Model}
Standard Model of Particle Physics~(SM) is currently our best theory to describe the nature of elementary particles and their interactions~(see Appendix~\ref{sm}). It has six leptons, six quarks, twelve gauge mediators, and one Higgs particle. However, DM is electrically neutral; therefore, only neutrinos, Z bosons, and Higgs bosons remain. The DM stability requirement rules out Higgs and Z bosons. Though neutrinos are massless in SM, it is now a fact that neutrinos have a tiny mass, according to several experiments~\cite{PhysRevLett.9.36,Kajita:2016cak,doi:10.1146/annurev-nucl-102115-044600,deSalas:2020pgw}. Cosmological observations suggest that the sum of the active neutrino masses should be $\sum m_{\nu}\leq 0.12$ eV~\cite{Aghanim:2018eyx,Lattanzi:2017ubx}, which is a hot DM candidate and contributes to the relic density $\leq4.5\times10^{-3}$. The hot neutrino relic is too low to explain the current DM abundances but is good for explaining small-scale problems. Hence, no particle is left within SM that produces the correct DM abundance. However, in a beyond-standard model (BSM) case, it is possible that some heavy-colored quark can bind with itself or SM quark to form a coloured neutral state, which might constitute some fraction of DM~\cite{De_Luca_2018}.
\subsubsection{Thermal dark matter}
These are DM candidates, which were produced thermally in the early universe when it was in the plasma state. One can do such a study and calculate the DM relic density using the Boltzmann equation. A popular mechanism following thermal DM is the freeze-out mechanism~\cite{Kolb:1990vq}. It assumes that DM is in thermal equilibrium with the SM bath in the very early stages of the universe, but as the universe expands and the interaction rate becomes ineffective against the expansion rate, DM number density or relic density freezes out. It is found that if DM is in the GeV~-~TeV mass range and couples weakly to SM, it can easily satisfy the correct relic density. This is called the WIMP miracle, and such particles are named weakly interacting massive particles (WIMPs). A brief on the freeze-out mechanism is described here~\cite{Lisanti:2016jxe,Kolb:1990vq},

The most general Boltzmann equation for the particle density of a given species ($n_i$) using the FRW metric (Appendix-\ref{Friedman}) is given by the following expression:
\begin{eqnarray}\nonumber
\frac{dn_i}{dt} \ = \ -3Hn_i \ - \ \sum_{j=1}^{N}\langle\sigma_{ij}v_{ij}\rangle(n_in_j \ - \ n_{i}^{eq}n_{j}^{eq}) \ - \ \sum_{j{\neq}i}[\Gamma_{ij}(n_i-n_{i}^{eq}) \ - \ \Gamma_{ji}(n_{j}-n_{j}^{eq})] \\ - \ \sum_{j{\neq}i}[\langle\sigma_{Xij}v_{ij}\rangle(n_in_X \ - \ n_{i}^{eq}n_{X}^{eq} ) \ - \langle\sigma_{Xji}v_{ji}\rangle(n_jn_X \ - \ n_{j}^{eq}n_{X}^{eq})]. \nonumber
\end{eqnarray}
In the above expression, we have included all the necessary processes, with the first term being due to the expansion of the universe; the second term includes the interaction (both scattering and annihilation) of two particle species; the third term expresses the decay of $i_{th}$ particles into $j_{th}$ and vice versa; and the final term is due to scattering off the thermal background.
Now if the decay rate is much faster than the age of the universe, then all those particles that can decay to DM have decayed; hence, we say the abundance of DM is the sum of all $(n_i)$, say it is n. This implies that the third and fourth terms cancel each other and we have,
\begin{eqnarray}
\frac{dn}{dt} \ = \ -3Hn \ - \ \langle\sigma_{eff}\text{v}\rangle(n^2 - n^2_{eq}),
\end{eqnarray} 
where $\langle\sigma_{eff}\text{v}\rangle$ is the thermal average of effective annihilation cross section times relative velocity.
\begin{eqnarray}
\langle \sigma \text{v}\rangle = \frac{1}{8m^4_{\chi}TK^2_2(m_{\chi}/T)} \int_{4m^2_{\chi}}^{\infty} \sigma \sqrt{s}(s-4m^2_{\chi}) K_1(\sqrt{s}/T) ds
\end{eqnarray}
We can define $Y = n/s$ and $x = m_\chi/T $ and using relation $\dot{x} \ = \ H(m)/x^2$ and equation for entropy density $\dot{s}= -3sH$, one can rewrite the Boltzmann equation as follow,
\begin{eqnarray}
\frac{dY}{dx} = -\frac{xs\langle\sigma_{eff}\text{v}\rangle}{H(m_{\chi})}(Y^2 - Y^2_{eq})
\end{eqnarray}
where $H(m_{\chi})=1.67\sqrt{g_\star}\frac{m^2_{\chi}}{M_{Pl}}$ with $g_{\star}=\sum_{i=\text{bosons}} g_i \frac{T^4_i}{T^4} + \frac{7}{8} \sum_{i=\text{fermions}} g_i \frac{T^4_i}{T^4} $. The above equation can be solved numerically. 
\begin{figure}[htbp]
\centering
\includegraphics[width=12cm,height=8.5cm]{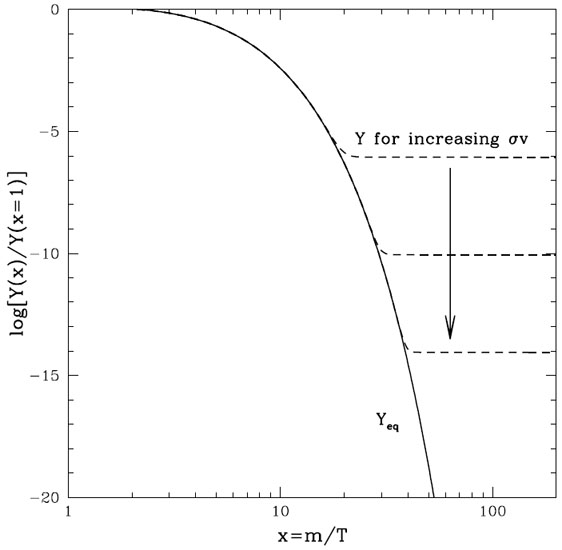}
\caption{The comoving number density (Y$_{\chi}$) as a function for three different $\langle \sigma \text{v}\rangle$ is shown. Figure has been taken from~\cite{Garrett_2011}}
\label{relic1}
\end{figure}
The final DM abundance or relic density, is given by:
\begin{eqnarray}
\Omega_{\chi}h^2  = \frac{\rho_{\chi}}{\rho_{cr}} = \frac{h^2 m_{\chi}s_0Y_0 }{\rho_{cr}} \ \approx \ 0.1 \left(\frac{0.01}{\alpha}\right)^2 \left(\frac{m_{\chi}}{100~\text{GeV}}\right)^2 
\end{eqnarray}
where $\rho_{cr}(t)=3 H(t) M^2_{Pl}$ is critical density and we have taken $\langle\sigma \text{v}\rangle \sim \alpha^2/m^2$, if one chooses $\alpha\sim 0.01$ and $m_{\chi}\sim 100$ GeV, which gives us the correct DM abundance today, measured by Planck and WMAP. The fact that a massive, weakly interacting particle can give the correct relic is known as the WIMP miracle.

Strongly interacting massive particles (SIMPs) is another interesting DM candidate that lies in the MeV mass range and interacts strongly with both particles~\cite{Choi:2021yps,Ho:2022erb}. These thermal DM candidates are being searched in several experiments.
\subsubsection{Non-thermal dark matter}
\label{non_thermal_DM}
These are the candidates that are not in thermal equilibrium with SM in the early universe. Many mechanisms can produce a non-thermal DM, e.g. Freeze-in~\cite{Belanger:2018ccd}, exponential production~\cite{Bringmann_2021}, misalignment mechanism~\cite{Di_Luzio_2020}, gravitational production, etc.

Freeze-in is a popular mechanism to produce DM non-thermally in the early universe. It starts with a negligible abundance of DM in the early universe, but it grows and freezes in via a tiny coupling with the bath particles from the hidden sector~(dark sector). DM mass or coupling can be tuned to have the correct relic. Such DM candidates are called feebly interacting massive particles (FIMPs). A FIMP candidate is hard to detect in the lab due to its tiny coupling with bath particles~\cite{Belanger:2018ccd}.

Some popular non-thermal DM candidates lie in all mass ranges, e.g. fuzzy dark matter or Ultra-light dark matter~(ULDM) having mass $\sim 10^{-20}$ eV~\cite{Ferreira_2021}, QCD axions with mass $\sim 10^{-6}$ eV~\cite{Di_Luzio_2020}, KeV sterile neutrinos~\cite{Boyarsky_2019}, Klauza-Klein particle~(extra-dimensional theories)~\cite{Hooper_2007}, Massive Compact Halo Objects~(MACHOs)~\cite{Alcock_1998} etc.

In particular, MACHOs are astrophysical objects like isolated planets, brown dwarfs (failed stars), neutron stars, black holes, etc. These objects have been ruled out by a survey done by the MACHO collaboration and the EROS-2 group using the microlensing technique~\cite{Alcock_1998}.
\section{Dark matter detection}
\label{detection}
In this section, we describe the most popular methods to detect DM.
\subsection[Direct detection]{Direct detection}
This method was proposed by Goodman and Witten and developed by Drukier and others in the mid-1980s. The key idea behind the direct detection (DD) technique is simple: if DM is omnipresent, then it must be falling on us (earth) all the time, and it may occasionally collide with the nucleus of the detector material. If our detector is sensitive enough, one can measure the energy of recoil nuclei through the kinematics of the reaction.
Assuming a WIMP DM, with mass $m_{\chi}$, hits the detector nuclei of mass $m_N$ with velocity v and q is momentum transferred to nuclei, then nuclear recoil energy is,
\begin{equation}
E_r \ = \ \frac{q^2}{2m_n} \ \approx \ 50 \ \text{KeV} \Big(\frac{m_{\chi}}{100~\text{GeV}}\Big)^2\Big(\frac{100~\text{GeV}}{m_N}\Big)
\end{equation}
Here we have substituted $q \sim m_{\chi} \text{v}$ and $\text{v} \sim 10^{-3}$. A direct detection signal from a few hundred GeV DM is of the order of KeV energy. Therefore, to detect such a signal, DD experiments are usually protected from radioactive or cosmic ray backgrounds.

The interaction of WIMP with the detector material can be classified into two categories:
\begin{itemize}
\item Elastic and inelastic scattering: When WIMP interacts with the nucleus as a whole, it causes the nucleus to recoil, and the change in energy is of the order of KeV. It is elastic scattering, whereas in the inelastic case, the nucleus is excited and releases gamma photons (MeV). Therefore, the two events can be easily separated in the detector module.
\item Spin-dependent and spin-independent scattering: If DM's spin couples with the spin contents of the nucleon, then such scattering events are called spin-dependent (SD) scattering; otherwise, spin-independent (SI) scattering. SI cross section is usually larger than SD due to coherence and therefore can be measured easily.
\end{itemize}
Recoil events due to elastic scattering can be categorised as follows:
\begin{itemize}
\item Vibration: Recoil nuclei can cause a vibration in the lattice by its movement, which can be detected as a rise in temperature by placing a highly sensitive thermometer
\item Ionization: An incident particle can ionize the atoms of the detector material, which creates free electrons in the material that can be detected by placing an electric field.
\item Excitation: Another possibility is that an incident particle excites the atom of the detector material, which spontaneously emits photons that can be collected in a photomultiplier tube and converted into an electric signal for analysis.
\end{itemize} 
The interesting quantity here is the scattering rate of DM particles to the nuclear target, which is given by
\begin{equation}
\frac{dR}{dE_{r}} \ = \ \frac{\rho_{\chi}}{m_{\chi}m_N} \int_{\text{v}_{min}}^{\text{v}_{max}}d^3\text{v} \ \text{v} \ \tilde{f}(\mathbf{v},t)\frac{d\sigma}{dE_r}
\end{equation}
here $\tilde{f}(\mathbf{v},t)$ is the DM velocity distribution in the lab frame, $\text{v}_{min}$ is the minimum velocity needed to cause a nucleus to scatter with energy $E_r$, and $\text{v}_{max}$ is the escape velocity. $\frac{d\sigma}{dE_r}$ is the differential cross section and can be determined from the model. $\rho(x)$ is the density profile for DM, which can be determined from astrophysical observations.

\subsubsection[Density profile]{Density profile}
\label{density}

The Standard Halo Model~(SHM)~\footnote {See, note by M.Lisanti~\cite{Lisanti:2016jxe}} is the simplest model to describe the distribution of DM in our galaxy. It assumes that DM is distributed smoothly, homogeneously, and isotropically.
If f(x,v) is the distribution function, then the DM density is given by
\begin{equation}
\rho(x) = \int d^3\text{v} f(x,\text{v})
\label{rho} 
\end{equation}
In SHM, the distribution function f(x,v) is given by
\begin{equation}
f(x,\text{v})= \frac{\rho_0}{(2\pi\sigma^2)^{3/2}} \exp\left(\frac{\psi - \text{v}^2/2}{\sigma^2} \right)
\label{distfunc}
\end{equation}
Here, $\rho_0$ and $\sigma$ are constants. $\psi$ is the potential field and related with $\rho$ via Poisson equation.
\begin{eqnarray}
\nabla^2\psi \ = \ -4\pi G\rho 
\label{poission}
\end{eqnarray}
Solving equations ~\ref{rho}, ~\ref{distfunc}, and ~\ref{poission}, one can get.
\begin{eqnarray}
\rho(r) = \frac{\sigma^2}{2{\pi}Gr^2}
\end{eqnarray}
Similarly, the distribution function can be found,
\begin{equation}
f(x,\text{v}) \propto \exp\left(\frac{- \text{v}^2}{2\sigma^2} \right)
\label{distfunc1}
\end{equation}
which is a well-known Maxwell-Boltzmann~(MB) distribution, and the velocity is proportional to a constant $\sigma$. Thus, SHM predicts pretty well the observed DM distribution. \\
However, galaxies are always merging with some small or big astrophysical objects, which perturbs the galactic disk and distorts the current structure; therefore, the smooth isotropy assumption of the DM profile does not fit well. N-body simulations have shown a significantly different density profile from the MB distribution. \\
Some popular DM density profiles are commonly used in studies, such as the NFW profile,
\begin{equation}
\rho_{NFW}(r) \ = \ \frac{\rho_0}{r/r_s(1+r/r_s)^2}
\end{equation}
where $r_s = 20 kpc$ and $\rho_0=0.259$. Another commonly used density profile is the Einasto profile, given by
\begin{equation}
\rho_{Einasto}(r) \ = \ \rho_{0}\Big{[}-\frac{2}{\gamma}\left(\left(\frac{r}{r_s}\right)^{\gamma}-1 \right)\Big{]}
\end{equation}   
Here, $\gamma = 0.17$, $r_s = 20 kpc$, and $\rho_0 = 0.061$.
\subsubsection{Effective Interaction}
Any DM model can have the following effective interaction between DM~($\chi$) and SM quarks~(q),
\begin{equation}
\mathcal{L}_{eff} \ = \ g \ (\bar{\chi}\Gamma_{\small{\chi}}\chi) \ (\bar{q}\Gamma_{\small{q}} q)
\end{equation}
Here g is an effective coupling and the nature of the interaction: $ \Gamma \ = \ \lbrace I, \gamma^{\mu}, \gamma^{5}, \gamma^{\mu}\gamma^{5}, \sigma^{\mu\nu}\rbrace $. Based on these operators, one can determine the scattering rate. The SI differential scattering cross section has the following generic form:
\begin{equation}
\frac{d\sigma}{dE_r} \ = \ \frac{2m_N}{\pi \text{v}^2}[Zf_p + (A-Z)f_n]^2F^2(k).
\end{equation}
Here $f_p$ and $f_n$ are the proton and neutron fractions, and F(k) is the nucleon form factor. One can see that the SI differential cross section is proportional to $A^2$ for $f_p \ = \ f_n$. However, if the couplings are momentum-dependent or interaction-dependent and $\Gamma = \ \gamma^{\mu}\gamma^5$ have the following form, then one has a spin-dependent cross-section, which is challenging to observe experimentally.
  
There are many experiments dedicated to direct searches, such as LUX~\cite{LUX:2016ggv}, XENON1T~\cite{XENON:2018voc}, XENONnT~\cite{Aprile:2020vtw,Aprile_2023}, LUX-ZEPLIN~(LZ)~\cite{Akerib:2018lyp,LZ:2022lsv}, etc. These searches have put very tight constraints on the DM-nucleon cross-section. However, the above-mentioned experiments are not sensitive to DM mass below a few GeV.

\begin{figure}[htbp]
\centering
\includegraphics[width=0.55\textheight]{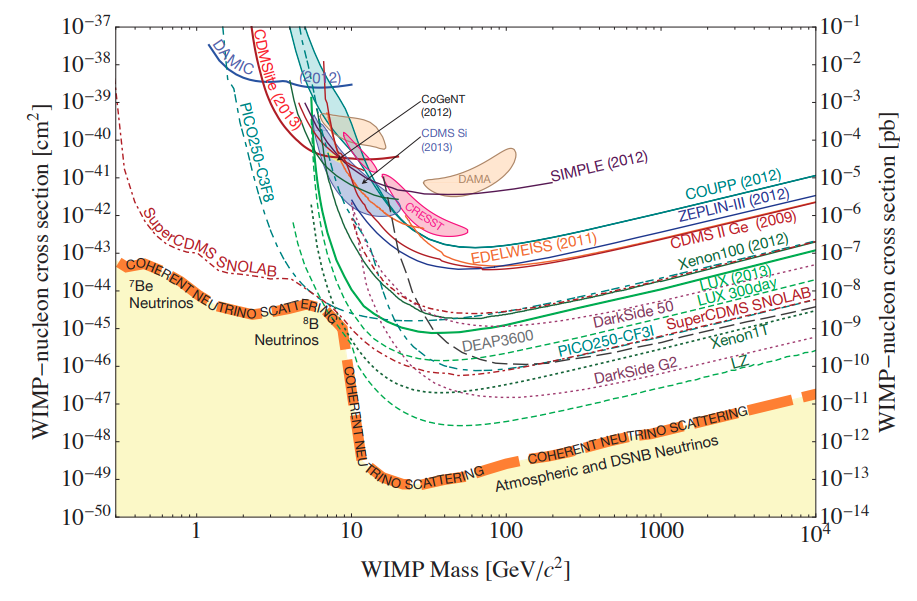}
\caption{Upper bounds on DM-nucleus cross-section as a function of DM mass by several experiments. Figure has been taken from~\cite{Cooley_2014}}
\label{DD}
\end{figure}

In figure~\ref{DD}, the limit on DM~-~nucleon cross-section as a function of DM mass is shown. The strong bounds are by XENON1T. New limits from LZ~\cite{LZ:2022lsv} and XENONnT~\cite{Aprile_2023} have further put strong bounds on the DM-Nucleon scattering cross-section. These bounds get weaker at low masses due to the energy threshold of the experiments. The orange region is a neutrino background~\cite{O_Hare_2021}. One can see that there is a very small parameter space left for searches in the GeV–TeV mass range.
\subsection{Indirect detection}
\begin{figure}[t!]
\centering
\includegraphics[width=15cm,height=12cm]{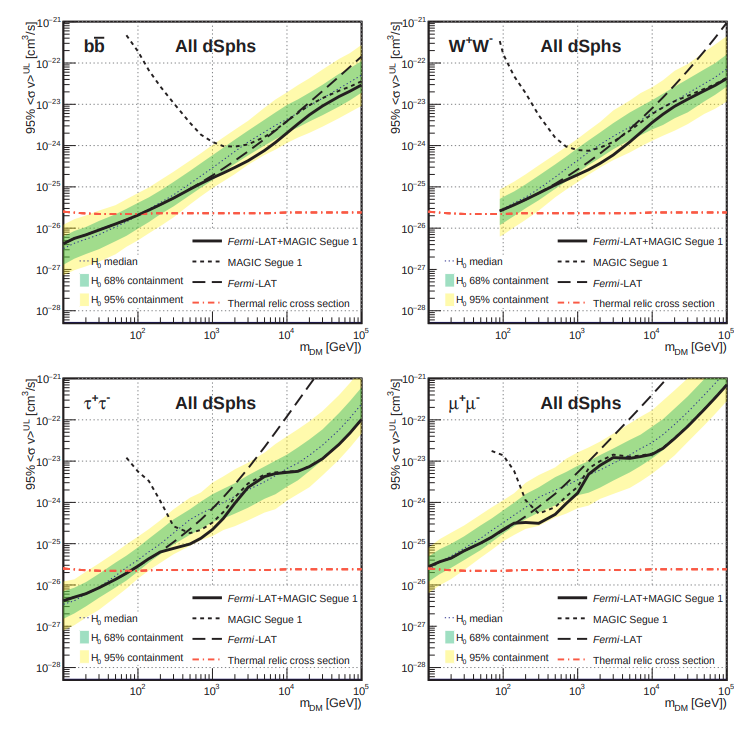}
\caption{Upper bounds on the thermal average annihilation cross section for DM annihilate to $b\bar{b}, W^+W^-, \tau\bar{\tau}, \mu\bar{\mu}$ states as a function of DM mass is shown. Figure has been taken from~\cite{2016}}
\label{idd}
\end{figure}

The indirect detection method searches for DM annihilation products such as $\gamma, e^{\pm}, \nu, \bar{\nu}$, etc. in DM-rich places like Earth, Sun, Galaxies, etc. Here, it is assumed that the DM couples with electrons, photons, or neutrinos via an effective operator. Photon and neutrino signals are heavily searched for in the detection of dark matter. Photon can be produced via direct annihilation or cascade channels, which is being searched in experiments such as HESS~\cite{HESS:2018cbt}, FermiLAT~\cite{Ackermann:2013uma} collaboration, etc. Neutrino experiments, e.g., ICECUBE~\cite{PhysRevLett.102.201302}, ANTARES~\cite{AGERON201111}, etc., also search for DM signals. An unknown 3.5 keV X-ray signal from a galaxy cluster could have arisen from a keV sterile neutrino dark matter mixing with active neutrinos~\cite{De_Luca_2018}. Experiments such as PAMELA~\cite{PAMELA:2008gwm} and AMS~\cite{Elor:2015bho} are searching for positrons and other charge particle signals. Figure~\ref{idd} shows the current limit on DM annihilation into $b\bar{b}, W^+W^-, \tau\bar{\tau}, \mu\bar{\mu}$ final states~\cite{2016}.
\subsection{Dark matter at colliders}
Dark matter can be produced at colliders such as the LHC, which is the largest particle collider at CERN. DM can be searched here by using two different strategies. One can start searching for an ultraviolet (UV) complete model, e.g., SUSY or GUT (Grand Unified Theory), or search for a new physics signal that arises from an effective field theory (EFT) operator or extra-dimensional theory. However, no candidate for DM has been found. A DM candidate might exist at a much higher scale of energy than the LHC current run at 13 TeV. It is also possible that DM interacts very weakly with SM; therefore, it has not been produced with the current luminosity of the LHC~\cite{Profumo:2009tb,Bauer:2017qwy}.
%
%%%%%%%%%%%%%%%%%%%%%%%%%%%%
\section{Outline of the thesis}
%%%%%%%%%%%%%%%%%%%%%%%%%%%%%
To explain DM, we introduce a few simple WIMP models that can bypass present constraints and have observable signatures in the future. The main challenge in any WIMP-like DM model is to explain relic density and stringent direct detection bounds simultaneously. In later chapters, we describe a few models to counter these challenges.

In chapter~\ref{chap_ALP_RHN}, we study a model where we extended SM via three right-handed neutrinos~(RHNs) and one axion-like particle~(ALP) field. We introduce all-dimensional five couplings between ALP, SM, and RHN fields. RHNs and active neutrinos mix following Type I seesaw structure~(see Appendix \ref{seasaw}). We introduce a $\mathbb{Z}_2$ symmetry under which one of the RHN is odd and all other particles are even. $\mathbb{Z}_2$ odd RHN can act as a DM candidate, while ALP acts as a mediator between SM and DM. We use all the available constraints on masses and couplings on ALP and RHNs to analyze relic density bounds on parameter space. Direct detection bounds can be easily satisfied due to the pseudo-scalar nature of the mediator. Lastly, we checked for a di-photon signal from the model and compared it with experimental bounds by Fermi-LAT, HESS, AMS, MAGIC, etc.

In chapter~\ref{chap_U1X}, we study a generic $U(1)_X$ extension of the SM, which includes three RHNs and two additional scalars. The model is anomaly-free. Two $\mathbb{Z}_2$ symmetry groups are introduced in the model. The lightest RHN is odd under one $\mathbb{Z}_2$ group, and all other fields are even. Similarly, one of the additional scalars is odd under second $\mathbb{Z}_2$. Hence, both scalars and fermions can act as DM in this model. Then we study the theoretical and experimental constraints on the model, which we later employed in analyzing relic density constraints on both DM parameter spaces. We study the direct detection constraints while respecting the relic density bound. Lastly, we study which $U(1)_X$ model is favorable for DM study by checking $U(1)_X$ charge dependence on the relic.

In this chapter~\ref{chap_U1X2}, we have the particle and symmetry content similar to our previous model described in chapter~\ref{chap_U1X}, but with no Z2 symmetry. The U(1)$_X$ symmetry breaking results in a massive pseudo-scalar particle, which is the proposed candidate for dark matter. We study the parameter space where this massive pseudo-scalar particle has a lifetime much larger than the age of the universe. We constrain the allowed parameter space by the lifetime bound using perturbative unitarity, invisible Higgs width, and relic density bounds. The main advantage of having a pseudo-scalar DM is that it implies a vanishing scattering amplitude for direct detection at tree level in a non-relativistic limit. However, the one-loop level contribution could be finite, which we have studied and plotted against the current direct detection bounds.

%%%%%%%%%%%%%%%%%%%%%%%%%%%%%%%%%%%%%
\chapter{Dark matter through ALP portal}
\label{chap_ALP_RHN}
%%%%%%%%%%%%%%%%%%%%%%%%%%%%%%%%%%%%%%%
This chapter deals with fermion dark matter interacting with SM via the ALP portal. The results are based on the work: Shivam Gola, Sanjoy Mondal, and Nita Sinha, "ALP portal majorana dark matter, Int.J.Mod.Phys.A 37 (2022) 22, 2250131''.

\section{Introduction}

%%%%%%%%%%%%%%%%%%%%%%%%%%%%%%%%%%%%%%%%%%%%%%%%%%%%%%%%%%%%%%%%
The discovery of neutrino oscillations confirming the existence of at least two non-vanishing neutrino mass-squared differences necessitates physics beyond the Standard Model (BSM). In principle, neutrino mass could be simply generated by the addition of right-handed neutrinos (RHNs) to the SM particle content. These RHNs interact with SM fields via mixing with active neutrinos. Since RHNs are SM singlets, they allow the Majorana mass term along with the usual Dirac mass term. This is known as type-I seesaw mechanism~\cite{Minkowski:1977sc,Schechter:1980gr,Mohapatra:1979ia,Schechter:1981cv}. The mass of these RHNs could range from eV to GUT scale, depending on the models~\cite{Dorsner:2006fx,Bajc:2007zf,deGouvea:2006gz,deGouvea:2007hks}. RHNs can also play the role of warm dark matter (WDM), which is singlet under the SM gauge symmetry and has tiny mixing with the SM neutrinos, leading to a long lifetime~\cite{Abada:2014vea,Borah_2016,Das:2019kmn}. Also, KeV-scale RHN has been studied as a viable DM candidate~\cite{Merle:2017jfn,Adhikari:2016bei,Abada:2014zra}. In this work, we have instead focused on the prospects of having GeV-scale RHNs as WIMP DM candidates.

%%%%%%%%%%%%%%%%%%%%%%%%%%%%%%%%%%%%%%%%%%%%%%%%%%%%%%%%%%%%%%%%%%%%%%
Axion~\cite{Peccei:1996ax} was postulated in the Peccei-Quinn (PQ) mechanism to solve the strong CP problem~\cite{Peccei:2006as,Kim:2008hd,Hook:2018dlk,Lombardo:2020bvn} of quantum chromodynamics (QCD). This axion can be identified as a (pseudo) Nambu-Goldstone boson associated with the spontaneous breaking of the $U(1)_{\text{PQ}}$ global symmetry~\cite{Weinberg:1977ma,Wilczek:1977pj,Berezhiani:1989fp}. This QCD axion gets a tiny mass from the explicit breaking of this global symmetry due to a QCD anomaly. Astrophysical and experimental searches have not favoured the PQ model. To resolve issues with the PQ model, other popular solutions like KSVZ~\cite{Kim:1979if,Shifman:1979if}, DFSZ~\cite{Dine:1981rt}, etc. invoking axion were also proposed and studied afterward. The magnitude of the couplings of axions to ordinary matter is inversely proportional to the axion decay constant $f_a$, which is associated with the $U(1)_{\text{PQ}}$ symmetry breaking scale. Hence, the couplings are highly suppressed if $f_a$ is sufficiently large, and these features make the axion suitable to be a DM candidate. Many BSM extensions that feature spontaneously broken global U(1) symmetry predict massless Nambu-Goldstone bosons whose couplings are not constrained, unlike the original QCD axion. These kinds of particles are known as axion-like particles (ALPs). The mass of these ALPs is not related to their symmetry-breaking scale, unlike the PQ axion. In general, they are not supposed to solve the strong CP problem, but with the introduction of Planck scale operators, they could solve the strong CP problem~\cite{Hook:2019qoh,Kelly:2020dda}. Here we will consider the most general $SU(2)_L\otimes U(1)_Y$ invariant formulation of ALP interactions developed in Refs.~\cite{Georgi:1986df,Brivio:2017ije,Salvio:2013iaa}. This generic effective ALPs Lagrangian allows ALPs to couple with all the SM gauge bosons as well as with all the SM fermions. In addition to this effective ALP Lagrangian, we introduce three RHNs that can generate light neutrino masses through a type-I seesaw. We invoke a $\mathbb{Z}_2$ symmetry under which all SM and BSM particles are even except the lightest RHN. Further, we introduced the RHN-ALP coupling and showed that the lightest RHN~(Singlet majorana fermion), which is odd under $\mathbb{Z}_2$, can play the role of DM candidate. Note that the phenomenology of this ALP-mediated DM~\cite{Hochberg:2018rjs} will be similar to pseudoscalar-portal DM. Also, DM interacting via the exchange of a light ALP can induce observable signals in indirect detection experiments while evading the strong bounds of direct DM searches. Note that ALPs with mass $M_a\sim\mathcal{O}(\text{GeV})$ may also show up at colliders~\cite{Mimasu:2014nea,Jaeckel:2015jla,Alves:2016koo} or in rare meson decays~\cite{Dolan:2014ska,Izaguirre:2016dfi,CHOI1986145}.

Chapter~\ref{chap_ALP_RHN} is organised as follows: In Sec.~\ref{sec:Model}, we introduced our model, detailing the new interactions present. In Sec.~\ref{sec:Constraints}, we summarise the existing constraints on ALP parameter space coming from various observable and collider searches. In Sec.~\ref{sec:dark-matter}, we have explored and discussed the feasible parameter space coming from DM analyses, such as relic density and direct and indirect detection. Finally, we give our summary in Sec.~\ref{summary_alp}.

%%%%%%%%%%%%%%%%%%%%%%%%%%%%%%%%%%%%%%%%%%%%%%%%%%%%%%%%%%%%%%%%%%
\section{Model}
\label{sec:Model}
%%%%%%%%%%%%%%%%%%%%%%%%%%%%%%%%%%%%%%%%%%%%%%%%%%%%%%%%%%%%%%%%%%

The model that we consider is the minimal combination of type-I seesaw and effective ALP interaction with additional $\mathbb{Z}_2$ symmetry apart from the SM gauge symmetry~\cite{Salvio:2021puw,Salvio:2015cja}. The matter content of the model is shown in Table.~\ref{tab:quantun-numbers}. 

\begin{table}[!h]
\centering
\begin{tabular}{|c||c|c|c|c|c|c||c|c||c|c|}
\hline
        & \multicolumn{6}{|c||}{Standard Model} &  \multicolumn{2}{|c||}{New Fermions}  & \multicolumn{1}{|c|}{New Scalar}  \\
        \cline{2-10}
        & \hspace{0.2cm} $\ell_L$  \hspace{0.2cm} &  \hspace{0.2cm} $e_R$ \hspace{0.2cm} & \hspace{0.2cm} $q_L$ \hspace{0.2cm} & \hspace{0.2cm} $u_R$ \hspace{0.2cm} & \hspace{0.2cm} $d_R$  \hspace{0.2cm} &  \hspace{0.2cm}$H$ \hspace{0.2cm}  &  \hspace{0.2cm} $N_1$ \hspace{0.2cm} &  \hspace{0.1cm} $N_{2,3}$  \hspace{0.1cm}  &  \hspace{0.01cm}$a$   \hspace{0.01cm}\\
\hline     
% types & 3 & 3 & 1 & 1 & 1 & 1  \\                            
$SU(2)_L$ &  2    &  1  & 2  & 1 & 1  &    2    &     1    &    1       \\
\hline
$U(1)_Y$  & -1/2    &  -1 & 1/6  & 2/3  & -1/3    &    1/2    &     0    &  0   &   0    \\
\hline
$\mathbb{Z}_2$   &  $+$  &  $+$  & $+$ &  $+$ & $+$  &  $+$  &  $-$   & $+$  &  $+$     \\
\hline
\end{tabular}
\caption{Matter content and charge assignment of the considered model.}
\label{tab:quantun-numbers}
\end{table}

Let's first briefly discuss the features of a generic ALP Lagrangian. We extend the SM particle content by adding ALP, which is a singlet under SM charges and is a pseudo-Nambu-Goldstone boson of a spontaneously broken symmetry at some energy, which is higher than the electroweak scale $v$. In effective theory, the operators will be weighted by powers of $a/f_a$, where $f_a$ is the scale associated with the physics of the ALP, $a$. Effective linear Lagrangian with one ALP has already been discussed in great detail in Ref.~\cite{Georgi:1986df,Brivio:2017ije}. For linear EWSB realisations, the most general linear bosonic Lagrangian involving $a$ is given by,
%%%%%%%%%%%%%%%%%%%%%%%%%%%%%%%%%%%%%%%%%%%%%%%%%%%%%%%%%%%%%%%%%%%%
\begin{align}
\mathcal{L}  =  \mathcal{L}_{\text{SM}} +  \mathcal{L}_{\text{ALP}}
\end{align}
%%%%%%%%%%%%%%%%%%%%%%%%%%%%%%%%%%%%%%%%%%%%%%%%%%%%%%%%%%%%%
where the leading order effective Lagrangian $\mathcal{L}_{\text{SM}}$ is same as the SM one and with
%%%%%%%%%%%%%%%%%%%%%%%%%%%%%%%%%%%%%%%%%%%%%%%%%%%%%%%%%%%%
\begin{align}
\mathcal{L}_{\text{ALP}} \ &= \ \frac{1}{2}\partial_{\mu}a \partial^{\mu}a \ - \ \frac{1}{2} M_a^2 a^2  -  \frac{C_{\tilde{G}}}{f_a} \ a G_{a\mu\nu}\tilde{G}^{\mu\nu a} -  \frac{C_{\tilde{B}}}{f_a} \ a B_{\mu\nu}\tilde{B}^{\mu\nu} - \frac{C_{\tilde{W}}}{f_a} \ a W_{a\mu\nu}\tilde{W}^{a\mu\nu} \nonumber \\
& + i C_{a\Phi}\times \big[(\bar{Q}_L Y_U \tilde{H}u_R - \bar{Q}_L Y_D H d_R - \bar{L}_L Y_E H e_R )\frac{a}{f_a} + h.c. \big] 
\end{align}
%%%%%%%%%%%%%%%%%%%%%%%%%%%%%%%%%%%%%%%%%%%%%%%%%%%%%%
Here H and $a$ are the Higgs and ALP fields, respectively. $C_{i}$, where $i=G,B,W,a\Phi$, are the corresponding Wilson coefficients for ALP-gauge boson and ALP-matter interactions. Parameters $M_a$ and $f_a$ are the ALP mass and energy scales associated with ALP physics. $Y_D$, $Y_U$, and $Y_E$ are couplings for down-type quarks, up-type quarks, and charged leptons, respectively. We see that the ALP Lagrangian has a much more generic form than that of the QCD axion, where the mass of ALP is not restricted by the new physics scale. The remaining fields and parameters are the same as in SM. References \cite{alves2020probing, Brivio:2017ije, Atre:2009rg} have discussed the phenomenology of the various pieces of the model. Now let's introduce the type-I seesaw Lagrangian with three additional SM singlet RHNs:
%%%%%%%%%%%%%%%%%%%%%%%%%%%%%%%%%%%%%%%%%%%%%%%%
\begin{align}
\mathcal{L}_{\text{RHN}} \ =  \ i\sum^3_{i=1}\bar{N}_i\gamma^{\mu}\partial_{\mu} N_i \ - \sum^3_{j=2} Y_{\alpha j} \bar{L}_\alpha \tilde{H} N_j  \ - \ \sum^3_{i,j=2}M_{ij} \bar{N}^c_i N_j \ - \ M_{N_1} \bar{N}^c_1 N_1 \ + \ \text{h.c.} 
\end{align}
%%%%%%%%%%%%%%%%%%%%%%%%%%%%%%%%%%%%%%%%%%%%%%
where $N_i$ are the SM singlet RHNs, $Y_{\alpha j}$ is the Dirac Yukawa coupling, and $M_{ij}$ is the Majorana mass term. As $N_1$ is odd under $\mathbb{Z}_2$ symmetry, the Dirac type of Yukawa interaction is forbidden for it, unlike that for $N_{2,3}$. As a result of this, one of the light neutrinos will remain massless, and we have enough parameters to describe the neutrino oscillation data. In addition to this, $\mathbb{Z}_2$ symmetry stabilises the $N_1$, and it can play the role of DM candidate if one allows the following ALP-RHN interaction:
%%%%%%%%%%%%%%%%%%%%%%%%%%%%%%%%%%%%%%%%%%%%%%%%%%%
\begin{align}
\mathcal{L}_{\text{ALP-RHN}} \  =  - \sum^3_{i=1} \frac{C_{aN_{i}}}{f_{a}} (\bar{N_{i}}\gamma^{\mu}\gamma^{5}N_i) \partial_{\mu}a,
\end{align}
%%%%%%%%%%%%%%%%%%%%%%%%%%%%%%%%%%%%%%%%%%%%%%%%%%%%%%%%%%%%%%%%
where $C_{aN_i}$ denotes the ALP-RHNs Wilson coefficient, and through this DM particle, $N_1$ can communicate with the ALP sector. We call this an ALP-portal RHN DM. We consider ALP mass to be in the order of a few hundreds of MeV to GeV. We have considered $N_1$ to have a mass of up to a few TeV. In the next section, we discuss the allowed range for the several parameters of the model from the various phenomenological bounds.
%%%%%%%%%%%%%%%%%%%%%%%%%%%%%%%%%%%%%%%%%%%%%%%%%55
\section{Existing constraints on ALP parameter space}
\label{sec:Constraints}
%%%%%%%%%%%%%%%%%%%%%%%%%%%%%%%%%%%%
Before diving into the details of DM analyses, let us first recall the existing experimental bounds on the couplings of ALPs to gluons, photons, fermions, and also from collider searches with $f_a\sim\mathcal{O}(1\,\text{TeV})$~\cite{Mimasu:2014nea,Jaeckel:2015jla,Dolan:2014ska,Izaguirre:2016dfi,Agashe:2014kda,Vinyoles:2015aba,Raffelt:2006cw,Friedland:2012hj,Ayala:2014pea,Khachatryan:2014rra,Aad:2015zva,Krnjaic:2015mbs,Clarke:2013aya,Aprile:2014eoa,Viaux:2013lha,Rodriguez-Tzompantzi:2020kuc}. ALP photon coupling is the primary parameter through which astrophysical and cosmological bounds are set on such particles. The ALP and photon coupling can be deduced
%%%%%%%%%%%%%%%%%%%%%%%%%%%%%%%%%%%%
\begin{equation}
\mathcal{L}_{a\gamma} = - C_{a\gamma} aF_{\mu\nu}\tilde{F}^{\mu\nu},\,\,\text{with}\,\,C_{a\gamma}= \frac{(C_{\tilde{B}}\cos^2\theta_w + C_{\tilde{W}}\sin^2\theta_w)}{f_a} 
\end{equation}
Particularly experiments like CAST, using the primakoff process, constrained the parameter space~($M_a-C_{a\gamma}$) heavily for $M_a$ smaller than few eV~\cite{Anastassopoulos:2017ftl}. For $M_a\sim 1$~MeV, the best present constraint comes from beam dump experiments: $C_{a\gamma}/f_a \leq 10^{-5}\,\text{GeV}^{-1}$~\cite{Mimasu:2014nea}~\footnote{For very low masses, tighter constraint exist, but we do not discuss them here as we are interested in mass range $M_a\sim\text{MeV-few GeV}$.}. A slightly higher mass range is constrained by collider experiments such as LEP and LHC~\cite{Mimasu:2014nea}. In LEP, the process like $e^-e^+\to\gamma a \to 3\gamma$ is being analysed to constrain the coupling, whereas in LHC, it is $pp \to \gamma a$ that searches for mono, di, or tri photon signals. All these constraints are described in the references~\cite{Bauer:2018uxu,Brivio:2017ije,vinyoles2015new}. A limite on axion gauge boson coupling $C_{\tilde{W}}/f_a < 10^{-5}$ GeV$^{-1}$ is obtained for $0.175 \le M_a \le 4.78$ GeV by analyzing the process $B^{\pm} \rightarrow K^{\pm} a, a \rightarrow \gamma\gamma$ at the BABAR experiment~\cite{BaBar:2021ich}. The constraint on $C_{\tilde{G}}$ is set by monojet 8 TeV LHC analysis. Unlike $C_{a\gamma}$, here it is much more complicated to put a bound on $C_{\tilde{G}}$ due to the large number of diagrams involved in the process. However, the dominant diagram was found to be $gg\to ag$, which is complicated due to hadronization that leads to jets in the final state. These constraints have been analysed in the reference~\cite{Mimasu:2014nea,Khachatryan:2014rra}. The limit from this study reads as $C_{\tilde{G}}/f_a \leq 10^{-4}\,\text{GeV}^{-1}$ for $M_a\leq 0.1\,\text{GeV}$. Also, the bound on $\text{BR}(K^{+}\to\pi^{+}+\text{nothing})$~\cite{Adler:2004hp} can be used to constrain the process $K^{+}\to\pi^{+}+\pi^0(\pi^0\to a)$, which can be reinterpreted in terms of ALP-gluon coupling $C_{\tilde{G}}$, yielding $C_{\tilde{G}}/f_a \leq 10^{-5}\,\text{GeV}^{-1}$ for $M_a\leq 60\,\text{MeV}$. Constraints on the ALP matter coupling $C_{a\phi}$ are studied less compared to gauge boson coupling; however, several processes involving flavours have been studied to put bounds on it~\cite{Dolan:2014ska}. The constraints on ALP-fermion coupling $C_{a\Phi}$ depend on the ALP mass. The higher mass range is tested through rare meson decays. Rare meson decay at beam dump experiments~(CHARM) sets tight constraints on $C_{a\Phi}$ for mass range $1\,\text{MeV}\leq M_a\leq 3\,\text{GeV}$ as $C_{a\Phi}/f_a<(3.4\times 10^{-8}-2.9\times 10^{-6}) \ \text{GeV}^{-1}$~\cite{Bergsma:1985qz}. For our interested mass range, we summarise the existing tightest constraints in Table.~\ref{tab:constraint}. 
%%%%%%%%%%%%%%%%%%%%%%%%%%%%%%%%%%%%%%%%%%%%%%%%%%%%%
\begin{table}[h]
\begin{tabular}{|c|c|c| }
 \hline
  Bound on Coupling  & ALP Mass Range  &  Observable \\
 \hline
 $\frac{C_{\tilde{G}}}{f_a}  \leq 10^{-4}$ GeV$^{-1}$   &   $M_a \leq 0.1$~GeV  &  mono-jet 8 TeV@LHC\\
$\frac{C_{a\gamma}}{f_a}  \leq 10^{-5}$ GeV$^{-1}$   &   $M_a \sim$ 1 MeV   &  Beam Dump\\
 $\frac{C_{\tilde{W}}}{f_a} \sim 10^{-5}$ GeV$^{-1}$   &  0.175 GeV $ \le M_a \le $ 4.78 GeV   &  BABAR Exp.\\ 
 $\frac{C_{a\Phi}}{f_a} \sim  10^{-8} - 10^{-6}$ GeV$^{-1}$   &  1 MeV $< M_a <$ 3 GeV   &  Rare meson decay\\
\hline
\end{tabular}
\caption{Summary of existing constraint on ALP couplings.}
\label{tab:constraint}
\end{table}
%%%%%%%%%%%%%%%%%%%%%%%%%%%%%%%%%%%%%%%%%%%%%%%%%%%%%%%%%

\underline{\bf{Benchmark:}} To start analyzing the model, we choose the ALP mass ($M_a$ = 10 GeV). ALP-Gauge boson couplings $C_{\tilde{B}}=C_{\tilde{W}}$, since ALP is mostly constrained through $C_{a\gamma}, C_{\tilde{G}}$ and we also choose $\frac{C_{\tilde{G}}}{f_a}, \frac{C_{a\gamma}}{f_a}$ of the order of $10^{-4}$ GeV$^{-1}$ and ALP-fermion coupling $\frac{C_{a\Phi}}{f_a} \sim 10^{-6}$ GeV$^{-1}$, which satisfy all of the above constraints.

\underline{\bf{ALP decay:}} ALP can decay into SM final states when kinematically accessible, such as leptons, gauge bosons~($W^{\pm},Z,\gamma,g$), and hadrons. The analytical form of these decay widths is calculated. We found that for the range of couplings considered by us, the ALP decay width is small enough to use the narrow width approximation.

\begin{align*}
& \Gamma_{agg}= \frac{2 C_{\tilde{G}} M_a^3}{\pi f_a^2},\ \ \
 \Gamma_{a\gamma\gamma}=\frac{M^3_a(C_{\tilde{B}}\cos^2{\theta_w} + C_{\tilde{W}}\sin^2{\theta_w})^2}{4\pi f_a^2},\\
& \Gamma_{aW^+W^-}= \frac{C_{\tilde{W}}^2(M_a^2 - 4 M_W^2)^{\frac{3}{2}}}{2\pi f_a^2}, \nonumber \ \ \
\Gamma_{aZZ}=\frac{(C_{\tilde{W}}\cos^2{\theta_w} + C_{\tilde{B}}\sin^2{\theta_w})^2(M_a^2 - 4 M_Z^2)^{\frac{3}{2}}}{4\pi f_a^2},\\ 
& \Gamma_{a\gamma Z}=\frac{\sin^2{\theta_w}\cos^2{\theta_w}(C_{\tilde{B}}-C_{\tilde{W}})^2(M^2_a-M^2_z)^3}{2\pi f_a^2 M^3_a}, \\
& \Gamma_{af\bar{f}}=\frac{N_c C_{a\phi}^2 M_f^2 \sqrt{M_a^2-4M_q^2}}{8\pi f_a^2}, \nonumber \ \ \
\Gamma_{aN_i}=\frac{C_{aN_i}^2 m^2_{N_{i}} \sqrt{M_a^2 - 4 m_{N_{i}}^2}}{\pi f_a^2}
\end{align*}
where $N_c=3~(1)$ for quark~(lepton).

%%%%%%%%%%%%%%%%%%%%%%%%%%%%%%%%%%%%%%%%%%%%%%%%%%%%%%%%%%%
\begin{figure}[htbp]
\includegraphics[height=5cm, width=7.2cm]{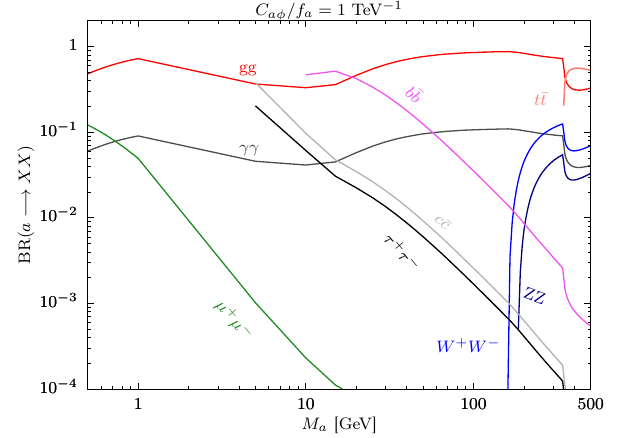}
\includegraphics[height=5cm, width=7.2cm]{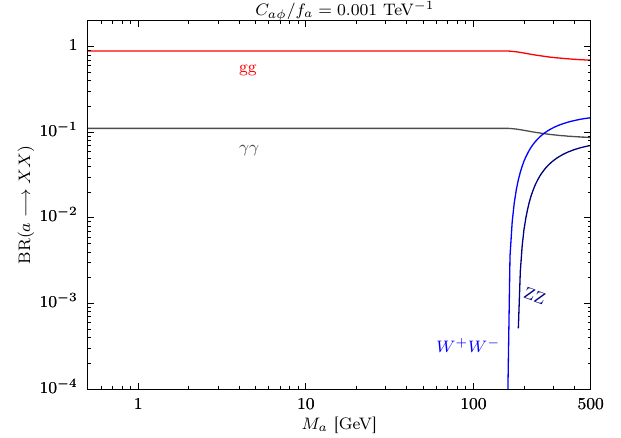}
\caption{The left and right panels show the variations of branching ratios for various decay channels concerning ALP mass ($M_a$). Different colours stand for different final states. In the right panel, we choose the smaller value of $C_{a\phi}/f_a=10^{-3}$ TeV$^{-1}$ to show that ALP decays dominantly to gauge bosons.}
\label{branching ratio}
\end{figure}
%%%%%%%%%%%%%%%%%%%%%%%%%%%%%%%%%%%%%%%%%%
In Fig.~\ref{branching ratio}, we show the various branching ratios of ALP decay to SM final states. For the left and right panels, we choose $C_{\tilde{G}}/f_a=C_{a\gamma}/f_a=0.1\,\text{TeV}^{-1},\,C_{a\Phi}/f_a=1\,\text{TeV}^{-1}$ and $C_{\tilde{G}}/f_a=C_{a\gamma}/f_a=0.1\,\,\text{TeV}^{-1},C_{a\Phi}/f_a=0.001\,\text{TeV}^{-1}$, respectively. We see from Fig.~\ref{branching ratio} that ALP mostly decays to gluon and photon pairs. Also, the $W^+W^-$ and $ZZ$ pairs contribute significantly as ALP mass crosses respective threshold values. Comparing the left panel with the right panel, one sees that lepton channels only contribute when $C_{a\Phi}$ is relatively large.

%%%%%%%%%%%%%%%%%%%%%%%%%%%%%%%%%%%%%%%%%%%%%%
\section{Dark matter analysis}
\label{sec:Dark matter}
%%%%%%%%%%%%%%%%%%%%%%%%%%%%%%%%%%%%%
So far, we have discussed the constraints on ALP couplings in the SM field. We collect the results of our analyses of DM phenomenology here. In our case, $N_1$ plays the role of DM due to $\mathbb{Z}_2$ symmetry protection. To calculate all the vertices, the model is implemented in the FeynRules package~\cite{Alloul:2013bka}. All DM observables, such as the relic abundance and DM-nucleon cross sections, are determined using micrOMEGAS~\cite{Belanger:2018ccd}, which relies on the CalcHEP~\cite{Belyaev:2012qa} model file obtained from FeynRules. In the scanning of the relevant parameters of the model, we have imposed the phenomenological bounds we discussed earlier. The mass of $N_{2,3}$ is always chosen to be greater than $N_1$ for further analysis.
%%%%%%%%%%%%%%%%%%%%%%%%%%%%%%%%%%%%%%%
\begin{figure}[htbp]
\includegraphics[height=5cm, width=14cm]{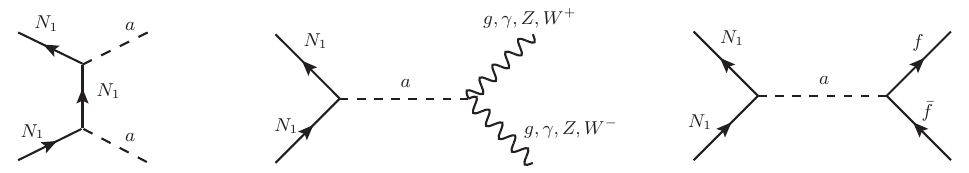}
\caption{Annihilation diagrams contributing to the relic abundance of $N_1$.}
\label{annihilation diagram}
\end{figure}
%%%%%%%%%%%%%%%%%%%%%%%%%%%%%%%%%%%%%%%

\subsection{Relic density}
\label{subsec:relic density}
\begin{figure}[htbp]
\includegraphics[height=5cm, width=7.2cm]{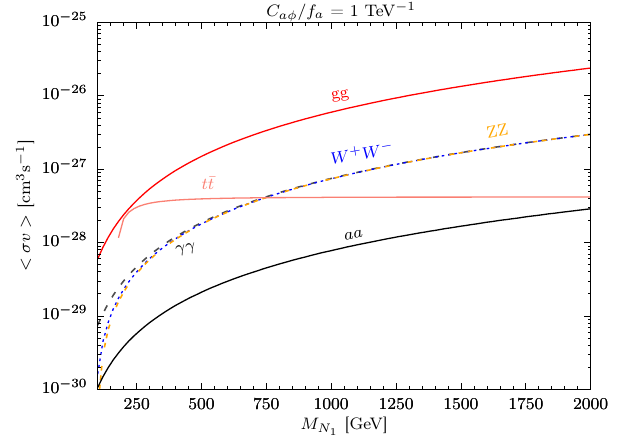}
\includegraphics[height=5cm, width=7.2cm]{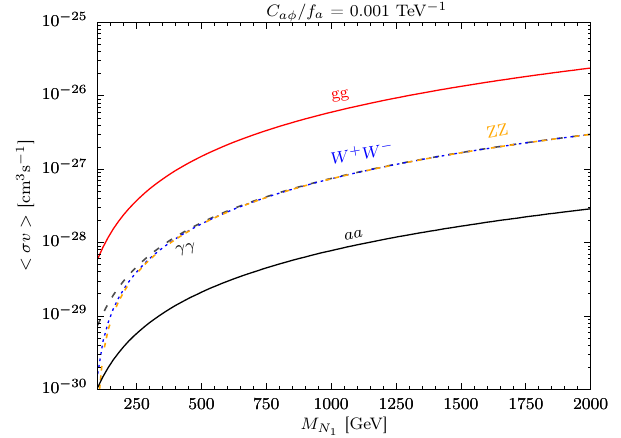}
\caption{Annihilation cross sections for different channels as a function of DM mass $M_{N_1}$. In the right panel, we choose a smaller value of $C_{a\Phi}/f_a = 10^{-3}\,\text{TeV}^{-1}$ to illustrate that annihilation to gauge bosons dominates. In both panels, we fix $C_{aN_1}=0.1\,\text{TeV}^{-1}$.}
\label{annihilation cross section}
\end{figure}

%%%%%%%%%%%%%%%%%%%%%%%%%%%%%%%%%%%
The relevant processes responsible for the freeze-out of DM in the early universe are shown in Fig.~\ref{annihilation diagram}. All together, they determine the relic abundance of our assumed DM, $N_1$. The thermal average annihilation cross section $\braket{\sigma \text{v}}$ and thus the relic abundance scale straightforwardly with the parameters of the model. The analytical expressions for all the annihilation cross-sections are computed. For our considered benchmark, only gauge boson channels are relevant. We see that the annihilation cross section for the process $N_1 ~N_1 \to f \bar{f}$ is negligible due to a very small value of $C_{a\Phi}$. On the other hand, with the limit $M_{N_1} > M_a$, the $N_1N_1\to aa$ annihilation channel opens up, but it is $\text{v}^2$ suppressed.
\begin{align*}
&(\sigma \text{v})_{gg}\approx \frac{256 C^2_{aN_1}C^2_{\tilde{G}}M^6_{N_1}}{\pi f^4_a(M^2_a-4M^2_{N_1})^2},\,\,
(\sigma \text{v})_{\gamma\gamma}\approx \frac{32 C^2_{aN_1}[C_{\tilde{B}} \cos^2\theta_w + C_{\tilde{W}} \sin^2\theta_w]^2 M^6_{N_1}}{\pi f^4_a(M^2_a-4M^2_{N_1})^2},  \nonumber \\
&(\sigma \text{v})_{ZZ}\approx \frac{32 C^2_{aN_1}[C_{\tilde{B}} \sin^2\theta_w + C_{\tilde{W}} \cos^2\theta_w]^2 M^3_{N_1}(M^2_{N_1}-M^2_Z)^{3/2}}{\pi f^4_a(M^2_a-4M^2_{N_1})^2}, \nonumber \\
&(\sigma \text{v})_{W^+W^-}\approx \frac{32 C^2_{aN_1} C_{\tilde{W}}^2 M^3_{N_1}(M^2_{N_1}-M^2_W)^{3/2}}{\pi f^4_a(M^2_a-4M^2_{N_1})^2},\,
(\sigma \text{v})_{f\bar{f}}\approx \frac{2N_c C^2_{aN_1} C^2_{a\phi} M^3_{N_1}M^2_f \sqrt{M^2_{N_1}-M^2_f}}{\pi f^4_a(M^2_a-4M^2_{N_1})^2}, \nonumber \\
&(\sigma \text{v})_{aa} \approx \frac{8C^4_{aN_1}M_{N_1}v^2\sqrt{M^2_{N_1}-M^2_a}(32 M^8_{N_1} - 64 M^2_a M^6_{N_1} + 48 M^4_a M^4_{N_1} - 16 M^6_a M^2_{N_1} + 3 M^8_a)}{3 \pi f^4_a (2M^2_{N_1}-M^2_a)^4} 
\end{align*}
In Fig.~\ref{annihilation cross section}, we show the annihilation cross section for different channels as a function of DM mass $M_{N_1}$. For the left and right panels, we choose $C_{a\Phi}/f_a=1\,\text{TeV}^{-1}$ and $C_{a\Phi}/f_a=0.001\,\text{TeV}^{-1}$, respectively. For both panels, we fix $C_{\tilde{G}}/f_a=C_{a\gamma}/f_a = C_{aN_1}=0.1\,\text{TeV}^{-1}$. For a relatively small value of $C_{a\Phi}/f_a$, annihilation cross-section to gauge boson dominates over fermionic channels.
%
%%%%%%%%%%%%%%%%%%%%%%%%%%%%%%%%%%%%%%%%%%%%%%%%%%%%%%%%%%%
%%%%%%%%%%%%%%%%%%%%%%%%%%%%%%%%%%%%%%%%%

%%%%%%%%%%%%%%%%%%%%%%%%%%%%%%%%%%%%%%%%%%%%
\begin{figure}[htbp]
\includegraphics[height=5cm, width=7.2cm]{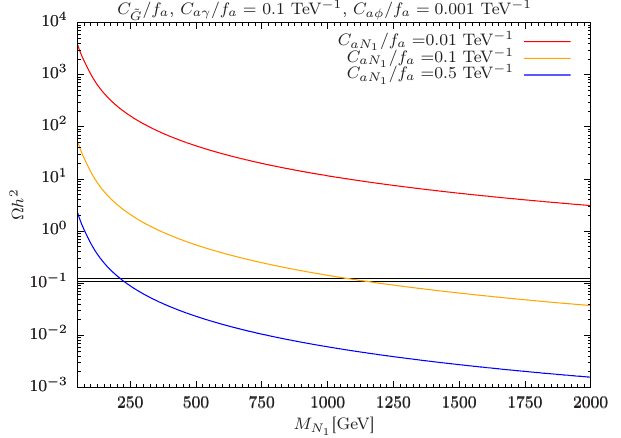}
\includegraphics[height=5cm, width=7.2cm]{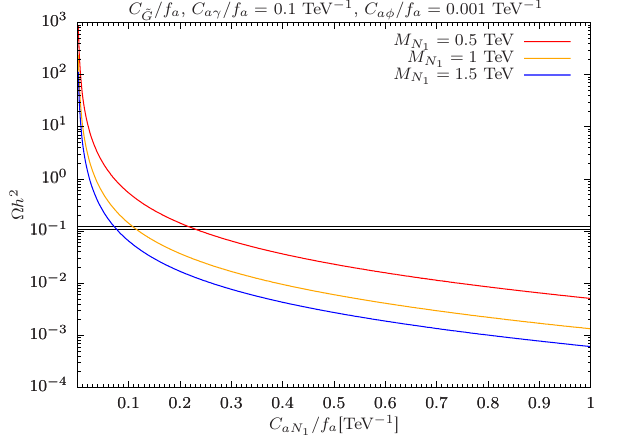}
\caption{The left panel shows the relic density~($\Omega_{N_1} h^2$) behaviour with DM mass $M_{N_1}$ for the chosen ALP parameters, labeled at the top. The three coloured curves are due to three discrete choices of ALP-$N_1$ coupling. The region inside the horizontal black lines stands for the measured $3\sigma$ relic density range given by Planck satellite data, Eq.~\ref{eq:omega}. The right panel is done using the same analysis but with ALP-$N_1$ coupling varying continuously on the horizontal axis, whereas the mass of $N_1$ has been chosen discretely.}
\label{fig:relic density1}
\end{figure}
%%%%%%%%%%%%%%%%%%%%%%%%%%%%%%%%%%%%%%%%%%
The left and right panels of Fig.~\ref{fig:relic density1} show the relic density behaviour as a function of DM mass $M_{N_1}$ and ALP-RHN coupling $C_{aN_1}$, respectively. In the left and right panels, three curves stand for three choices of ALP-RHN coupling and DM masses, respectively. The two closed horizontal lines are the $3\sigma$ range for cold DM from the Planck satellite data~\cite{Aghanim:2018eyx}:
%%%%%%%%%%%%%%%%%%%%%%%%
\begin{align}
  \label{eq:omega}
\Omega_{\text{DM}} h^2 = 0.12\pm 0.001
\end{align}
%%%%%%%%%%%%%%%%%%%%%%%
The solutions that fall exactly within this band can explain the total relic abundance of $N_1$ dark matter. We see from Fig.~\ref{fig:relic density1} that for a smaller value of DM mass, $M_{N_1}$, the required values of $C_{aN_{1}}/f_a$ are relatively large to explain the correct relic density.
%%%%%%%%%%%%%%%%%%%%%%%%%%%%%%%%%%%%%%%%%%%%%
\begin{figure}[htbp]
\includegraphics[height=5cm, width=7.2cm]{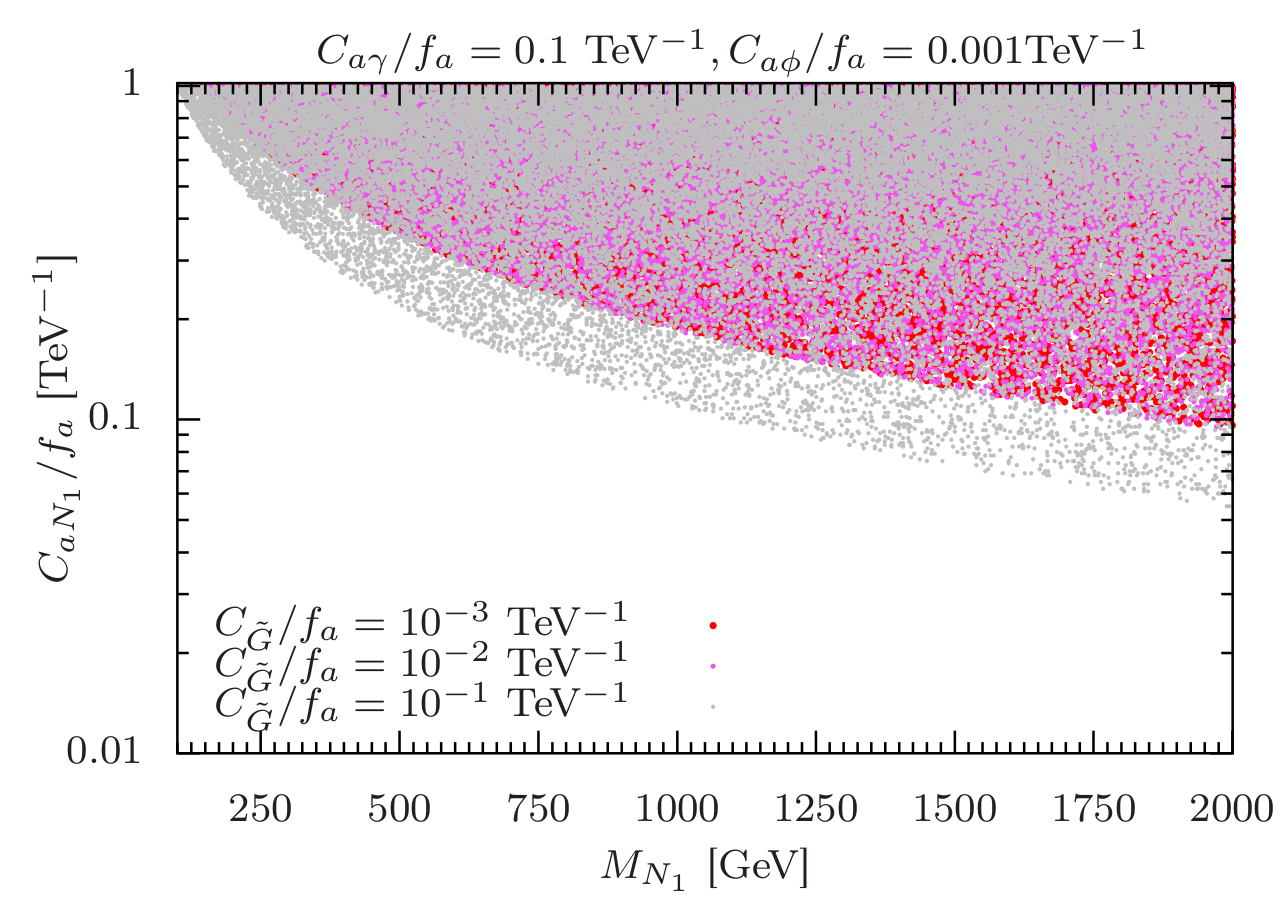}
\includegraphics[height=5cm, width=7.2cm]{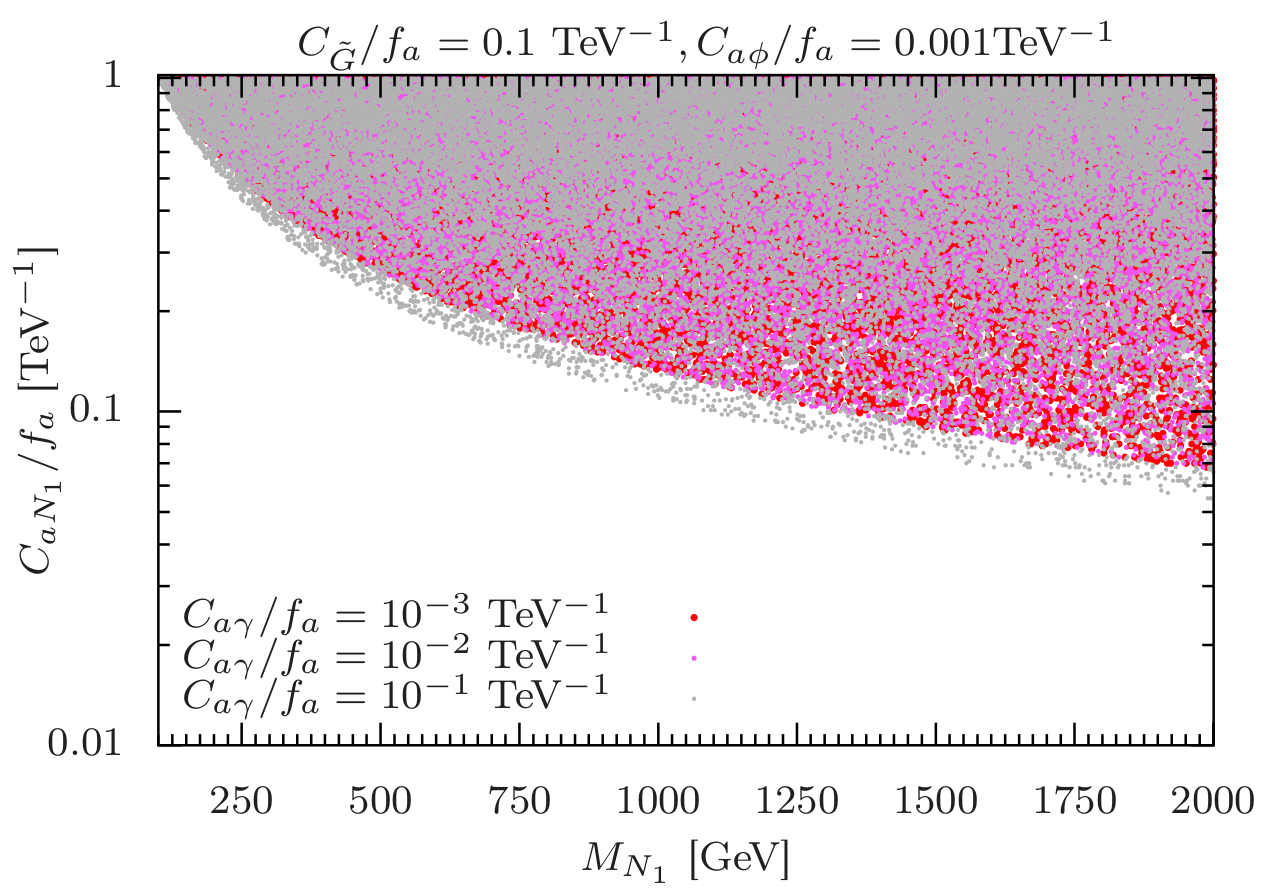}
\caption{The left panel describes the coloured regions where the relic density bound~($\Omega_{N_1} h^2\leq 0.12$) holds on the $m_{N_1}$-$C_{aN_1}$ plane. We have chosen the three discrete values for the ALP-gluon coupling~($C_{\tilde{G}}$) labeled by the corresponding colours. In the right panel, the same analysis is done but now with three discrete values for ALP-photon coupling~($C_{a\gamma}$).}
\label{fig:relic density2}
\end{figure}
%%%%%%%%%%%%%%%%%%%%%%%%%%%%%%%%%%%%%%%%%%%%%%%%

In Fig.~\ref{fig:relic density2}, we show the allowed region~(colored) where the relic density bound $\Omega_{N_{1}}h^2\leq 0.12$ holds on the $M_{N_1}-C_{aN_1}$ plane. In the left panel, we have chosen three discrete values for the ALP-gluon coupling: $C_{\tilde{G}}/f_a=10^{-3}\,\text{TeV}^{-1}$~(red), $10^{-2}\,\text{TeV}^{-1}$~(pink) and $10^{-1}\,\text{TeV}^{-1}$~(grey) by fixing other couplings as $C_{a\gamma}/f_a=10^{-1}\,\text{TeV}^{-1}$ and $C_{a\Phi}/f_a=10^{-3}\,\text{TeV}^{-1}$. In the right panel, the same analysis is done, but now we have fixed $C_{\tilde{G}}/f_a=10^{-1}\,\text{TeV}^{-1}$ and choose three discrete values of ALP-photon coupling: $C_{a\gamma}/f_a=10^{-3}\,\text{TeV}^{-1}$~(red), $10^{-2}\,\text{TeV}^{-1}$~(pink), and $10^{-1}\,\text{TeV}^{-1}$~(grey). Note that in the left (right) panel, for relatively smaller values of ALP-gluon (ALP-photon) couplings, the red and pink regions overlap. This happens due to the subdominant contribution to the relic density of the annihilation channel $N_1N_1\to gg$~($N_1N_1\to\gamma\gamma$). We find that when the gg channel dominates compared to other annihilation channels, the required value of $C_{aN_1}/f_a$ is smaller to satisfy the relic density.
%%%%%%%%%%%%%%%%%%%%%%%%%%%%%%%%%%%%%%%%%%%%%%%%%
\subsection{Direct detection}
\label{subsec:direct detection}
%%%%%%%%%%%%%%%%%%%%%%%%%%%%%%%%%%%%%%%%%%%%
The XENON1T~\cite{Aprile:2018dbl} experiments have a strong sensitivity for spin-independent and spin-dependent DM-nucleon interactions in our interested mass range of DM. However, recent data from LZ~\cite{LZ:2022lsv} and XENONnT~\cite{Aprile_2023} have further put stronger bounds on scattering cross-section. The interaction between DM ($N_1$) and a quark (q) can be described by the following effective Lagrangian:
%%%%%%%%%%%%%%%%%%%%
\begin{align}
\mathcal{L}=\frac{C_{a\Phi}C_{aN_1}}{f_a^2 M_{a}^2}m_q M_{N_1} \bar{q}\gamma_5 q\,\bar{N_1}\gamma_5 N_1.
\end{align}
%%%%%%%%%%%%%%%%%%%%%%%%%%%%%%%%%%
Note that this is only valid when the mediator ALP mass ($M_a$) is relatively large compared to the momentum transferred involved in the scattering process. Following Ref.~\cite{Boehm:2014hva,Freytsis:2010ne,Cheng:2012qr,Dolan:2014ska,Banerjee:2017wxi} we found that in the non-relativistic limit, differential scattering cross section to the scatter of a nucleus is $d\sigma/dE_R\propto q^4$, where $q^2=2m_N E_R$ is the momentum transfer, $m_N$ is the mass of the nucleus, and $E_R$ is the nuclear recoil energy. In direct detection experiments, typical recoil energy is $\mathcal{O}(10\,\text{KeV})$, hence the direct detection cross section is heavily suppressed.
%%%%%%%%%%%%%%%%%%%%%%%%%%%%%%%%%%%%%%%%%%%%%%%%%%%%%%%%%
\subsection{Indirect detection}
\label{subsec:indirect detection}
%%%%%%%%%%%%%%%%%%%%%%%%%%%%%%%%%%%%%%%%%%%%%%%%%%%%%%%%%%
If DM $N_1$ annihilates to SM final states with annihilation cross section near the thermal relic benchmark value $\braket{\sigma \text{v}}\sim 3\times 10^{-26}\,\text{cm}^3/s$, it may be detected indirectly. Perhaps gamma rays are the best messengers since they proceed almost unaffected during their propagation, thus carrying both spectral and spatial information. These $\gamma$-rays can be produced from DM annihilation, either mono-energetically from direct annihilation $N_1 N_1\to\gamma\gamma$, $\gamma X$, or with continuum spectra from decays of the annihilation products $N_1 N_1\to X\bar{X}(X=\text{SM state})$.
%%%%%%%%%%%%%%%%%%%%%%%%%%%%%%%%%%
%\begin{align}
%E_{\gamma}=M_{N_1}\big(1-\frac{m_X^2}{4M_{N_1}^2}\big)\,\,\,\text{and}\,\,\,E_\gamma=M_{N_1}\,\,\text{for}\,\,X=\gamma .
%\end{align}

%%%%%%%%%%%%%%%%%%%%%%%%%%%%%
\begin{figure}[htbp]
\begin{center}
\includegraphics[width=0.9\textwidth]{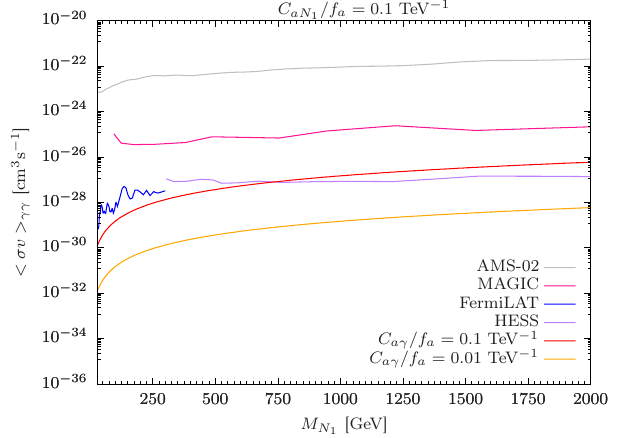}
\caption{The above plot shows the thermally averaged annihilation cross section for di-photon emissions. Experimental limite obtained from AMS-02 (Grey)~\cite{Elor:2015bho}, MAGIC (Dark-Pink)~\cite{HESS:2018cbt}, FermiLAT (Blue)~\cite{Ackermann:2013uma}, and HESS (Purple)~\cite{HESS:2018cbt} data are shown. The red and orange lines are the model predictions for different choices of ALP-photon couplings $C_{a\gamma}/f_a$.}
\label{fig:indirect}
\end{center}
\end{figure}
%%%%%%%%%%%%%%%%%%%%%%%%%%%%%%%%%%%%%%%%%%%%%%%%%%%%%%%%%%

These gamma rays would be produced preferentially in regions of high DM density and can be best detected by Fermi-Lat~\cite{Ackermann:2013uma}, HESS~\cite{HESS:2018cbt}. The integrated gamma-ray flux from the DM annihilation in a density distribution $\rho(\bf r)$ is given by
%%%%%%%%%%%%%%%%%%%%%%%%%%%%%%%%%%%%%%%%%%%%%
\begin{align}
\Phi_\gamma(\Delta \Omega)=\frac{1}{4\pi}\frac{\braket{\sigma \text{v}}}{2 M_{N_1}^2}\int_{E_{\text{min}}}^{E_{\text{max}}}\frac{dN_\gamma}{dE_\gamma}dE_{\gamma}.J,
\end{align}
%%%%%%%%%%%%%%%%%%%%%%%%%%%%%%%%%%%%%%%%%%%%%%%%%%%
where $J=\int_{\Delta\Omega}{\int_{\text{l.o.s}}\rho^2(\bf r)d\ell}d\Omega'$ is the line-of-sight~(l.o.s) integral through the DM distribution integrated over a solid angle $\Delta \Omega$. The integral $\int^{\text{ROI}}\frac{dJ}{d\Omega}d\Omega$ represents the astrophysical component of the DM flux calculation in particular Region of Interest (ROI). AMS-02 looks for excess positron flux in the positron energy range of 1 GeV to 500 GeV over the cosmic positron background. This positron can give rise to two photons in the final state after the subsequent process. A model-independent study of such bounds is studied\cite{Elor:2015bho}. FermiLAT collaboration looks for direct DM annihilation of two photons from dwarf spheroidal galaxies of the Milky Way in photon energy from a few GeV to a few hundreds of GeV. A higher energy range is being explored by the Magic and HESS collaboration. We have considered the current upper limit from AMS-02, MAGIC, HESS, and annihilation data from FermiLAT, respectively, in Fig.~\ref{fig:indirect} on the thermally averaged annihilation cross section of the di-$\gamma$ final state, along with our model predictions for our chosen benchmark in Fig.~\ref{fig:indirect}. The model prediction for ALP-photon coupling $C_{a\gamma}/f_a=0.1\,\text{TeV}^{-1}$ lies very close to the Fermi-LAT and HESS upper limits. This suggests that future sensitivities of Fermi-LAT or HESS can either probe or exclude the large parameter space of the model considered by us.
%%%%%%%%%%%%%%%%%%%%%%%%%%%%%%%%%%%%%%%%
\section{Summary}
\label{summary_alp}
We have analyzed heavy neutrino DM candidates in a minimal extension of SM, which features three RHNs and one ALP. This model is well motivated, as it not only accounts for DM but also explains neutrino oscillations. Hence, ALP-mediated RHN DM is interesting from both the model-building and phenomenological perspectives. We have considered the lightest RHN as DM, which is odd under $\mathbb{Z}_2$ symmetry, and identified the region of parameters where DM predictions are in agreement with DM relic abundance.  In addition, this model also quite naturally explains the null results of LUX and XENON1T due to the pseudoscalar nature of interactions with quarks. We have highlighted the importance of complementary searches, for instance, via indirect detection with single and di-photons. Although the current limits from Fermi-LAT lie above the predicted signals for our choice of parameter space, future sensitivities of Fermi-LAT might offer promising prospects to probe both the low and high DM mass regions.
%%%%%%%%%%%%%%%%%%%%%%%%%%%%%%%%%%%%%%%%%

%%%%%%%%%%%%%%%%%%%%%%%%%%%%%%%%%%%%%%%%%
\chapter{A two component dark matter model in a generic \texorpdfstring{$U(1)_X$}{U(1)X} extension of SM}
\label{chap_U1X}
%%%%%%%%%%%%%%%%%%%%%%%%%%%%%%%%%%%%%%%%%
In this chapter, we study a two-component DM model interacting with SM via Higgs and Z portals. The results are based on the work: Arindam Das, Shivam Gola, Sanjoy Mondal, Nita Sinha, "Two-Components Scalar and Fermionic Dark Matter candidates in a generic U(1)$_X$ model, Phys.Lett.B 2022.137117''.  

\section{Introduction}
Underpinning the origin of neutrino mass and elucidating the nature of DM would constitute a major step forward in particle physics. Several simple extensions of the SM that can account for the DM have already been studied \cite{Jungman:1995df,Bertone:2004pz,Hooper:2007qk,McDonald:1993ex,Burgess:2000yq,LopezHonorez:2006gr,Barbieri:2006dq,Lopez-Honorez:2012tov}. In these models, the SM particle content is extended by additional fields, and a discrete symmetry is usually introduced to guarantee the stability of the DM particle in cosmological scale. In recent years, a class of models has been proposed to incorporate the neutrino mass generation and the existence of DM in a unified framework. Motivated by this, people have studied well-motivated BSM framework based on the gauged $U(1)_X$ model \cite{Okada:2016tci,Bandyopadhyay:2017bgh,Das:2019pua}. The most intriguing aspect of this model is that including three generations of right-handed neutrinos, as in the type-I seesaw process for creating light neutrino masses, is no longer an option, but emerges as the simplest solution to eliminate the gauge and mixed gauge-gravity anomalies \cite{Das:2016zue}. The scalar DM can be inherently stable in such models due to its $U(1)_X$ charge, but the fermionic DM cannot be realized in the simplest $U(1)_X$ model. Additional discrete symmetries can be introduced, which can stabilize one of the right-handed neutrinos to play the role of DM, while the other two neutrinos participate in the type I seesaw process to generate the required light neutrino masses and flavor mixing. Also, there are many models proposed in the literature, where neutrino mass generation is intimately connected with DM \cite{Ma:2006km,Hirsch:2013ola,Merle:2016scw,Avila:2019hhv,Mandal:2021yph,Mandal:2019oth}. In these types of models, DM is a mediator of neutrino mass generation.

A single-particle DM model may not be sufficient to account for the relic density of DM observed in the universe. Many such models face strong constraints from direct detection experiments and other observations. Therefore, it is reasonable to consider multi-particle DM scenarios, where two or more particles contribute to the DM abundance.~\cite{Aprile_2023,Boyarsky_2019,vinyoles2015new}. Multi-component DM refers to a situation in which two or more particles contribute to the measured DM density. This has been already studied in many BSM scenarios \cite{Profumo:2009tb,Feldman:2010wy,Baer:2011hx,Aoki:2012ub,Bhattacharya:2013hva,Bian:2013wna,Kajiyama:2013rla,Esch:2014jpa,Bhattacharya:2016ysw,Bhattacharya:2019fgs,Bhattacharya:2019tqq,Bhattacharya:2017fid,Choi:2021yps,DiazSaez:2021pmg,Mohamadnejad:2021tke,Ho:2022erb}. The multi-component DM model has also some benefits over the single dark matter scenario. For instance, it can avoid some stringent constraints arising from various experiments that probe the properties and interactions of dark matter and other particles. A multi-component DM model can also accommodate different observational features of dark matter, such as its distribution and abundance in the universe. 

In this article we consider a generic $U(1)_X$ model for two-component DM and discuss its phenomenological implications in this study. The general charge assignment of the particles in the model is obtained after the gauge and mixed gauge gravity anomalies. It further allows us to study the dependence of the DM relic abundance on these charges which will appear at different interactions between the potential DM candidates, mediators, and the other particles present in the model. In the context of the two-component DM system such properties have not been discussed in the previous literature. Among these two DM candidates, one is scalar $\chi$ and the other one is fermion $N_R^3$. $N_R^3$ and $\chi$ are SM singlets but charged under $U(1)_X$. $N_R^3$ is odd under a discrete symmetry $\mathbb{Z}_2$ and $\chi$ is odd under another discrete symmetry $\mathbb{Z}_2'$. The model is not the same as the singlet fermionic \cite{Lopez-Honorez:2012tov} plus scalar \cite{Burgess:2000yq} DM due to the presence of new gauge boson $Z'$ which couples with $N_R^3$ or $\chi$ as both are charged under $U(1)_X$. Hence, there will be many new processes that contribute to the relic density. Also, this model contains additional features such as the annihilation of one type of DM into the other, which we investigate in some depth. We further examine the direct detection prospects of this model. We find that there is parameter space where both these particles can produce observable signals in future direct detection experiments. In this analysis, we employ the latest bounds on the $U(1)_X$ coupling from $Z^\prime$ mediated dilepton and dijet scenarios for different combinations of $U(1)_X$ charges. We estimate the results by taking the allowed couplings after considering a variety of experimental searches. In addition to that, we consider the mixing between the SM Higgs doublet and SM singlet $U(1)_X$ scalar. Considering the allowed parameter of this mixing angle we estimate the relic abundance of a viable DM candidate in the model. 
%%%%%%%%%%%%%%%%%%%%%%%%%%%%%%%%%%%%%%%%%
Chapter~\ref{chap_U1X} is organised as follows. In sec.~\ref{sec:model2} we introduce the model and discuss the details of the new fields and their interactions. In sec.~\ref{sec:constraints2} we discuss different theoretical and experimental constraints on the model parameters. In sec.~\ref{sec:dark-matter2} we discuss in detail the relic density and direct detection properties coming from our two-component DM candidates. In sec.~\ref{sec:relic-dependence-xH} we study the relic density dependence on $U(1)_X$ charge in the case of $Z'$-portal DM. Finally in sec.~\ref{summary_tdm} we will summarise.

%%%%%%%%%%%%%%%%%%%%%%%%%%%%%%%%%%%%%%%%%%%
\section{Model}
\label{sec:model2}

%%%%%%%%%%%%%%%%%%%%%%%%%%%%%%%%%%%%%%%%%%%%%%%
The considered model is a general but minimal U$(1)_X$ extension of the SM. Three generations of right-handed neutrinos~($N_R^i$) and two U$(1)_X$ complex scalar fields~($\Phi,\chi$) are included in addition to the SM particles. The SM as well as new particles and their charges are given in Table~\ref{tab2}, where the family index $i$ runs from 1 to 3. The model has been constructed such that all SM fields are even under the discrete symmetry $\mathbb{Z}_2\otimes \mathbb{Z}'_2$, whereas $N_R^3$ is odd under $\mathbb{Z}_2$ and $\chi$ is odd under $\mathbb{Z}'_2$. 

%%%%%%%%%%%%%%%%%%%%%%%%%%%%%%%%%%%%%%%%%%%%%%%
\begin{table}[t]
\begin{center}
\begin{tabular}{||c|ccc||c||c|c||}
\hline
\hline
            & SU(3)$_c$ & SU(2)$_L$ & U(1)$_Y$ & U(1)$_X$  & $\mathbb{Z}_2$  & $\mathbb{Z}'_2$\\[2pt]
\hline
\hline
&&&&&&\\[-12pt]
$q_L^i$    & {\bf 3}   & {\bf 2}& $\frac{1}{6}$ & 		 $x_q= \frac{1}{6}x_H + \frac{1}{3}x_\Phi$  & + & +  \\[2pt] 
$u_R^i$    & {\bf 3} & {\bf 1}& $\frac{2}{3}$ &  	  $x_u= \frac{2}{3}x_H + \frac{1}{3}x_\Phi$ & + & + \\[2pt] 
$d_R^i$    & {\bf 3} & {\bf 1}& $-\frac{1}{3}$ & 	 $x_d=-\frac{1}{3}x_H + \frac{1}{3}x_\Phi$ & + & + \\[2pt] 
\hline
\hline
&&&&&&\\[-12pt]
$\ell_L^i$    & {\bf 1} & {\bf 2}& $-\frac{1}{2}$ & 	 $x_\ell =- \frac{1}{2}x_H - x_\Phi$ & + & +  \\[2pt] 
$e_R^i$   & {\bf 1} & {\bf 1}& $-1$   &		$x_e= - x_H - x_\Phi$ & + & +  \\[2pt] 
\hline
\hline
$N_R^{1,2}$   & {\bf 1} & {\bf 1}& $0$   &	 $x_\nu =- x_\Phi$ & + & + \\[2pt] 
$N_R^{3}$   & {\bf 1} & {\bf 1}& $0$   & 	 $x_\nu=- x_\Phi$  & - & + \\[2pt] 
\hline
\hline
&&&&&&\\[-12pt]
$H$         & {\bf 1} & {\bf 2}& $\frac{1}{2}$  &  	 $\frac{x_H}{2}$ & + & + \\ 
$\Phi$      & {\bf 1} & {\bf 1}& $0$  & 	 $2 x_\Phi$  & +  &  +  \\ 
$\chi$      & {\bf 1} & {\bf 1}& $0$  & 	 $-x_\Phi$  & +  & -  \\ 
\hline
\hline
\end{tabular}
\end{center}
\caption{
Particle and symmetry content of the minimal model where $i(=1, 2, 3)$ represents the family index. The U$(1)_X$ charges $x_H$, $x_\Phi$ are the real parameters.}
\label{tab2}
\end{table}
%%%%%%%%%%%%%%%%%%%%%%%%%%%%%%%%%%%%%%%%%%%%%%%%%%%%%%%%%%%

The $U(1)_X$ charges of the particles are controlled by two parameters only, $x_H$ and $x_\Phi$, as seen in Table~\ref{tab2} and can be defined as a linear combination of the SM $U(1)_Y$ and the $U(1)_{B - L}$ which can also be obtained after solving the gauge and mixed gauge-gravity anomaly cancellation equations obtained from~\ref{Yukawa2} and given here. Note that the $B-L$ case can be obtained with the choice $x_H=0$ and $x_\Phi=1$. For simplicity, we fix $x_\Phi=1$ in our analysis throughout the paper. In addition to that, we find if $x_H=-2$, then the left-handed fermions have no interactions with the $Z^\prime$ leading to an $U(1)_R$ scenario. For $x_H=-1$ and $1$ the interactions of $d_R$ and $e_R$ with $Z^\prime$ are switched off.

%%%%%%%%%%%%%%%%%%%%%%%%%%%%%%%%
\begin{align*}
{U}(1)_X \otimes \left[ {SU}(3)_c \right]^2&\ :&
			2x_q - x_u - x_d &\ =\  0, \nonumber \\
{U}(1)_X \otimes \left[ {SU}(2)_L \right]^2&\ :&
			3x_q + x_\ell &\ =\  0, \nonumber \\
{U}(1)_X \otimes \left[ {U}(1)_Y \right]^2&\ :&
			x_q - 8 x_u - 2x_d + 3x_\ell - 6x_e &\ =\  0, \nonumber \\
\left[ {U}(1)_X \right]^2 \otimes {U}(1)_Y&\ :&
			{x_q}^2 - {2x_u}^2 + {x_d}^2 - {x_\ell}^2 + {x_e}^2 &\ =\  0, \nonumber \\
\left[ {U}(1)_X \right]^3&\ :&
			{6x_q}^3 - {3x_u}^3 - {3x_d}^3 + {2x_\ell}^3 - {x_\nu}^3 - {x_e}^3 &\ =\  0, \nonumber \\
{U}(1)_X \otimes \left[ {\rm grav.} \right]^2&\ :&
			6x_q - 3x_u - 3x_d + 2x_\ell - x_\nu - x_e &\ =\  0. 
%\label{anom}
\end{align*}
%%%%%%%%%%%%%%%%%%%%%%%%%%%%%%%%

In the next subsections, we discuss various parts of the lagrangian of the model, 
%%%%%%%%%%%%%%%%%%%%%%%%%%%%%%%%%%%%%%%%%%%%%%%%
\subsection{Scalar Sector}
%%%%%%%%%%%%%%%%%%%%%%%%%%%%%%%%%%%%%%%%%%%%%%%%

We begin by writing down the Lagrangian of the scalar sector. Apart from the SM Higgs doublet $H$ we have two complex scalars $\Phi$ and $\chi$, both charged under $U(1)_X$, but with $\mathbb{Z}_2$ even parity. The most general renormalizable and gauge gauge-invariant scalar sector can be written as
\begin{align}
 \mathcal{L}_{s}=(D^{\mu}H)^{\dagger}(D_{\mu}H)+(D^{\mu}\Phi)^{\dagger}(D_{\mu}\Phi)+(D^{\mu}\chi)^{\dagger}(D_{\mu}\chi)-V(H,\Phi,\chi),
\end{align}
where the covariant derivative is defined as, $D_{\mu}=\partial_{\mu}-ig_{s}T^{a}G^{a}_{\mu}-igT^{a}W^{a}_{\mu}-ig_{1}YB_{\mu}-ig_{1}^{'}Y_{X}B_{\mu}^{'}$. $B_{\mu}^{'}$
is the $U(1)_{X}$ gauge fields. The $U(1)_X$ gauge coupling $g_{1}^{'}$ is a free parameter. The scalar potential $V(H,\Phi,\chi)$ is given by,
\begin{align}
 V(H,\Phi,\chi) = &-\mu_{H}^{2}H^{\dagger}H-\mu_{\Phi}^{2}\Phi^{\dagger}\Phi + m_{\chi}^{2}\chi^{\dagger}\chi +\lambda_{H}(H^{\dagger}H)^{2}+\lambda_{\Phi}(\Phi^{\dagger}\Phi)^{2}
 +\lambda_{\chi}(\chi^{\dagger}\chi)^{2}\nonumber \\
 &+\lambda_{H\Phi}(H^{\dagger}H)(\Phi^{\dagger}\Phi)\nonumber 
 +\lambda_{\Phi\chi}(\Phi^{\dagger}\Phi)(\chi^{\dagger}\chi)+\lambda_{H\chi}(H^{\dagger}H)(\chi^{\dagger}\chi)+(\lambda_{\Phi\chi\chi}\Phi\chi\chi + \text{H.c.})
\end{align}
The breaking of the electroweak and the $U(1)_X$ gauge symmetries are driven by the vacuum expectation values~(vev) of the scalar fields $H$ and $\Phi$ as the field $\chi$ does not get any vev due to $\mathbb{Z}'_2$ symmetry protection. Denoting the vevs of $H$ and $\Phi$ by $v_H$ and $v_\Phi$, the fields $H,\Phi$ and $\chi$ can be written in unitary gauge after symmetry breaking in the form
%%%%%%%%%%%%%%%%%%%%%%%%%%%%%%%%%%%%%%%%%%%%%%%%%%%%%%%%%%%%%%%%%%%%%%%%%%%%%%%
\begin{align}
 H=\frac{1}{\sqrt{2}}
 \begin{pmatrix}
  \phi^+ \\
  v_H+R_1+i I_1  \\
 \end{pmatrix},\hspace{0.2cm}
 \Phi=\frac{1}{\sqrt{2}}(v_\Phi+R_2+i I_2),
 \hspace{0.2cm}
 \chi=\frac{1}{\sqrt{2}}(\chi_{R}+i\chi_{I})
\end{align}
%
%%%%%%%%%%%%%%%%%%%%%%%%%%%%%%%%%%%%%%%%%%%%%%%%%%%%%%%
$\Phi^{\pm}$ are the would be Goldstone boson of $W^{\pm}$, while $I_1$ and $I_2$ will mix to give the Goldstone bosons of the $Z$ and $Z^{'}$ bosons, respectively. The mass matrix of CP-even Higgs scalars in the basis $(R_1 , R_2 )$ reads as

%%%%%%%%%%%%%%%%%%%%%%%%%%%%%%%%%%%%%%%%%%%%%%%%%%%
\begin{align}
M_R^2=\begin{bmatrix}
2\lambda_H v_H^2 &  \lambda_{H\Phi} v_H  v_\Phi \\
\lambda_{H\Phi} v_H v_\Phi  &  2\lambda_{\Phi}v_\Phi^2 \\
\end{bmatrix}
\end{align}
%%%%%%%%%%%%%%%%%%%%%%%%%%%%%%%%%%
%
with the mass eigenvalues given by
\begin{align}
m_{h_{1}}^{2}=\lambda_{H}v_H^{2}+\lambda_{\Phi}v_\Phi^{2}-\sqrt{(\lambda_{H}v_H^{2}-\lambda_{\Phi}v_\Phi^{2})^{2}+(\lambda_{H\Phi}v_H v_\Phi)^{2}} \\
m_{h_{2}}^{2}=\lambda_{H}v_H^{2}+\lambda_{\Phi}v_\Phi^{2}+\sqrt{(\lambda_{H}v_H^{2}-\lambda_{\Phi}v_\Phi^{2})^{2}+(\lambda_{H\Phi}v_H v_\Phi)^{2}}
\end{align}
%%%%%%%%%%%%%%%%%%%%%%%%%%%%%
where the scalars $h_1$ and $h_2$ have masses $m_{h_1}$ and $m_{h_2}$ respectively, and by convention $m_{h_1}^2 \leq m_{h_2}^2$ throughout this work. We have identified $h_1$ as the SM Higgs discovered at LHC, with mass $m_{h_1}=125$~GeV. The two mass eigenstates $h_i$ are related with the $(R_1, R_2)$ fields through the rotation matrix $O_R$ as,
%%%%%%%%%%%%%%%%%%%%%%%%%%%%%%%%%%%%%%%%%
\begin{align}
\begin{bmatrix}
h_1 \\
h_2 \\
\end{bmatrix}=O_R
\begin{bmatrix}
R_1 \\
R_2 \\
\end{bmatrix}=\begin{bmatrix} \cos\alpha  &  \sin\alpha \\
-\sin\alpha  &  \cos\alpha  \\  \end{bmatrix}
\begin{bmatrix}
R_1 \\
R_2 \\
\end{bmatrix},
\label{eq:rotation}
\end{align}
%%%%%%%%%%%%%%%%%%%%%%%%%%%%%%%%%%%%%%%%%
where $\alpha$ is the mixing angle. The rotation matrix satisfies
\begin{align}
O_R M_R^2 O_R^T=\text{diag}\left(m_{h_1}^2,m_{h_2}^2\right).
\label{eq:diag}
\end{align}
We can use Eq.~\eqref{eq:rotation} and \eqref{eq:diag} to solve for the potential parameters $\lambda_H$, $\lambda_\Phi$ and $\lambda_{H\Phi}$ in terms of the mixing angle $\alpha$ and the scalar masses $m_{h_i}$ as
%%%%%%%%%%%%%%%%%%%%%%%%%%%%%%%%%%%%%%%%%%%%%%%
\begin{align}
 &\lambda_{H}=\frac{m_{h_{2}}^{2}}{4v_H^{2}}(1-\text{cos}~2\alpha)+\frac{m_{h_{1}}^{2}}{4v_H^{2}}(1+\text{cos}~2\alpha)\\
 & \lambda_{\Phi}=\frac{m_{h_{1}}^{2}}{4v_\Phi^{2}}(1-\text{cos}~2\alpha)+\frac{m_{h_{2}}^{2}}{4v_\Phi^{2}}(1+\text{cos}~2\alpha)\\
 & \lambda_{H\Phi}=\text{sin}~2\alpha\left(\frac{m_{h_{1}}^{2}-m_{h_{2}}^{2}}{2v_\Phi v_H}\right)
 \label{coup}
\end{align}
The real and imaginary components of $\chi$ have the following masses
%%%%%%%%%%%%%%%%%%%%%%%%%%%%%%%%%%%%%%%%%%%%%%%%%
\begin{align}
 M_{\chi_R}^2=m_{\chi}^{2}+v_\Phi^{2}\frac{\lambda_{\Phi\chi}}{2}+v_H^{2}\frac{\lambda_{H\chi}}{2}+\sqrt{2}v_\Phi\lambda_{\Phi\chi\chi}\\
M_{\chi_I}^2=m_{\chi}^{2}+v_\Phi^{2}\frac{\lambda_{\Phi\chi}}{2}+v_H^{2}\frac{\lambda_{H\chi}}{2}-\sqrt{2}v_\Phi\lambda_{\Phi\chi\chi}
\end{align}
%%%%%%%%%%%%%%%%%%%%%%%%%%%%%%%%%%%%%%%%%%%%
The difference $M_{\chi_R}^2-M_{\chi_I}^2$ depends only on the parameter $\lambda_{\Phi\chi\chi}$. The lightest of the two eigenstates $\chi_R$ and $\chi_I$ can be a viable DM candidate from the conservation of $\mathbb{Z}'_2$ symmetry.
%%%%%%%%%%%%%%%%%%%%%%%%%%%%%%%%%%%%%%%%
\subsection{Gauge sector}
%%%%%%%%%%%%%%%%%%%%%%%%%%%%%%%%%%%%%%%%%%
To determine the gauge boson spectrum, we have to expand the scalar kinetic terms and replace 
\begin{align}  
H=\frac{1}{\sqrt{2}}
\begin{pmatrix}
0   \\
v_H+R_1  \\
\end{pmatrix},\,\,\,\,\text{and  }\,\Phi=\frac{v_\Phi + R_2}{\sqrt{2}},
\end{align}
%%%%%%%%%%%%%%%%%%%%%%%%%%%%%% 
With this above replacement, we can expand the scalar kinetic terms $(D^{\mu}H)^{\dagger}(D_{\mu}H)$ and $(D^{\mu}\Phi)^{\dagger}(D_{\mu}\Phi)$ as follows
\begin{align}
 (D^{\mu}H)^{\dagger}(D_{\mu}H)&\equiv\frac{1}{2}\partial^{\mu}R_1\partial_{\mu}R_1+\frac{1}{8}(R_1+v_H)^{2}\Big(g^{2}|W_{1}^{\mu}-iW_{2}^{\mu}|^{2}+(gW_{3}^{\mu}-g_{1}B^{\mu}-\tilde{g}B^{'\mu})^{2}\Big)\\
 (D^{\mu}\Phi)^{\dagger}(D_{\mu}\Phi)&\equiv\frac{1}{2}\partial^{\mu}R_2\partial_{\mu}R_2+\frac{1}{2}(R_2+v_\Phi)^{2}(2g_{1}^{''} B^{'\mu})^{2}.
\end{align}
where we have defined $\tilde{g}=g_{1}^{'}x_H$ and $g_{1}^{''}=g_{1}^{'}x_\Phi$. SM charged gauge boson $W^{\pm}$ can be easily recognised with mass $M_{W}=\frac{gv_H}{2}$. Linear combination of $B^{\mu}$, $W_{3}^{\mu}$ and $B^{'\mu}$ gives definite mass eigenstates $A^{\mu}$, $Z^{\mu}$ and $Z^{'\mu}$,
%%%%%%%%%%%%%%%%%%%%%%%%%%%%%%%%%%
\begin{align}
 \begin{pmatrix}
  B^{\mu}\\
  W_{3}^{\mu}\\
  B^{'\mu}
 \end{pmatrix}
=
\begin{pmatrix}
 \text{cos}~\theta_{w}  & -\text{sin}~\theta_{w}~\text{cos}~\theta^{'}  &  \text{sin}~\theta_{w}~\text{sin}~\theta^{'} \\
 \text{sin}~\theta_{w}  & \text{cos}~\theta_{w}~\text{cos}~\theta^{'}  &  -\text{cos}~\theta_{w}~\text{sin}~\theta^{'} \\
 0                      & \text{sin}~\theta^{'}                        &  \text{cos}~\theta^{'} \\
\end{pmatrix}
\begin{pmatrix}
 A^{\mu}\\
 Z^{\mu} \\
 Z^{'\mu} \\
\end{pmatrix}
\end{align}
%%%%%%%%%%%%%%%%%%%%%%%%%%%%%%%%%%%%%%%%%
where $\theta_{w}$ is the Wienberg mixing angle and,
\begin{align}
 \text{tan}~2\theta^{'}=\frac{2\tilde{g}\sqrt{g^{2}+g_{1}^{2}}}{\tilde{g}^{2}+16\left(\frac{v_\Phi}{v_H}\right)^{2}g_{1}^{''2}-g^{2}-g_{1}^{2}}.
\end{align}
Masses of $A,\,Z$ and $Z^{'}$ are given by,
\begin{align}
 M_{A}=0,
 \hspace{0.5cm}
 M_{Z,Z^{'}}^{2}=\frac{1}{8}\left(Cv_H^{2}\mp\sqrt{-D+v_H^{4}C^{2}}\right),
\end{align}
where,
\begin{align}
 C=g^{2}+g_{1}^{2}+\tilde{g}^{2}+16\left(\frac{v_\Phi}{v_H}\right)^{2}g_{1}^{''2},
 \hspace{0.5cm}
 D=64v_H^{2}v_\Phi^{2}(g^{2}+g_{1}^{2})g_{1}^{''2}.
\end{align}
%%%%%%%%%%%%%%%%%%%%%%%%%%%%%%%%%%%%%%%
\subsection{Yukawa sector}
%%%%%%%%%%%%%%%%%%%%%%%%%%%%%%%%%%%%%%%
The Yukawa sector of the model can be written in a gauge-invariant way as
\begin{align}
 \mathcal{L}_{y} & =-\sum_{i,j=1}^3 y_{u}^{ij}\overline{q_{L}^{i}}\tilde{H}u^{j}_{R} - \sum_{i,j=1}^3 y_{d}^{ij}\overline{q_{L}^{i}}H d^{j}_{R} - \sum_{i,j=1}^3 y_{e}^{ij}\overline{\ell_{L}^{i}} H e^{j}_{R}- \sum_{i=1}^3\sum_{j=1}^2 y_{\nu}^{ij}\overline{\ell_{L}^{i}}\tilde{H} N^{j}_{R} \nonumber \\
& -\frac{1}{2}\sum_{i,j=1}^2 y_{M}^{ij}\Phi\overline{N_{R}^{ic}} N^{j}_{R}
  -\frac{1}{2} y_{M}^{3}\Phi\overline{N_{R}^{3c}}N^{3}_{R} +\text{H.c.}
\label{Yukawa2}
\end{align}

The last two terms will give the Dirac and Majorana contributions to the neutrino mass generation. We have assumed a basis in which $y_M^{ij}$ is diagonal $y_M=\text{diag}(y_M^1,y_M^2)$, without the loss of generality. Relevant light neutrino masses will come from the fourth and fifth term of Eq.~\ref{Yukawa2}. After the electroweak symmetry breaking we can write the mass terms as,
 \begin{align}
  -\mathcal{L}_{M}=\sum_{j=1}^3\sum_{k=1}^2\overline{\nu_{jL}}m^{jk}_{D} N_{kR}+\frac{1}{2}\sum_{j,k=1}^2\overline{(N_{R})_{j}^{c}}M_R^{jk} N_{kR}+\text{H.c.},
 \end{align}
where $m_D^{jk}=\frac{y_{\nu}^{jk}v_H}{\sqrt{2}}$ and $M_R^{jk}=\frac{y_{M}^{jk}}{\sqrt{2}}v_\Phi$. Now we can write the $\mathcal{L}_{M}$ in the following matrix form,
\begin{align}
 -\mathcal{L}_{M}=\frac{1}{2}
 \begin{pmatrix}
  \overline{\nu_{L}} & \overline{(N^1_{R})^{c}} & \overline{(N^2_{R})^{c}} \\
 \end{pmatrix}
 \begin{pmatrix}
  0_{3\times 3}   &  (m_{D})_{3\times 2} \\
  (m_{D}^T)_{2\times 3}  &  (M_{R})_{2\times 2} \\
 \end{pmatrix}
\begin{pmatrix}
 (\nu_{L})^{c}\\
 N^1_{R} \\
 N^2_{R} \\
\end{pmatrix}
\end{align}
From this mass matrix, using the assumption $m_{D}\ll M_{R}$, it is easy to recover the seesaw formula for the light Majorana neutrinos as, $\mathcal{M}_\nu \approx m_D M_R^{-1}m_D^T$ and the heavy neutrino mass as $M_{N}\approx M_{R}$. We emphasize that the right-handed neutrino $N_R^3$ is decoupled by construction from this seesaw mechanism as it is odd under the $\mathbb{Z}_2$ symmetry.
%
%%%%%%%%%%%%%%%%%%%%%%%%%%%%%%%%%%%%%%%%%%%%%%%%%%%%%%%%%
\section{Theoretical and experimental constraints}
\label{sec:constraints2}
%%%%%%%%%%%%%%%%%%%%%%%%%%%%%%%%%%%%%%%%%%%%%%%%%%
We discuss different constraints on the model parameters such as $U(1)_X$ gauge coupling and scalar mixing angle. To estimate the constraints we consider vacuum stability, perturbative unitarity, and collider searches of BSM Higgs and $Z^\prime$ boson respectively. 

%%%%%%%%%%%%%%%%%%%%%%%%%%%%%%%%%%%%%%%%
\subsection{Vacuum Stability}
%%%%%%%%%%%%%%%%%%%%%%%%%%%%%%%%%%%%%%%%%
\vspace*{-0.5cm}
The above scalar potential must be bounded from below.  To determine the conditions for $V(H,\Phi,\chi)$ to be bounded from below, we need to check the following symmetric matrix which comes from the quadratic part of the potential,
%%%%%%%%%%%%%%%%%%%%%%%%%%%%%%%%%%%%%%%%%%%%%%%%%%%%
\begin{align}
 V^4_s=
\begin{pmatrix}
 \lambda_{H}    &   \frac{\lambda_{H\Phi}}{2}    &   \frac{\lambda_{H\chi}}{2}  \\
 \frac{\lambda_{H\Phi}}{2}  & \lambda_{\Phi}  &    \frac{\lambda_{\Phi\chi}}{2}  \\
 \frac{\lambda_{H\chi}}{2}  &  \frac{\lambda_{\Phi\chi}}{2}  & \lambda_{\chi}  \\
\end{pmatrix}
\end{align}
%%%%%%%%%%%%%%%%%%%%%%%%%%%%%%%%%%%%%%%%%%%%%%%%%%%%%%%
Requiring such a matrix to be positive-definite, we obtain the following conditions,
\begin{align}
 &\lambda_{H} > 0,
 \hspace{1cm}
 4\lambda_{H}\lambda_{\Phi}-\lambda_{H\Phi}^{2}>0 ,\nonumber \\
 &(-\lambda_{H}\lambda_{\Phi\chi}^{2}+\lambda_{H\Phi}\lambda_{\phi\chi}\lambda_{H\chi}-\lambda_{\Phi}\lambda_{H\chi}^{2}+4\lambda_{H}\lambda_{\Phi}\lambda_{\chi}-\lambda_{H\Phi}^{2}\lambda_{\chi})>0.
\label{eq:stability2}
\end{align}
%%%%%%%%%%%%%%%%%%%%%%%%%%%%%%%%%%%%%%%%%%%%
One needs to satisfy the above conditions at every energy scale to have a stable vacuum. To ensure perturbativity, we take a conservative approach that $\lambda_i\leq 4\pi$ and $g_i (x_H, x_\Phi) \leq \sqrt{4\pi}$ where $g_i$ denotes the respective gauge couplings of the model.
%%%%%%%%%%%%%%%%%%%%%%%%%%%%%%%%%%%%%%%%%%%%%%%%%%
\subsection{Higgs Invisible decay}
\label{sec:inv-higgs}
%%%%%%%%%%%%%%%%%%%%%%%%%%%%%%%%%%%%%%%%%%%%%%%%%%%%%%%
In this section, we discuss the invisible Higgs constraints on the relevant parameter space of Higgs bosons which follow from searches performed at LHC. 
If either of $M_{\chi_R}$ or $M_{\chi_I}$ is smaller than $m_{h_{1,2}}/2$ then these two channels will also contribute to the invisible mode. The partial decay width to $\chi_{R}\chi_R$ and $\chi_I\chi_I$ are given as follows:
%%%%%%%%%%%%%%%%%%%%%%%%%%%%%%%%%%%%%%%%%%%%%%%%%%%%%%%%%%%%%
\begin{align}
\Gamma(h_1\to\chi_R\chi_R)&=\frac{1}{32\pi m_{h_1}}\big(\lambda_{H\chi}v_H \cos\alpha + \lambda_{\Phi\chi}v_\Phi\sin\alpha + \sqrt{2}\lambda_{\Phi\chi\chi}\sin\alpha \big)^2\sqrt{1-\frac{4M_{\chi_R}^2}{m_{h_1}^2}}\\
\Gamma(h_1\to\chi_I\chi_I)&=\frac{1}{32\pi m_{h_1}}\big(\lambda_{H\chi}v_H \cos\alpha + \lambda_{\Phi\chi}v_\Phi\sin\alpha - \sqrt{2}\lambda_{\Phi\chi\chi}\sin\alpha \big)^2\sqrt{1-\frac{4M_{\chi_I}^2}{m_{h_1}^2}}
\end{align}
%%%%%%%%%%%%%%%%%%%%%%%%%%%%%%%%%%
Hence the total invisible decay width of SM Higgs boson $h_1$ is given as
%%%%%%%%%%%%%%%%%%%%%%%%%%%%%%%%%
\begin{align}
\Gamma^{\text{inv}}(h_1)=\Gamma(h_1\to\chi_R\chi_R)+\Gamma(h_1\to\chi_I\chi_I)
\end{align}
%%%%%%%%%%%%%%%%%%%%%%%%%%%%%%%%%%%%%%%%%%%%%%%%
Accordingly, the invisible branching ratio for $h_1$ is given by 
%%%%%%%%%%%%%%%%%%%%%%%%%%%%%%%%%%%%
\begin{align}
\text{BR}^{\text{inv}}(h_1)=
\frac{\Gamma^{\text{inv}}(h_1)}{\cos^2\alpha\Gamma^\text{SM}(h_1)+\Gamma^{\text{inv}}(h_1)},
\label{eq:inv-BR-higgs}
\end{align}
%%%%%%%%%%%%%%%%%%%%%%%%%%%%%%%%%%%%%%%%%%%%%%%%%%%%%%%%%%
where $\Gamma^\text{SM}(h_1)=4.1$ MeV. The upper bound on the invisible branching ratio from CMS experiment~\cite{CMS:2018yfx}~\footnote{The present bound from ATLAS for invisible Higgs decays is $\text{BR}^{\text{inv}}(h_1) \leq 0.26$~\cite{ATLAS:2019cid}.},
%%%%%%%%%%%%%%%%%%%%%%%%%%%%%%%%%%%%%%%
\begin{align}
\text{BR}^{\text{inv}}(h_1) \leq 0.19.
\label{eq:invisible3}
\end{align}
%%%%%%%%%%%%%%%%%%%%%%%%%%%%%%%%%%%%%%%%%%%%%%%%%
In the case of $M_\chi < m_{h_1}/2$ and $\sin\alpha\sim 0$, the invisible Higgs decay constraint can be translated as an upper bound on the quartic coupling $\lambda_{H\chi}$: 
%%%%%%%%%%%%%%%%%%%%%%%%%%%%%%%%%%%
\begin{align}
\lambda_{H\chi}\left(1-\frac{4M_\chi^2}{m_{h_1}^2}\right)^{\frac{1}{4}}\leq 9.8\times 10^{-3}.
\end{align}
%%%%%%%%%%%%%%%%%%%%%%%%%%%%%%%%%%%%%%%%%%%%%%%%%%%%%%%%%%
Notice that in case of $m_{\chi}<m_h/2$ and non-zero values of mixing parameter $\sin\alpha$, the exclusion limit on $\lambda_{H\chi}$ depends on other quartic couplings such as $\lambda_{\Phi\chi}$, $\lambda_{\Phi\chi\chi}$ and on the vev $v_\Phi$.

%%%%%%%%%%%%%%%%%%%%%%%%%%
\subsection{Collider constraints on \texorpdfstring{$g_1'-M_{Z'}$}{g1pmzp}}
\label{sec:collider-constraints}
%%%%%%%%%%%%%%%%%%%%%%%%%%%%%%%%%%%%%%%%%%%%%%%%

%%%%%%%%%%%%%%%%%%%%%%%%%%%%%%%%%%%%%%%%
\begin{figure}[t!]
\begin{center}
\includegraphics[height=5cm, width=7.2cm]{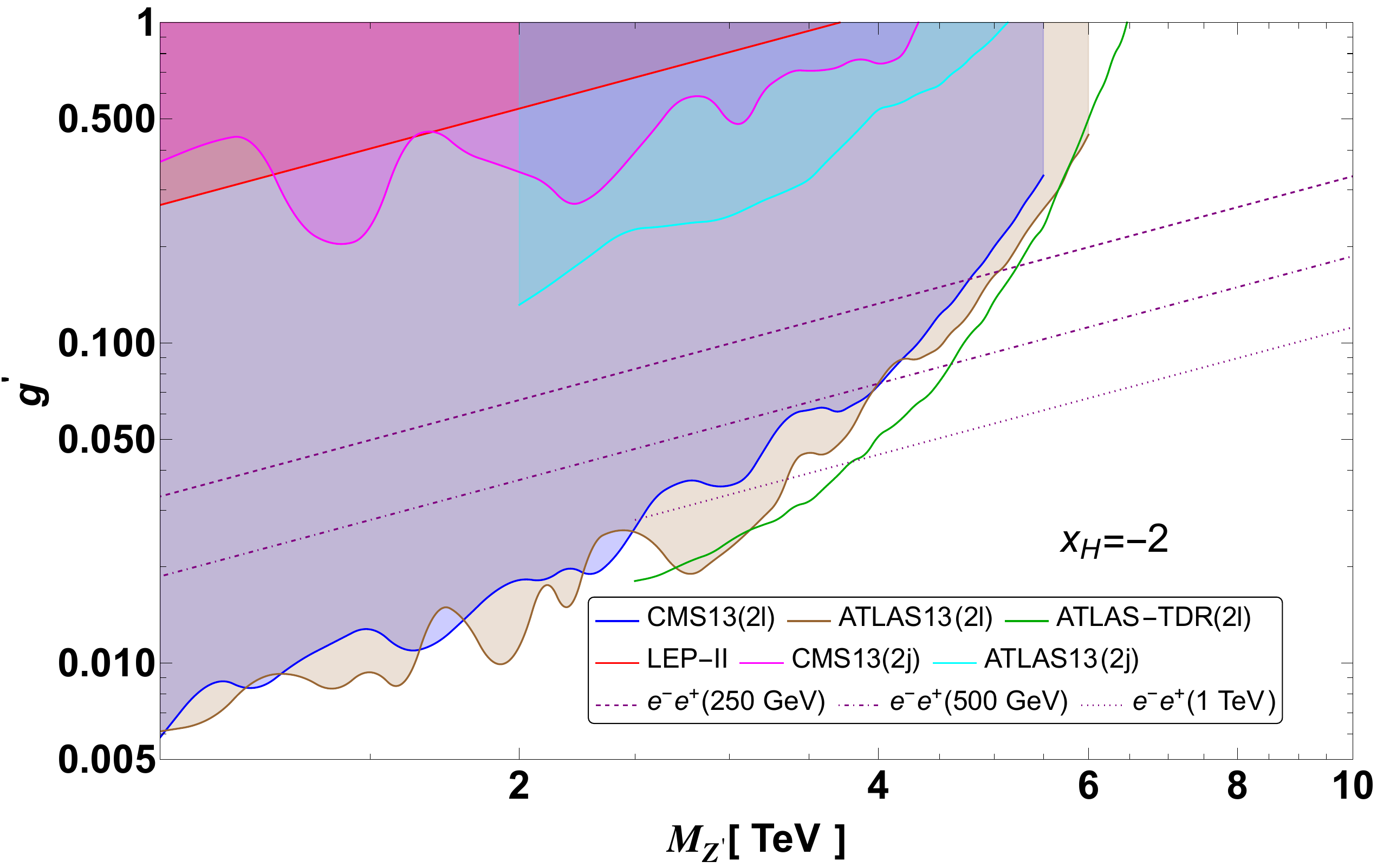}
\includegraphics[height=5cm, width=7.2cm]{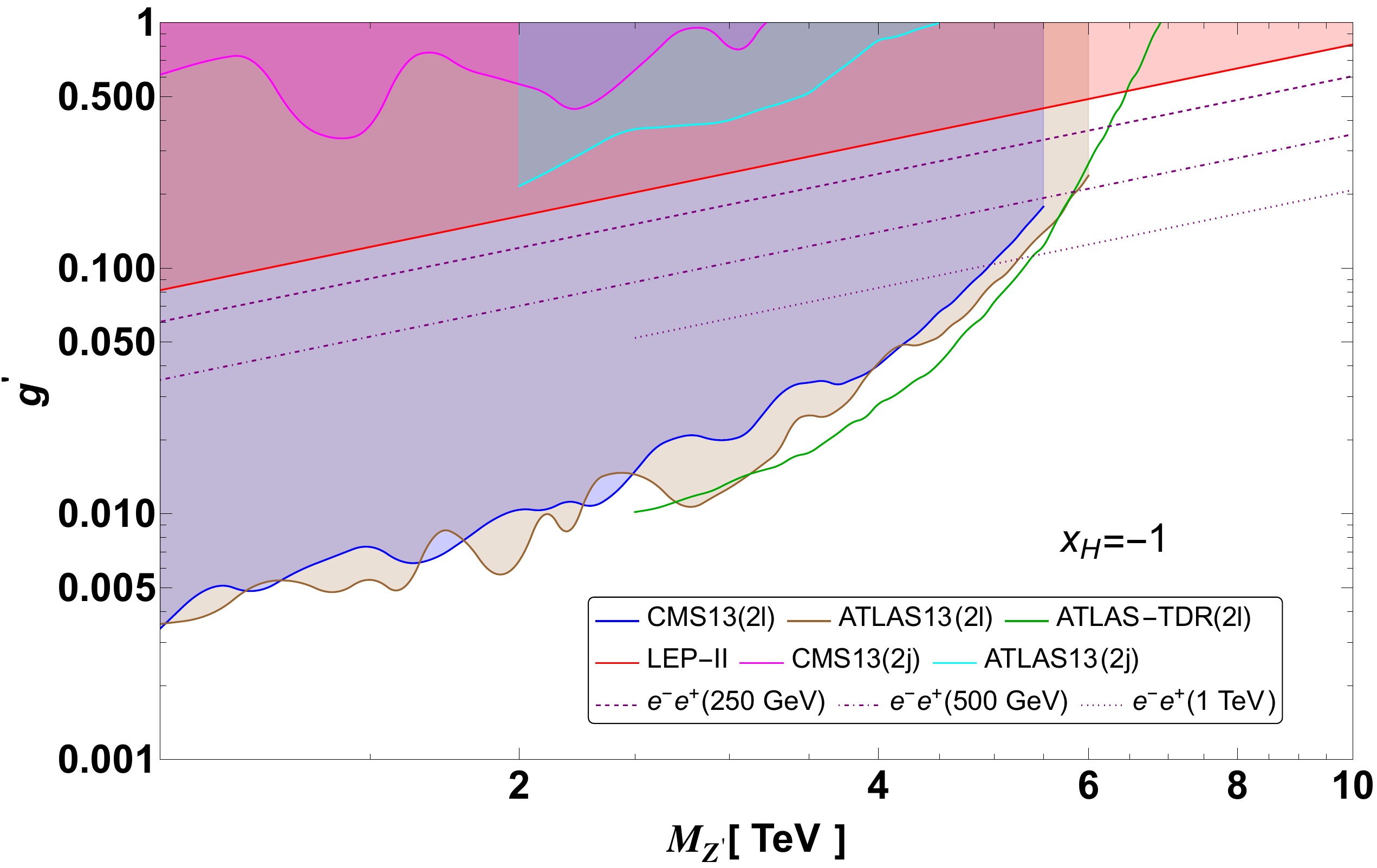}
\includegraphics[height=5cm, width=7.2cm]{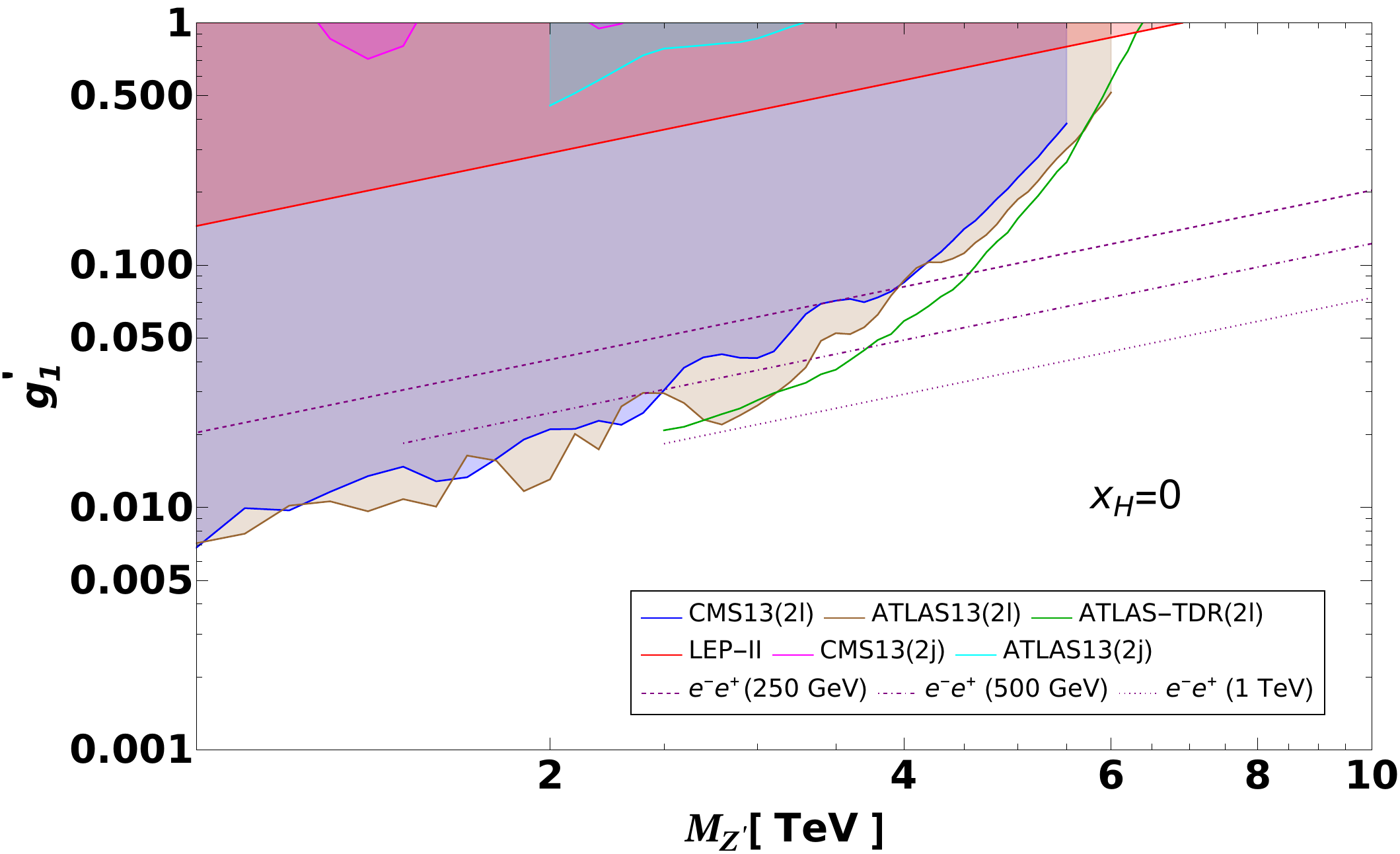}
\includegraphics[height=5cm, width=7.2cm]{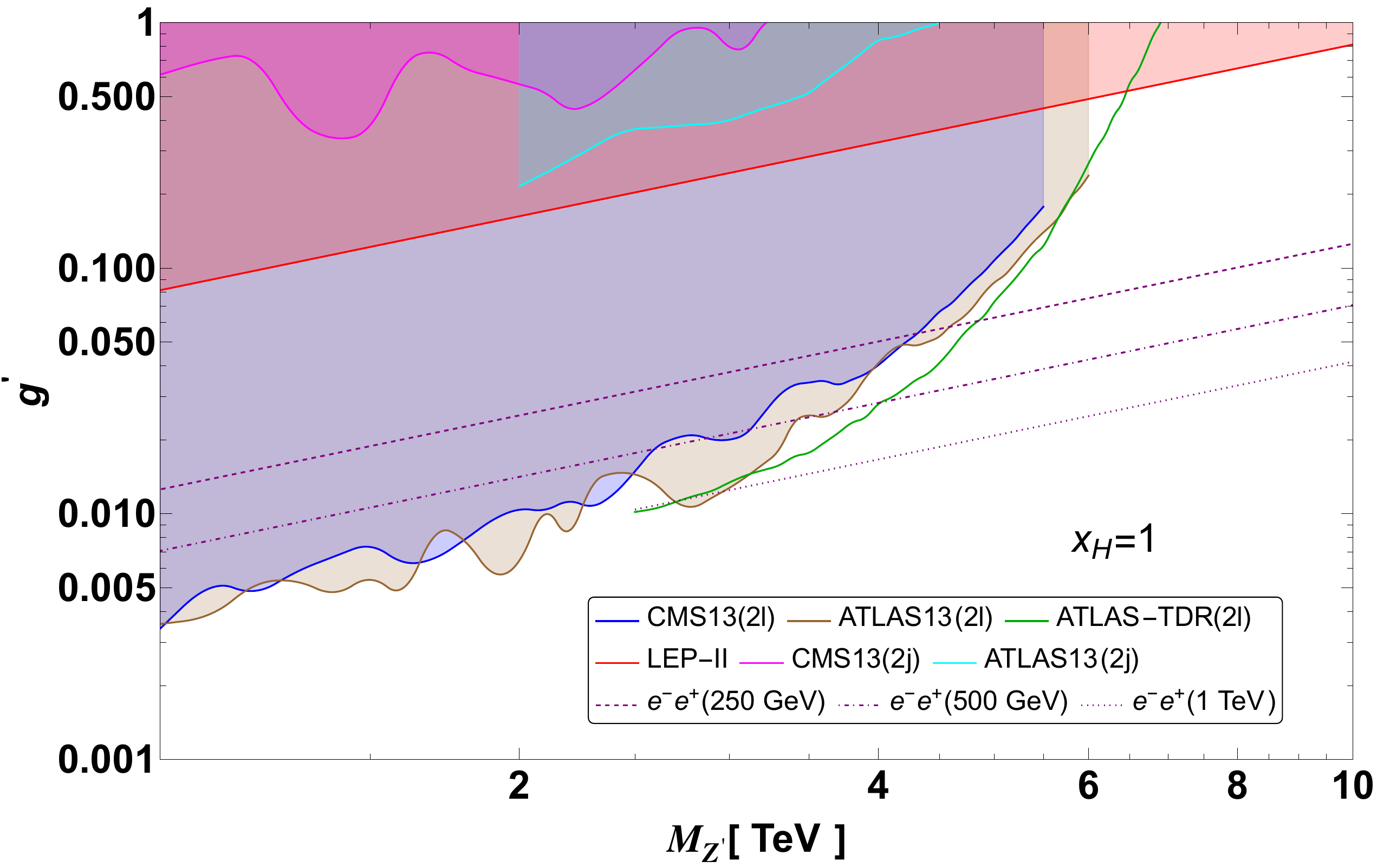}
\includegraphics[height=5cm, width=7.2cm]{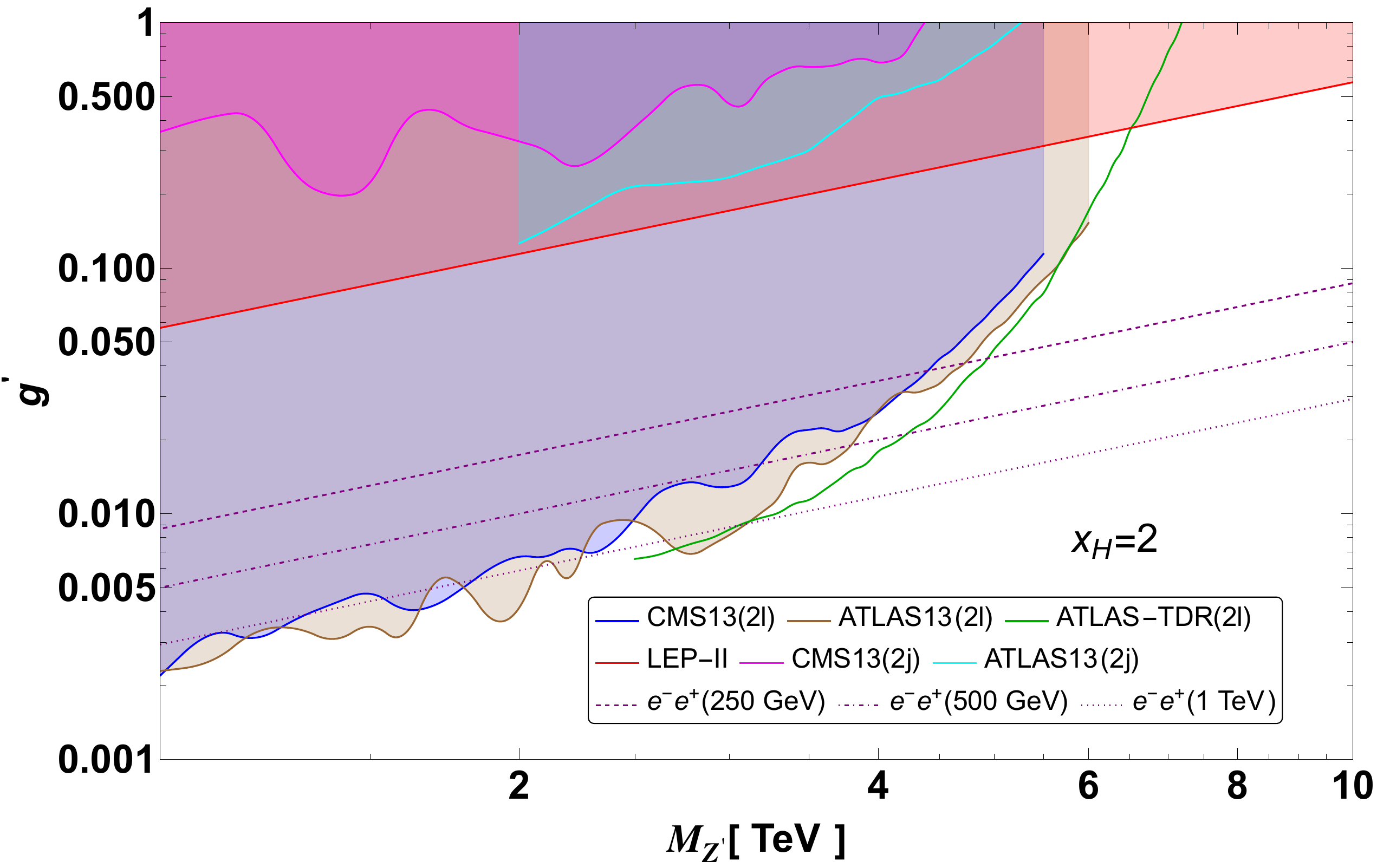}
\caption{Exclusion limits on $g_1'$ as a function of $M_{Z'}$ for various values of $x_H$ and with $x_{\Phi}=1$. The coloured regions are excluded by the experimental data from LEP-II~\cite{ALEPH:2013dgf}, and LHC dilepton~\cite{ATLAS:2019erb,CMS:2019tbu}, LHC dijet~\cite{ATLAS:2019bov,CMS:2018mgb} searches. Also, we have shown in unshaded curves the prospected limit coming from future HL-LHC as well as the ILC~\cite{Das:2021esm}.}
\label{fig:collider-constraints}
\end{center}
\end{figure} 
%%%%%%%%%%%%%%%%%%%%%%%%%%%%%%%%%%%%%%

In this section we will employ the most recent collider results to derive constraints on the model parameters such as $g_{1}', M_{Z'}$ and $x_H$. In the ATLAS and CMS collaborations analysis, sequential SM $Z'$ model~\cite{Barger:1980ti} has been considered as a reference model. We can easily translate the constraints of the sequential SM $Z'$ model to our $U(1)_X$ model parameters. 
For example, we can obtain limits on $M_{Z'}/g_{1}'$ for different values of $x_H$ with fixed $x_{\Phi}=1$, as shown in Fig.~\ref{fig:collider-constraints}. The various shaded regions in Fig.~\ref{fig:collider-constraints} show the excluded limit from ATLAS and CMS search for $Z'$ in both dilepton~\cite{CMS:2019tbu} and dijet~\cite{CMS:2018mgb} channels. The red-shaded region is excluded from LEP-II~\cite{ALEPH:2013dgf}, while the unshaded magenta dot-dashed, dashed and dotted lines are from the future ILC prospects~\cite{Das:2021esm} for $\sqrt{s}= 250$ GeV, 500 GeV and 1 TeV, respectively considering $M_{Z^\prime} >> \sqrt{s}$.
% CMS:2019tbu
For a detailed discussion of the methodology to derive these limits, see Ref.~\cite{Das:2021esm}. Note that the limit on $g_1'-M_{Z'}$ varies depending on the values of $x_H$ as couplings between SM particles and $Z'$ also vary. For example, with $x_H =-2$, $Z'$ interaction with left-handed quarks or leptons are absent, right-handed charged-leptons and $Z'$ interaction are absent in  $x_H =-1$ case. Similarly, for the $x_H = 1$ case, the right-handed down-type quarks have no interaction with $Z'$. For $x_H=0$ and 2, all the SM particles have non-trivial coupling with $Z'$.
%%%%%%%%%%%%%%%%%%%%%%%%%%%%%%%%%%%%%%%%%%%%%%%%%%%%%%%%%%%%%%%%%%%%%%%%%%%
From Fig.~\ref{fig:collider-constraints}, we see that the most stringent constraint up to $M_{Z'}= 6$ TeV comes from LHC dilepton channels. Above $M_{Z'}= 6$ TeV, the resonant $Z'$ production is kinematically limited at $\sqrt{s} = 13$ TeV LHC. Due to this same reason, one does not expect further improvement at HL-LHC. On the other hand, from the projected sensitivities we see that the lepton colliders do better for heavy $Z'$ bosons compared to the LHC limit. We will consider a specific benchmark value of $M_{Z'}$ and $g_1'$, which is allowed by the current limit, for the rest of this paper.
%%%%%%%%%%%%%%%%%%%%%%%%%%%%%%%%%%%%%%%%%%%%%%%%%%%%%%%%%%%%%%%%%%%%%%%%%%%%%
\subsection{Bounds on the mixing parameter between physical mass eigenstates}
\label{sec:scalar-mix}
%%%%%%%%%%%%%%%%%%

%%%%%%%%%%%%%%%%%%%%%%%%%%%
\begin{figure}[t]
\begin{center}
\includegraphics[height=10cm, width=14cm]{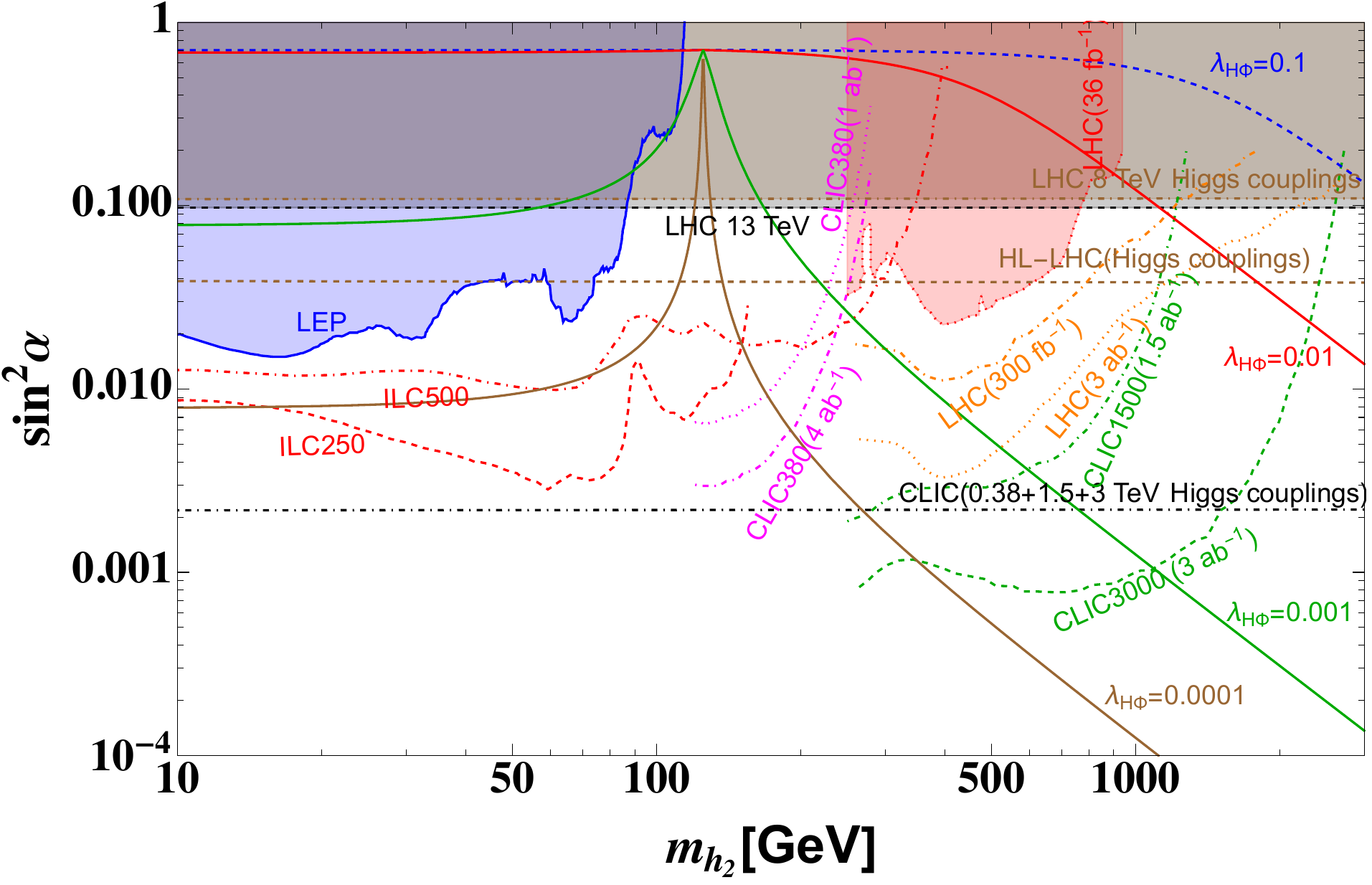}
\caption{Bounds on the mixing parameter $\sin^2\alpha$ between the as a function of $m_{h_2} $from LHC \cite{deBlas:2018mhx}, LEP \cite{LEPWorkingGroupforHiggsbosonsearches:2003ing}, prospective colliders like ILC \cite{Wang:2020lkq} and CLIC \cite{deBlas:2018mhx} respectively. Shaded regions are ruled out by corresponding experiments. Limits on the scalar mixing parameter from Eq.~\ref{coup} are also shown for different $\lambda_{H\Phi}=$0.1, 0.01, 0.001 and 0.0001 respectively.}
\label{fig:mix}
\end{center}
\end{figure} 
%%%%%%%%%%%%%%%%%%%%%%%%%%%%%%%%%%%%%%%%%%%

We summarize the bounds on the scalar mixing angle $\alpha$ from the LHC \cite{deBlas:2018mhx}, LEP \cite{LEPWorkingGroupforHiggsbosonsearches:2003ing} results, prospective colliders like ILC \cite{Wang:2020lkq} and CLIC \cite{deBlas:2018mhx} in Fig.~\ref{fig:mix}. The prospective limits on the mixing parameter $\sin\alpha$ from the $\sqrt{s}=250$ GeV and 2 ab$^{-1}$ luminosity are shown by the red dashed line with polarization effect $|P(e^+ e^-)|=(30\%, 80\%)$ from the $Z h_2$ mode where $Z$ boson decays to light charged leptons. Corresponding bounds from the $\sqrt{s}=500$ GeV are shown by the red dot-dashed line for $Z\to \mu^+\mu^-$ events. Whereas the LEP bounds shown by the blue solid line consider both $Z\to \mu^+\mu^-$ and $e^+e^-$ modes. For direct comparison with the LEP bounds ILC projections can be scaled by a factor of $\frac{1}{\sqrt{2}}$ assuming $Z \to e^-e^+$ channel is similar to $Z \to \mu^- \mu^+$ channel as analyzed in \cite{LEPWorkingGroupforHiggsbosonsearches:2003ing}. Bounds on the scalar mixing parameter from the Higgs couplings at the 8 TeV LHC and High Luminosity LHC (HL-LHC) are shown by the brown dot-dashed and dashed lines respectively. The bounds are taken from \cite{deBlas:2018mhx}. The bounds at the 13 TeV LHC are shown by the black dashed line using the signal strength at $95\%$ C.L. from the ATLAS results \cite{ATLAS:2020qdt}. The corresponding CMS signal strength can be found in \cite{CMS:2020gsy} which gives comparatively weaker bounds. The prospective bounds from the combined CLIC analysis art 380 GeV, 1.5 TeV and 3 TeV are shown by the black dot-dashed line. Considering the VEV of the $U(1)_X$ theory at 50 TeV we show the limits on the scalar mixing from the Eq.~\ref{coup} considering four different choices of $\lambda_{H\Phi}$ as $0.1$, $0.01$, $0.001$ and $0.0001$ respectively for $10$ GeV $\leq m_{h_2} \leq 3000$ GeV. 
%ATLAS:2020qdt
%CMS:2020gsy
Bounds on the mixing parameter at $95\%$ C. L. from the $h_2 \to Z Z$ mode at 36 fb$^{-1}$ luminosity is shown by the red dotted line \cite{Buttazzo:2018qqp}. The shaded regions with different colours are ruled out by different experiments. The prospective limits on the mixing parameter form the CLIC at $\sqrt{s}=380$ GeV are shown by the dotted (dot-dashed) magenta line for 1 (4) ab$^{-1}$ luminosity from \cite{Mekala:2021uvg}. The prospective limits on the mixing parameter from $h_2 \to hh (4b)$ from CLIC at $\sqrt{s}=$1.5 (3) TeV is shown by green dot-dashed (dashed) line from \cite{Buttazzo:2018qqp,deBlas:2018mhx}. The prospective bounds on the mixing parameter estimated from $h_2 \to ZZ$ mode at the LHC at 300 (3000) fb$^{-1}$ luminosity using orange dot-dashed (dotted) line from \cite{Buttazzo:2018qqp}.
%%%%%%%%%%%%%%%%%%%%%%%%%%%%%
\section{Phenomenology of dark matter}
\label{sec:dark-matter2}
%%%%%%%%%%%%%%%%%%%%%%%%%%%%%%%%%%%%%%%%%%%%%%%
We collect the results of our DM analysis in this section. In our model, a two-component DM scenario is possible due to the unbroken $\mathbb{Z}_2\otimes\mathbb{Z}'_2$ symmetry. We choose the $\mathbb{Z}_2$-odd fermion $N_3$ in set one and lightest of the $\mathbb{Z}'_2$-odd scalar $\chi_R$ or $\chi_I$ in set two as DM candidates. In our analysis, we assume scalar DM candidate as $\chi_R$, with the condition $\lambda_\Phi\chi\chi<0$~(the opposite scenario with $\lambda_{\Phi\chi\chi}>0$ would have $\chi_I$ as the DM particle). These two DM candidates must satisfy the following two experimental constraints:
\begin{itemize}
\item The relic density coming from Planck satellite data~\cite{Planck:2018vyg}
\begin{align}
\Omega_{\text{DM}} h^2 = 0.12\pm 0.001
\label{eq:relic-density2}
\end{align}
The total relic abundance of DM in our model is given by the sum of the scalar~($\chi$) and fermion~($N_3$) relic abundances:
\begin{align}
\Omega_{\text{DM}} h^2=\Omega_{\chi}h^2+\Omega_{N_3}h^2
\end{align}
Only for solutions falling exactly within the band given in Eq.~\eqref{eq:relic-density2} the totality of the DM can be explained by $\chi$ and $N_3$.
\item Direct detection cross-section of DM scattering of nucleon set by various experiments such as XENON1T~\cite{XENON:2018voc}, LUX~\cite{LUX:2016ggv} and PandaX-II~\cite{PandaX-II:2016vec}.
\end{itemize}

We implemented the model in the SARAH package~\cite{Staub:2015kfa} to calculate all the vertices, mass matrices, tadpole equations etc. The thermal cross sections and DM relic abundance are determined using micrOMEGAS-5.0.8~\cite{Belanger:2020gnr}. Even though the model introduces new free parameters, not all of them are important to DM analysis. For example, self-quartic coupling $\lambda_{\chi}$ does not play any role in DM phenomenology. Hence we choose to fix $\lambda_\chi=0.1$ in our analysis. The remaining free parameters relevant for DM analysis can be chosen as:
%%%%%%%%%%%%%%%%%%%%%%%%%%%%%%%%%%%%%%%%%%%%%%%%%%%%%%%%%
\begin{align}
m_{h_2},\,\, \sin\alpha,\,\, g_1',\,\, M_{Z'},\,\, x_H,\,\, \lambda_{H\chi},\,\, \lambda_{\phi\chi} \text{ and } \lambda_{\phi\chi\chi}.
\end{align} 
%%%%%%%%%%%%%%%%%%%%%%%%%%%%%%%%%%%%%%%%%%%%%%%%%%%%%%

In the next sections, we will study how the DM phenomenology of this model depends on the free parameters and to do that we choose the following benchmark points which are allowed from all the above-mentioned constraints:
%%%%%%%%%%%%%%%%%%%%%%%%%%%%%%%%%%%%%%%%%%%%%
\begin{align}
\textbf{BP: }m_{h_2}=1~\text{TeV}, \sin\alpha=0.01, g_1'=0.1, M_{Z'}=5~\text{TeV and } x_H=-1.
\label{eq:BP}
\end{align} 
%%%%%%%%%%%%%%%%%%%%%%%%%%%%%%%%%%%%%%%%
For simplicity, we further assume $\lambda_{H\chi}=\lambda_{\Phi\chi}\equiv\lambda$. We choose $\lambda_{\Phi\chi\chi}$ very small and negative throughout our DM analysis.
%%%%%%%%%%%%%%%%%%%%%%%%%%%%%%%%%%%%%%%%%
\subsection{Relic density}
\begin{figure}[htbp]
\begin{center}
\includegraphics[height=5cm, width=12cm]{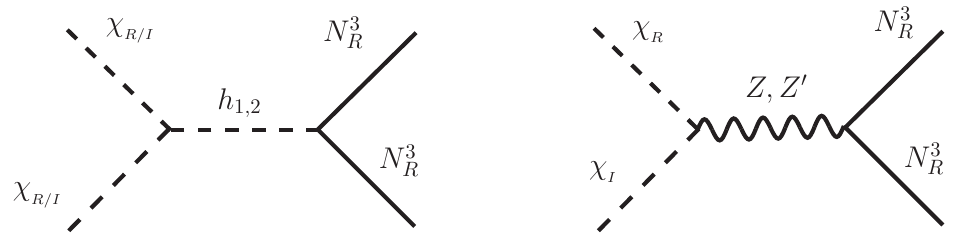}
\includegraphics[height=5cm, width=12cm]{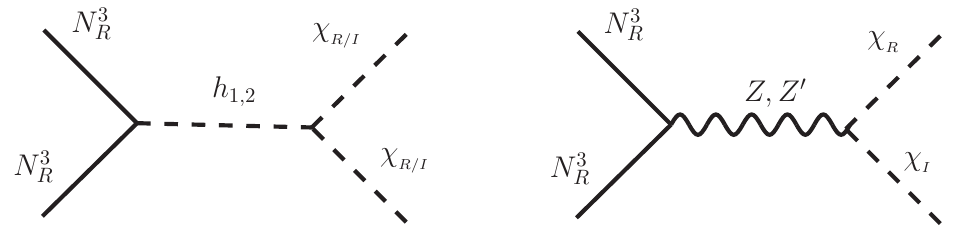}
\caption{The conversion channels which contribute to the two-component DM scenario.}
\label{fig:diagram-conversion-annihilation}
\end{center}
\end{figure}

%%%%%%%%%%%%%%%%%%%%%%%%%%%%%%%%%%%%%%%%%%%%%%%%%%%%%%%%%%%%%%%%%%%%%%%%%%%%%%
There are several annihilation and co-annihilation diagrams which contribute to the relic abundances of DM candidates, $\chi$ and $N_3$. We have listed the cubic and quartic scalar couplings in table.~\ref{tab:cubic-coupling} and \ref{tab:quartic-coupling}. We collect all the Feynman diagrams contributing to $\chi_R,\,N_3$ in figs.~\ref{fig:annihilation-diagram-scalar},~\ref{fig:annihilation-diagram-fermion} of Appendix~\ref{app:feynman diagram}. We find that the relic density of scalar DM $\chi_R$ is mostly determined by CP-even scalars~($h_{1,2}$) and gauge-bosons~($Z,Z'$)-mediated s-channel annihilation and co-annihilation to SM final states~($\ell^+\ell^-$, $q\bar{q}$, $W^+W^-$, $ZZ$, $h_1h_1$) as well as to $N_{1,2}N_{1,2}$, $h_2h_2$ and $Z'Z'$ final states. A sub-dominant role is played by annihilation into $h_1 h_1,h_2h_2$ and $ZZ,Z'Z'$ via the direct 4-point vertices $h_1^2\chi_{R/I}^2$, $h_2^2\chi_{R/I}^2$ and $Z^2\chi_{R/I}^2$, $Z'^{2}\chi_{R/I}^2$, respectively. Also, there could be an additional contribution from $\chi_{R/I}$ exchange in the t-channel. The fermionic DM $N_3$ relic density is determined by the $h_{1,2}/Z/Z'$-mediated s-channel annihilation to SM final states as well as to $N_{1,2}N_{1,2}$, $Z'Z'$, $Zh_{1,2}$ and $Z'h_{1,2}$. Besides these above DM annihilation channels, one also needs to take into account the possible conversion of one DM particle into the other, $\chi\chi\leftrightarrow N_3 N_3$. These are shown in Fig.~\ref{fig:diagram-conversion-annihilation}, which are mediated by s-channel $h_{1,2}/Z/Z'$.

%%%%%%%%%%%%%%%%%%%%%%%%%%%%%%%%%%%%%%%%%
\begin{table}[t]
\setlength\tabcolsep{0.25cm}
\centering
\begin{tabular}{| c || c |}
\hline
$\lambda_{abc}$ &  Couplings in terms of Lagrangian parameter \\
\hline
$h_1\chi_R\chi_R$ & $ \lambda_{\Phi\chi}v_{\Phi}\sin(\alpha)+\lambda_{H\chi}v_{H} \cos\alpha + \sqrt{2}\lambda_{\Phi\chi\chi}\sin\alpha $ \\
$h_2\chi_R\chi_R$ & $ \lambda_{\Phi\chi}v_{\Phi}\cos\alpha - \lambda_{H\chi}v_{H}\sin\alpha + \sqrt{2}\lambda_{\Phi\chi\chi}\cos\alpha$ \\
$h_1\chi_I\chi_I$ & $\lambda_{\Phi\chi}v_{\Phi}\sin\alpha+\lambda_{H\chi}v_{H} \cos\alpha-\sqrt{2}\lambda_{\Phi\chi\chi}\sin\alpha$ \\
$h_2\chi_I\chi_I$ & $\lambda_{\Phi\chi}v_{\Phi}\cos\alpha-\lambda_{H\chi}v_{H}\sin\alpha-\sqrt{2}\lambda_{\Phi\chi\chi}\cos\alpha$ \\
\hline
\hline
$Z\chi_I\chi_R$ &   $x_{\Phi}g^{\prime}_1\sin{\theta^\prime}(p^{\mu}_{\chi_R}-p^{\mu}_{\chi_I})$\\
$Z'\chi_I\chi_R$ &  $x_{\Phi}g^{\prime}_1\cos{\theta^\prime}(p^{\mu}_{\chi_R}-p^{\mu}_{\chi_I})$\\
\hline
\hline
$h_1N^3_RN^3_R$   &  $\frac{Y^3_{M}}{\sqrt{2}}\sin{\alpha}$  \\
$h_2N^3_RN^3_R$   &  $\frac{Y^3_{M}}{\sqrt{2}}\cos{\alpha}$ \\
$ZN^3_RN^3_R$   &  $x_{\Phi}g^{\prime}_1\sin{\theta^\prime}\gamma^{\mu}\gamma^{5}$\\
$Z'N^3_RN^3_R$  &  $x_{\Phi}g^{\prime}_1\cos{\theta^\prime}\gamma^{\mu}\gamma^{5}$\\
\hline
\end{tabular}
\caption{The cubic couplings of the DM scalar and fermion.}
\label{tab:cubic-coupling}
\end{table}
%%%%%%%%%%%%%%%%%%%%%%%%%%%%%%%%%%%%%%%%%%%%%%%%
\begin{table}[t]
\setlength\tabcolsep{0.25cm}
\centering
\begin{tabular}{| c || c |}
\hline
$\lambda_{abc}$ &  Couplings in terms of Lagrangian parameter \\
\hline
$h_1h_1\chi_{R(I)}\chi_{R(I)}$ & $\lambda_{\Phi\chi}\sin^2\alpha+\lambda_{H\chi}\cos^2\alpha $\\
$h_1h_2\chi_{R(I)}\chi_{R(I)}$ & $\sin\alpha\cos\alpha(\lambda_{\Phi\chi}-\lambda_{H\chi}) $ \\
$h_2h_2\chi_{R(I)}\chi_{R(I)}$ & $\lambda_{\Phi\chi}\cos^2\alpha+\lambda_{H\chi}\sin^2\alpha$\\
\hline
\hline
$ZZ\chi_{R(I)}\chi_{R(I)}$ & $2 x^2_{\Phi}(g^{\prime}_1)^2\sin^2{\theta^\prime} g_{\mu\nu}$\\
$ZZ'\chi_{R(I)}\chi_{R(I)}$ & $2 x^2_{\Phi}(g^{\prime}_1)^2\sin{\theta^\prime}\cos{\theta^\prime} g_{\mu\nu}$\\
$Z'Z'\chi_{R(I)}\chi_{R(I)}$ & $2 x^2_{\Phi}(g^{\prime}_1)^2\cos^2{\theta^\prime} g_{\mu\nu}$\\
%%%%%%%%%%%%%%%%%%%%%%%%%%%%%%%%%%%%%%%%%%%%%%%%%%%%%%%%%%%%%%%
\hline
\end{tabular}
\caption{The quartic couplings of the DM scalar.}
\label{tab:quartic-coupling}
\end{table}
%
%%%%%%%%%%%%%%%%%%%%%%%%%%%%%%%%%%%%%%%%%%%%%%%%%%%%%%%%%%%%
%%%%%%%%%%%%%%%%%%%%%%%%%%%%%%%%%%%%%%%%%%%%%

\begin{figure}[htbp]
\includegraphics[height=5cm, width=7.2cm]{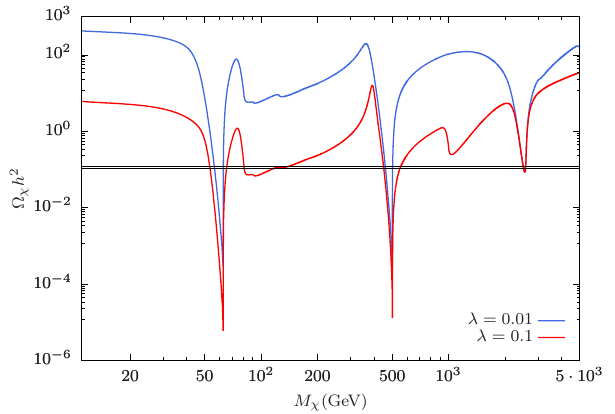}
\includegraphics[height=5cm, width=7.2cm]{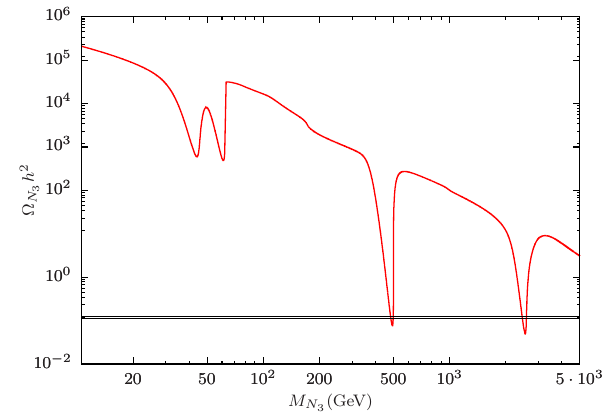}
\caption{Relic density of scalar (right) and fermion~(right) DM as a function of DM mass for one DM candidate. Here, we set the parameter space as \textbf{BP} given in Eq.~\eqref{eq:BP}. Blue and red lines stand for $\lambda=0.01$ and 0.1.}
\label{fig:relic-nocon}
\end{figure}

Firstly, we show the relic density of scalar and fermionic DM in the left and right panel of Fig.~\ref{fig:relic-nocon}, where we neglect the conversion $\chi\chi\leftrightarrow N_3N_3$. In the left panel of Fig.~\ref{fig:relic-nocon} we show the relic density of scalar DM $\chi_R$ for two benchmark points $\lambda=0.01$~(blue line) and 0.1~(red line). We see that there are a few dips in the plot. The dips at $M_{\chi_R}\sim m_{h_1}/2$ and $M_{\chi_R}\sim m_{h_2}/2$ occurs due to annihilation via s-channel $h_1$ and $h_2$ exchange, respectively.

%%%%%%%%%%%%%%%%%%%%%%%%%%%%%%%%%%%%
%%%%%%%%%%%%%%%%%%%%%%%%%%%%%%%%%%%%%%

The annihilation becomes very efficient when the Higgs bosons $h_1$ and $h_2$ are on-shell. For $M_{\chi_R}> 90$ GeV, annihilation are dominated by gauge boson final states $W^+W^-$ and $ZZ$, thus explaining the drop at $M_{\chi_R}\sim 90$~GeV. In the mass range $M_{\chi_R}\geq 125$~GeV, $\chi_R$ also annihilate also into SM-like Higgs bosons $h_1 h_1$. Also, when $M_{\chi_R}>m_t$, a new channel $\chi_R\chi_R\to t\bar{t}$ opens up. For DM mass $M_{\chi_R}\geq m_{h_2}$, $\chi_R\chi_R\to h_2 h_2$ channel opens up and this becomes dominant for large $\lambda$, hence the drop in relic density at $M_{\chi_R}\sim m_{h_2}$ for $\lambda=0.1$~(red line). When the DM mass $M_{\chi_R}\sim M_{Z'}/2$, annihilation through the $Z'$-portal becomes efficient, hence the drop at $M_{Z'}/2$. For very heavy DM mass $M_{\chi_R}$, annihilation cross section drops as $\sim 1/M_{\chi_R}^2$, hence the relic density increases. One more important point to note is that as $\chi_R$ and $\chi_I$ mass difference are small due to the small value of $\lambda_{\Phi\chi\chi}$, co-annihilation channels with $\chi_I$ occur in all regions of the parameter space, with the effect of lowering the relic density. In the right panel of Fig.~\ref{fig:relic-nocon}, we show the relic density for fermionic DM $N_3$. Again the drops at $m_{h_1,h_2}/2$ and $M_{Z'}/2$ due to s-channel annihilation through on-shell $h_{1,2}$ and $Z'$. Note the presence of annihilation dip at $M_{Z}/2$ for fermionic DM, unlike the scalar DM case. The reason behind no dip at $M_Z/2$ for scalar DM is that the $Z$-mediated dip is momentum suppressed. Note that without the conversion $\chi\chi\to N_3 N_3$, the annihilation cross section for $N_3$ has no dependence on $\lambda$, hence only one line instead of two lines in the right panel of Fig.~\ref{fig:relic-nocon}.

Next, we take into account the conversion of two-component DM $\chi\chi\leftrightarrow N_3 N_3$, which can be mediated by s-channel $h_{1,2}/Z/Z'$. In this case, we need to simultaneously follow their abundances in the early Universe. The coupled Boltzmann equations are given by~\cite{Esch:2014jpa,Belanger:2012vp},

%%%%%%%%%%%%%%%%%%%%%%%%%%%%%%%%%%%%%%%%%%%%%%%%%%%%%%%%%%%%%
\begin{align}
& \frac{dY_{\chi}}{dx} = -\sqrt{\frac{45}{\pi}}g^{1/2}_*M_{Pl}\frac{m}{x^2}\Bigg[ \langle \sigma \text{v}\rangle^{\chi\chi \rightarrow FF} \left(Y^2_{\chi}-\overline{Y}^2_{\chi}\right) + \langle \sigma \text{v}\rangle^{\chi\chi\rightarrow N_3 N_3} \left(Y^2_{\chi}-\overline{Y}^2_{\chi}\frac{Y^2_{N_3}}{\overline{Y}^2_{N_3}}\right) \Bigg] 
\end{align}
\begin{align}
& \frac{dY_{N_3}}{dx} = -\sqrt{\frac{45}{\pi}}g^{1/2}_*M_{Pl}\frac{m}{x^2}\Bigg[ \langle \sigma \text{v}\rangle^{N_3 N_3 \rightarrow FF} \left(Y^2_{N_3}-\overline{Y}^2_{N_3}\right) + \langle \sigma \text{v}\rangle^{N_3 N_3\rightarrow \chi\chi} \left(Y^2_{N_3}-\overline{Y}^2_{N_3}\frac{Y^2_{\chi}}{\overline{Y}^2_{\chi}}\right) \Bigg]
\end{align}
here $x=\frac{m}{T}$ and $m=\frac{M_{\chi}+M_{N_3}}{2}$. $\langle \sigma \text{v}\rangle$ is the thermally average annihilation cross-section.
%%%%%%%%%%%%%%%%%%%%%%%%%%%%%%%%%%%%%%%%%%%%%%%%%%%%%%%%%%%%%
which we can be utilized to compute the relic density of both the components, 

%%%%%%%%%%%%%%%%%%%%%%%%%%%%%%%%%%%%%%%%%%%%%%%%%%%%%%%%%%%%%%%%%%%
\begin{figure}[htbp]
\includegraphics[height=5cm, width=7.2cm]{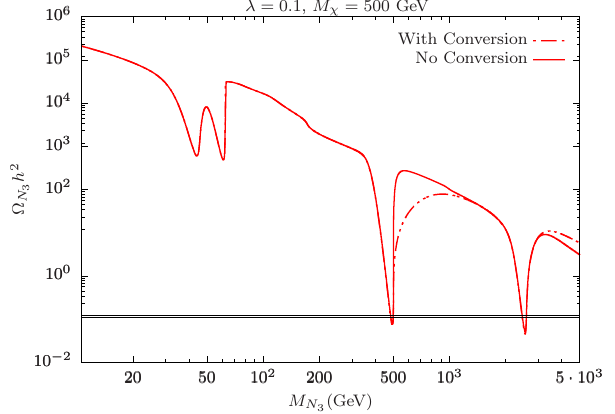}
\includegraphics[height=5cm, width=7.2cm]{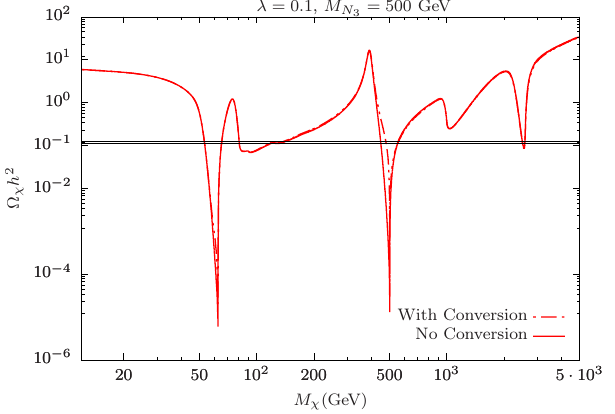}
\caption{Effect of conversion channel $\chi\chi\leftrightarrow N_3 N_3$ on the relic density. The solid and dashed lines stand for the case of without and with the conversion channel. Left(right) panel is for fermionic(scalar) DM with $M_{\chi}=500$~GeV~($M_{N_3}=500$~GeV) and $\lambda=0.1$. The other parameters are fixed as \textbf{BP} given in Eq.~\eqref{eq:BP}.}
\label{fig:relic-con}
\end{figure}
%%%%%%%%%%%%%%%%%%%%%%%%%%%%%%%%%%%%%%%%%%%%%%%%%%%%

In the above, $F$ can be any particle except $\chi$ and $N_3$. In these equations $g^{1/2}_*$, $M_{Pl}$ and $\overline{Y}$ are degrees of freedom, Planck mass and equilibrium value of $Y$, respectively. In the above equations, the most right term is responsible for the conversion process $\chi\chi \leftrightarrow N_3 N_3$. As $\chi\chi\to N_3 N_3$ and $N_3 N_3\to\chi\chi$ are determined by the same squared matrix elements~(see Fig.~\ref{fig:diagram-conversion-annihilation}), they are not independent but related to each other. These conversion processes are mediated by $h_{1,2}/Z/Z'$. Note that the couplings $\chi_R-\chi_I-Z$ and $N_3-N_3-Z$ are suppressed by $\sin\theta'$. Hence, the conversion process mediated by $h_{1,2}/Z'$ gives the dominant contribution, provided $\lambda_{H\chi}$, $\lambda_{\Phi\chi}$, $\sin\alpha$ and $g_1'$ are not too small. The effect of conversion on the $N_3/\chi$ relic density is shown in the left and right panels of Fig.~\ref{fig:relic-con}. The solid and dashed line represents the relic for without-conversion and with-conversion cases. In both panels, we fixed the other DM mass as $M_{\chi}(M_{N_3})=500$~GeV, whereas fixing other parameters same as before. For the fermionic DM, when $M_{N_3}>M_{\chi}$, the larger the quartic coupling $\lambda$, the larger the annihilation rate $N_3 N_3\to \chi\chi$ and hence smaller the relic density. Note that, when $M_{N_3}< M_{\chi}$, the effect of conversion $N_3 N_3\to\chi\chi$ is very small. Also as for the conversion case, $Z'\to \chi\chi$ process contributes to the total decay width of $Z'$, it can greatly enhance the total decay width of $Z'$, which causes the increase of $\Omega h^2$ above $M_{N_3}\sim M_{Z'}/2$ for a small mass of $M_{\chi}$. In addition to this, for $M_{\chi}<m_{h_2}/2$, there can be a modification in $\Omega h^2$ due to the change in the $h_2$ decay width. For the scalar DM, the annihilation channel $\chi\chi\to N_3 N_3$ via $Z$ and $Z'$ are momentum suppressed. The dominating contribution to the conversion comes from $\chi\chi\to N_3 N_3$ annihilation through $h_{1,2}$ mediated processes. But as $h_1-N_3-N_3$ coupling is suppressed by $\sin\alpha$, we see that only changes occurs around $\sim m_{h_2}/2$ for $M_{N_3}=500$~GeV.

%%%%%%%%%%%%%%%%%%%%%%%%%%%%%%%%%%%%%%%%%%%%%%%%%%%%%%%%%%%%%%%
\begin{figure}[htbp]
\includegraphics[height=5cm, width=7.2cm]{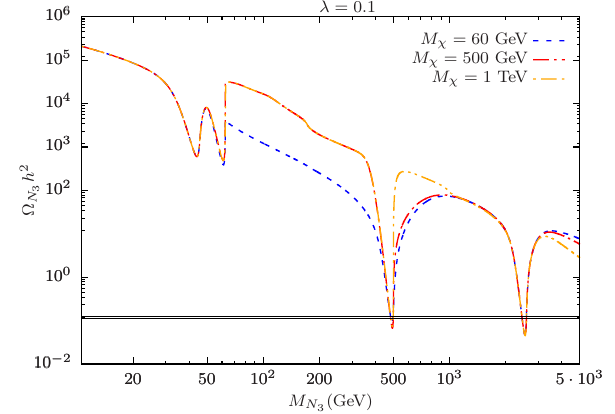} 
\includegraphics[height=5cm, width=7.2cm]{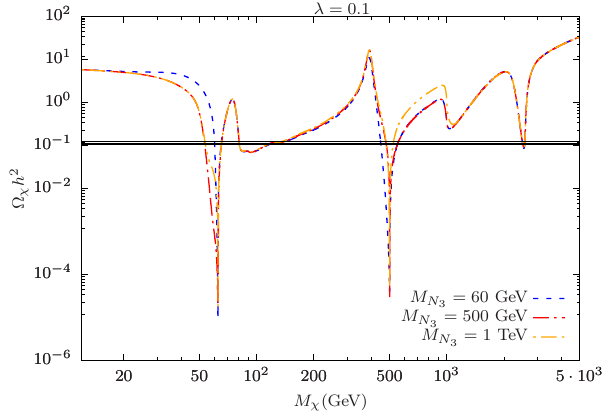}
\caption{The effect of two component DM conversion on $N_3$~(left panel) and $\chi$~(right panel) relic density for fixed $\lambda = 0.1$. The other parameters are fixed as \textbf{BP} given in Eq.~\eqref{eq:BP}.}
\label{fig:relic-mass}
\end{figure}
%%%%%%%%%%%%%%%%%%%%%%%%%%%%%%%%%%%%%%%%%

In Fig.~\ref{fig:relic-mass}, we extended the study of the conversion effect and two DM Boltzmann equations. The left~(right) of Fig.~\ref{fig:relic-mass} shows the $N_3(\chi)$ relic density for $M_{\chi(N_3)}=60, 500$~GeV and 1 TeV with fixed value of $\lambda=0.1$. In the case of fermionic DM, it is clear that for small $M_{\chi}$ the annihilation cross section $N_3 N_3\to \chi\chi$ is large and hence the smaller relic density for $M_{\chi}=60$~GeV compare to other masses. For relatively heavy scalar DM mass, such as $M_{\chi}=500$~GeV and 1 TeV, the conversion effects on the $N_3$ relic density are almost negligible when $M_{N_3}<m_{h_2}/2$. Since, the coupling $h_2-N_3-N_3$ is directly proportional to $M_{N_3}$, in the high mass region, the conversion effect is comparable to the $h_2-N_3-N_3$ coupling effect. This makes the dependence of relic density on $M_{\chi}$ nonlinear in the high $M_{N_3}$ region. On the other hand, for scalar DM the dependence of relic density on fermionic DM mass is simple. As we previously mentioned, the conversion effects mainly come from the channel $\chi\chi\to N_3 N_3$ via $h_1/h_2$. Hence we only see some changes for different $M_{N_3}$ masses either around $M_{\chi}\sim m_{h1}/2$ or around $M_{\chi}\sim m_{h_2}/2$. Also if $M_{N_3}<m_{h_{1,2}}/2$, the $h_{1,2}$ decay width changes, which in turn modifies the relic. In conclusion, $h_i-N_3-N_3$ and $h_i-\chi-\chi$ couplings play important roles in two-component DM conversion. When $M_{\chi}\sim M_{N_3}$, the conversion can take place in both directions, if not the case, only the conversion of heavier one into lighter one is important.

%%%%%%%%%%%%%%%%%%%%%%%%%%%%%%%%%%%%%%%%
\subsection{Direct detection}
%%%%%%%%%%%%%%%%%%%%%%%%%%%%%%%%%%%%%%%%%
The direct detection study of our DM candidates $\chi_R$ and $N_3$ are done here. The current experimental constraints on the DM direct detection assume the existence of only one DM candidate. As in our model, two-component DM candidates are predicted, and the contribution of each candidate to the direct detection cross-section should be rescaled by the fraction contributing to the total relic density. Hence it is convenient to define the fraction of the mass density of $i_{\text{th}}$ DM in the case of multi-component DM~\cite{Cao:2007fy,Wang:2015saa,Aoki:2012ub,Bhattacharya:2013hva}
%%%%%%%%%%
\begin{align}
\epsilon_i=\frac{\Omega_i h^2}{\Omega_{\text{DM}} h^2}
\label{eq:fraction}
\end{align}
%%%%%%%%%%%%%%%%%%%%%%%%%%%%%%%%%%%%%%%%%%%%%%%%%%%%%%
The upper limit on the direct detection now can be recast as 
%%%%%%%%%%%%%%%%%%%
\begin{align}
\frac{\epsilon_\chi}{M_\chi}\sigma_{\chi-N}+\frac{\epsilon_{N_3}}{M_{N_3}}\sigma_{N_3-N}<\frac{\sigma^{\text{exp}}}{M_{\text{DM}}}
\label{eq:direct-detection}
\end{align}
%%%%%%%%%%%%%%%%%%%%%%%%%%%%%%%%%
where, $\sigma_{\chi-N}$ and $\sigma_{N_3-N}$ are the scattering cross section of $\chi$ and $N_3$ with nucleon $N$.

\begin{figure}[htbp]
\centering
\includegraphics[height=5cm, width=7cm]{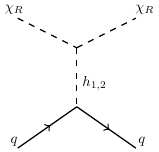}
\includegraphics[height=5cm, width=7cm]{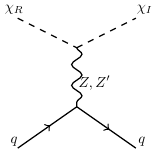}
\caption{The tree-level $\chi$-nuclei scattering Feynman diagrams mediated by Higgs, $Z$, and $Z'$.} 
\label{fig:scalar-dd-feynman}
\end{figure}

%%%%%%%%%%%%%%%%%%%%%%%%%%%%%%%%%%%%%%%%%%%%%%%%%%%%%%%%%%%%
\underline{\bf Estimation of $\sigma_{\chi-N}$: }
%%%%%%%%%%%%%%%%%%%%%%%%%%%%%%%%%%%%%%%%
In this model, the scattering of the scalar DM candidate $\chi_R$ with a nucleon happens via two t-channel diagrams with either $Z,Z'$ or $h_{1,2}$ as propagators, shown in Fig.~\ref{fig:scalar-dd-feynman}. Notice that, as the complex scalar $\chi$ has non-zero $U(1)_X$ charge, the $\chi_R$-nucleon spin-independent~(SI) cross-section can be mediated by the $h_{1,2},Z,Z'$. The non-zero $\lambda_{\Phi\chi\chi}$ implies a small mass splitting between $\chi_R$ and $\chi_I$, so that the interaction through the $Z,Z'$-boson is kinematically forbidden or leads to inelastic scattering. Therefore, the $\chi_R$-nucleon interaction via the Higgs~($h_{1,2}$) will be the dominant one. The effective lagrangian for nucleon-DM interaction can be written as
%%%%%%%%%%%%%%%%%%%%%%%%%%%%%%%%%%%%%
\begin{align}
\mathcal{L}_{\text{eff}}=a_N\bar{N}N\chi_R^2
\end{align}

%%%%%%%%%%%%%%%%%%%%%%%%%%%%%%%%%%%%%
\begin{figure}[htbp]
\centering
\includegraphics[height=5cm, width=7cm]{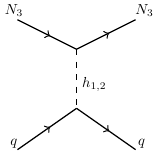}
\includegraphics[height=5cm, width=7cm]{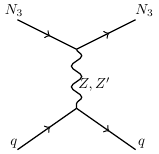}
\caption{\footnotesize{The tree-level $N_3$-nuclei scattering Feynman diagrams mediated by Higgs, $Z$, and $Z'$.} }
\label{fig:fermionic-dd-feynman}
\end{figure}

%%%%%%%%%%%%%%%%%%%%%%%%%%%%%%%%%%%%%%%%%%%%%%%%%%%%%%%%%%%%%
where $a_N$ is the effective coupling between DM and nucleon. The resulting spin-independent scattering cross-section is given by
%%%%%%%%%%%%%%%%%%%%%%%%%%%%%%%%%%%%%%%%%%%
\begin{align}
\sigma^{\text{SI}}_{\chi-N}=\frac{\mu_N^2 m_N^2 f_N^2}{4\pi M_{\chi_R}^2 v_H^2}\Big(\frac{\lambda_{h_1\chi_R\chi_R}}{m_{h_1}^2}\cos\alpha - \frac{\lambda_{h_2\chi_R\chi_R}}{m_{h_2}^2}\sin\alpha\Big)^2,
\label{eq:SI-scalar}
\end{align}
%%%%%%%%%%%%%%%%%%%%%%%%%%%%%%%%%
where $\mu_N=\frac{m_N M_{\chi_R}}{m_N+M_{\chi_R}}$ is the reduced mass for the nucleon-DM system. Here $f_N$ is the form factor, which depends on hadronic matrix elements. The trilinear couplings $\lambda_{h_1\chi_R\chi_R}$ and $\lambda_{h_2\chi_R\chi_R}$ are given as 
%%%%%%%%%%%%%%%%%%%%%%%%%%%%%%%%%%%%%%%%%%%%
\begin{align}
\lambda_{h_1\chi_R\chi_R}&=\lambda_{H\chi}v_H \cos\alpha + \lambda_{\Phi\chi}v_\Phi\sin\alpha + \sqrt{2}\lambda_{\Phi\chi\chi}\sin\alpha \\
\lambda_{h_2\chi_R\chi_R}&=-\lambda_{H\chi}v_H \sin\alpha + \lambda_{\Phi\chi}v_\Phi\cos\alpha + \sqrt{2}\lambda_{\Phi\chi\chi}\cos\alpha 
\label{eq:substitution2}
\end{align}
%%%%%%%%%%%%%%%%%%%%%%%%%%%%%%%%%%%%%%%%%%%%%%%%%%%%%%
The above formula in Eq.~\eqref{eq:SI-scalar} is an extension of the expression corresponding to the singlet scalar DM case~\cite{Cline:2013gha}. The relative negative sign between the $h_1$ and $h_2$ contributions arises in our considered model as the couplings get modified according to Eq.~\eqref{eq:substitution2}. Due to the presence of the two different channels, depending on the parameter space, we can have destructive interference between these two channels, and direct detection can be very small.\\
%%%%%%%%%%%%%%%%%%%%%%%%%%%%%%%%%%%%%%%%%%%%%%%%%%%%%%%%%%%%%%%%%%%%%%%%%%%%%%%%%%
\underline{\bf Estimation of $\sigma_{N_3-N}$: }
%%%%%%%%%%%%%%%%%%%%%%%%%%%%%%%%%%%%%%%%%%%%%%%%%%%%%%%%%%%
Again, for the fermionic case, there will be contributions from t-channel diagrams with either $Z,Z'$ or $h_{1,2}$ as propagators, as shown in the left and right panels of Fig.~\ref{fig:fermionic-dd-feynman}. The $Z,Z'$-mediated diagram contributions to spin-independent cross section are velocity suppressed and hence remain within the experimental bounds~\cite{Arcadi:2020aot}. The Higgs-mediated contribution to the spin-independent cross section can saturate the current experimental bounds.
%
%%%%%%%%%%%%%%%%%%%%%%%%%%%%%%%%%%%%%%%%%%%%%%%%%%%%%%%%%%
%%%%%%%%%%%%%%%%%%%%%%%%%%%%%%
This can be written as~\cite{Lopez-Honorez:2012tov}:
%%%%%%%%%%%%%%%%%%%%%%%%%%%%%%%
\begin{align}
\sigma^{\text{SI}}_{N_3-N}=\frac{(y_M^3)^2 \mu_N^2 m_N^2 f_N^2}{2\pi v_H^2}\sin^2(2\alpha)\Big(\frac{1}{m_{h_1}^2}-\frac{1}{m_{h_2}^2}\Big)^2
\end{align}
%%%%%%%%%%%%%%%%%%%%%%%%%%%%%%%%%%%%%%%%%%%%%
where $\mu_N=\frac{m_N M_{N_3}}{m_N+M_{N_3}}$ is the reduced mass for nucleon-DM system.

\begin{figure}[htbp]
\includegraphics[height=5cm, width=7.2cm]{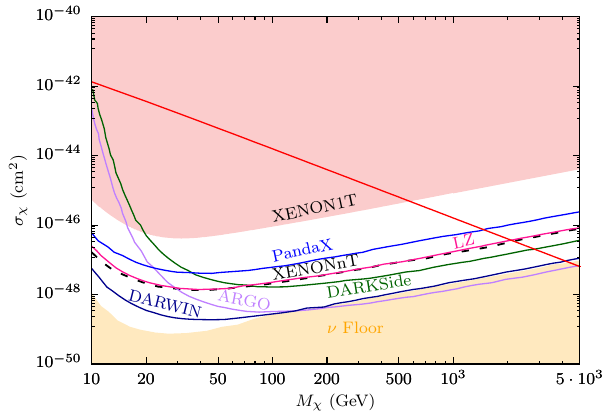}
\includegraphics[height=5cm, width=7.2cm]{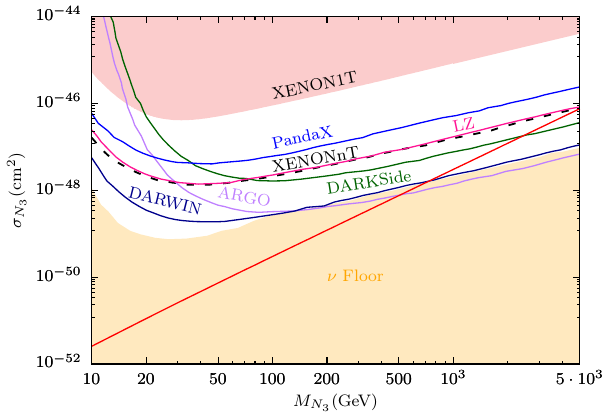}
\caption{The spin-independent DM-nucleon scattering cross section with respect to the DM mass is plotted here. The left and right panels stand for scalar and fermionic DM cases. The light-red shaded region denotes the excluded region coming from the XENON1T experiment~\cite{Aprile:2018dbl}. The light-orange region corresponds to the “neutrino floor” coming from coherent elastic neutrino scattering~\cite{Billard:2013qya}. We have also shown various projected sensitivities coming from experiments such as PandaX-4t~\cite{Zhang:2018xdp}, LUX-ZEPLIN(LZ)~\cite{Akerib:2018lyp,LZ:2022lsv}, XENONnT~\cite{Aprile:2020vtw,Aprile_2023}, DarkSide-20k~\cite{DS_ESPP}, DARWIN~\cite{Aalbers:2016jon} and ARGO~\cite{Billard:2021uyg}.}
\label{fig:dd}
\end{figure}

%%%%%%%%%%%%%%%%%%%%%%%%%%%%%%%%%%%%%%%%%%%%%%%%%%%%%%%%%%%
%
In Fig.~\ref{fig:dd}, the direct detection limit is shown for each DM separately. In both panels, axes are on a log scale, which implies a linear behavior as expected from the DM-nucleon cross-section for both DM. The DM-nucleon cross-section is dominated by Higgs-mediated channels. The fermionic cross section is smaller due to $\sin^2(2\alpha)$ dependence. The black lines in each panel denote the bound from the XENON1T collaboration~\cite{Aprile:2018dbl}. There are constraints from other experiments as well, such as LUX~\cite{Akerib:2016vxi} and PandaX-II~\cite{Tan:2016zwf}, but they are weaker when compared to the XENON1T limit. There are also the projected sensitivities for the PandaX-4t~\cite{Zhang:2018xdp}, LUX-ZEPLIN~(LZ)~\cite{Akerib:2018lyp,LZ:2022lsv}, XENONnT~\cite{Aprile:2020vtw,Aprile_2023}, DarkSide-20k~\cite{DS_ESPP}, DARWIN~\cite{Aalbers:2016jon} and ARGO~\cite{Billard:2021uyg} experiments, which we also show in Fig.~\ref{fig:dd}. The ``neutrino floor'' from coherent elastic neutrino scattering~\cite{Billard:2013qya} is indicated in the orange line. We see from Fig.~\ref{fig:dd} that there might be low-mass solutions with the correct DM relic density. We see that beyond 500 GeV, scalar DM satisfies the XENON1T bound, whereas fermionic DM satisfies the XENON1T bound for the whole DM mass we have considered, although the low fermionic DM mass conflicts with the Neutrino floor.
%DS_ESPP
%%%%%%%%%%%%%%%%%%%%%%%%%%%%%%%%%%%%%%%%%%%%%%%%%%%%%
%DS_ESPP
%%%%%%%%%%%%%%%%%%%%%%%%%%%%%%%%%%%%%%%%%%%%%%%%%%%%%%%%%%%%%%%%%
\begin{table}[ht!]
\setlength\tabcolsep{0.25cm}
\centering
\begin{tabular}{| c | c | c | c | c |}
\hline
$M_{\chi}$ (GeV) & $M_{N_3}$(GeV) & $\epsilon_{\chi}$  &  $ \epsilon_{N_3}$  & $\sum_{\{i=\chi,N_3\}} \epsilon_i \frac{\sigma_{i-N}}{M_i}$ (cm$^2$-GeV$^{-1}$)\\
\hline
500.14 &  499.0  &  0.001268 & 0.99874 &  3.138$\times10^{-51}$\\
62.17 &  2586.11  &  0.000083 & 0.99991 & 6.32$\times10^{-50}$\\
499.65 &  2585.96  &  0.00029 & 0.9997 & 8.27$\times10^{-51}$\\
500.78 &  2585.095  &  0.0043 & 0.9956 & 1.335$\times10^{-50}$\\
61.55 &  2585.348  &  0.00027 & 0.9997 & 1.936$\times10^{-49}$\\
\hline
\end{tabular}
\caption{Benchmark points where both relic abundance and direct detection limit for two-component DM are satisfied.}
\label{tab:DD limit}
\end{table}
%%%%%%%%%%%%%%%%%%%%%%%%%%%%%%%%%%%%%%%%%%%%%%%%%
In table.~\ref{tab:DD limit}, we have shown a few benchmark masses which satisfy both the relic abundance and the direct detection limit for our two components DM case. We have listed a few data points around which many points can be found that satisfy relic density bound also. In conclusion, we can have parameter space, especially around the resonance regions, where one can satisfy both the relic abundance and direct detection limit. Note that for our choice of benchmark $m_{h_2} = 1$ TeV, the mass fraction is very small, i.e $\epsilon_\chi = \mathcal{O}(10^{-3})$. This is due to the tight constraint coming from direct detection constraints for scalar dark matter, see left panel of Fig.~\ref{fig:dd}. This constraint becomes loose for large mass $M_\chi$, hence for a different benchmark such as $m_{h_2} = 2$ TeV, the mass fraction $\epsilon_\chi$ can be large near the resonance region $M_\chi\sim m_{h_2}/2$ where one can have correct relic density.
%%%%%%%%%%%%%%%%%%%%%%%%%%%%%%%%%%%%%%%%%%%%%%%%%%
\section{Relic density dependence on \texorpdfstring{$U(1)_X$}{U(1)X} charge \texorpdfstring{$x_H$}{xh}}
\label{sec:relic-dependence-xH}
%%%%%%%%%%%%%%%%%%%%%%%%%%%%%%%%%%%%%%%%%%%%%%%%%%%
In this model, there are two ways for the DM to interact with the SM particles. Either through the Higgs boson interactions or through the $Z'-$boson interactions as all particles in our model are charged under $U(1)_X$. Hence the relic density for both the DM will have some dependence on $U(1)_X$ charge $x_H$ when one considers the $Z'$-portal DM. Fig.~\ref{fig:relic-nocon}-\ref{fig:relic-con} and \ref{fig:relic-mass} also indicates that the $Z'$ boson resonance effect is very important in reproducing the known DM relic abundance and hence, $M_{\text{DM}}\sim M_{Z'}/2$. Hence, in the case of pure $Z'$-portal DM scenario~($\sin\alpha = 0$), the resultant DM relic abundance is controlled by four free parameters, namely, $g_1'$, $M_{Z'}$, $M_{\text{DM}}$ and $x_H$.

%%%%%%%%%%%%%%%%%%%%%%%%%%%%%%%%%%%%%%%%%%%%%%
\begin{figure}[htbp]
\includegraphics[height=5cm, width=7.2cm]{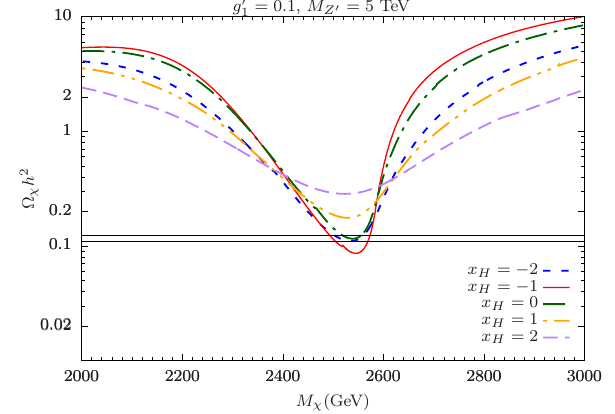}
\includegraphics[height=5cm, width=7.2cm]{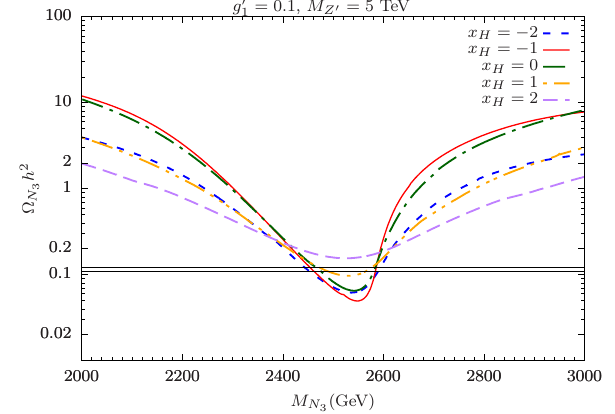}
\caption{The relic density variation is shown with respect to the mass of the scalar and fermion DM in the left and right panels respectively for different choices of $x_H$. We have fixed $M_{Z'}=5$ TeV and $g_1'=0.1$.}
\label{fig:relic-xH-scalar-fermion}
\end{figure}
%%%%%%%%%%%%%%%%%%%%%%%%%%%%%%%%%%%%%%%%%
%
In Fig.~\ref{fig:relic-xH-scalar-fermion}, we have chosen multiple $x_H=-2,-1,0,1,2$ etc, to see how relic behave in each of these cases for both scalar~(left panel) and fermion~(right panel) DM. We have fixed $M_{Z'}=5$ TeV and $g_1'=0.1$. We have chosen DM masses in the range from 2 to 3 TeV which is most suitable for studying $x_H$ behaviour as we fixed $M_{Z'}=5$ TeV. The dominating channels are here $\chi_R \chi_I/ N_3 N_3 \rightarrow Z'\rightarrow f\bar{f}$, where $f$ is the SM final states. Hence, the DM annihilation cross section for $M_{N_3},M_{\chi}\sim M_{Z'}/2$ is proportional to $1/\Gamma_{Z'}$. On the other hand, the $Z'$ decay width depends on the value of $x_H$ as $Z'$ interactions depend on $x_H$ when $x_\Phi$ is fixed. We found that $Z'$ becomes minimum at $x_{H}=-0.8$ and is a symmetric function of $x_H$ around this point. Due to this reason, $-2\leq x_H\leq 0$ are more suitable for both candidate to be DM in $Z'$ portal framework.
%
%%%%%%%%%%%%%%%%%%%%%%%%%%%%%%%%%%%%%%%%%%%%%%%%%%%%%%
\begin{figure}[htbp]
\includegraphics[height=5cm, width=7.2cm]{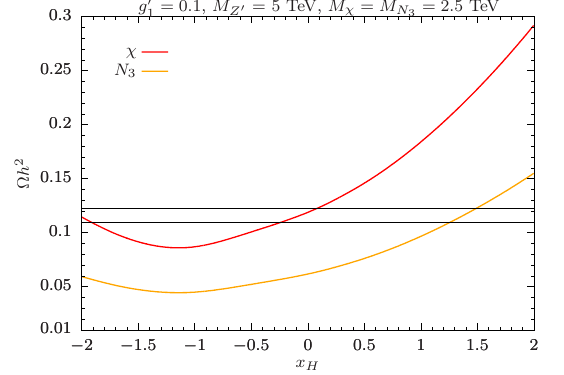}
\includegraphics[height=5cm, width=7.2cm]{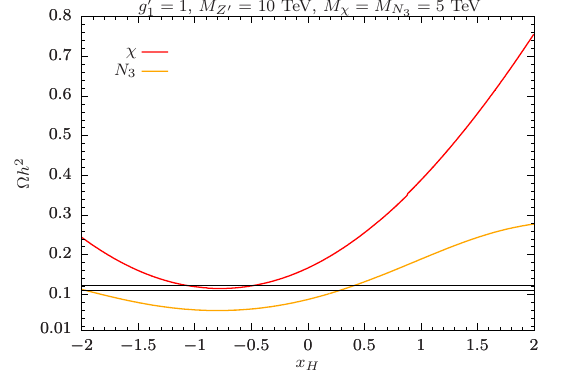}
\caption{The variation of relic density with $U(1)_X$ charge $x_H$ is shown here for $M_{Z'}=5$ TeV~(left panel) and $10$ TeV~(right panel). The red and orange curves are for scalar and fermion DM, respectively. We have set the mass for both the DM to be at $M_{Z'}/2$.}
\label{fig:relic-xH}
\end{figure}
%%%%%%%%%%%%%%%%%%%%%%%%%%%%%%%%%%%%%%%%%%%%%%%%%%%%%%%%%

In Fig.~\ref{fig:relic-xH}, we are showing the variation of scalar and fermion DM relic density with respect to $U(1)_X$ charge $x_H$ for $M_{Z'}=5$ TeV~(left panel) and $10$ TeV~(right panel). The mass of both DM candidates has been chosen at $M_{Z'}/2$ since only their relic abundance can have the correct order. For heavier $Z'$ mass $M_{Z'}=10$ TeV we choose relatively large coupling $g_1'=1$ which is allowed for any value of $x_H$ from the current collider constraints, see Fig.~\ref{fig:collider-constraints}. From Fig.~\ref{fig:relic-xH}, we again see that $-2\leq x_H\leq 0$ are preferable for both candidates to be DM in $Z'$ portal framework and for larger $M_{Z'}$ one need relatively large coupling $g_1'$ to have correct relic density.

%%%%%%%%%%%%%%%%%%%%%%%%%%%%%%%%%%%%%%%%
\section{Summary}
\label{summary_tdm}
%%%%%%%%%%%%%%%%%%%%%%%%%%%%%%%%%%%%%%%%

We have considered a generic $U(1)_X$ model which provides an economical extension of SM to accommodate DM and active neutrino masses. In this generic $U(1)_X$ model, the charges of the SM particles are defined as a linear combination of the SM $U(1)_Y$ and $U(1)_{B-L}$ charges. In addition to three generations of RHNs, we add two complex scalar $\Phi$, $\chi$ both are charged under the $U(1)_X$ gauge group. $U(1)_X$ symmetry breaking is driven by the VEV of $\Phi$. Note that, although Majorana neutrino masses for all three RHNs are generated through the VEV of $\Phi$, due to the $\mathbb{Z}_2$ charge assignment, only $N_R^{1,2}$ have Dirac Yukawa couplings with the SM lepton doublets. Hence, in this model, only two SM neutrinos are massive as the seesaw mechanism generates the SM neutrino mass matrix with only the two Majorana RHNs. As, additional fields such as $N_R^3$ and $\chi$ are odd under the discrete symmetries $\mathbb{Z}_2$ and $\mathbb{Z}_2'$, respectively, they both are stabilized and hence play the role of two-component DM. We have discussed the main features of this two-component DM scenario in the context of the generic $U(1)_X$ model.

Apart from the SM Higgs boson, we have one more heavy Higgs and an additional neutral gauge boson (Z). As a result, there are two ways for both the scalar and fermion DM to interact with the SM particles. One is through the $Z'$ boson interaction since all particles in this model are charged under the gauge group $U(1)_X$. The other is through Higgs boson interactions. In the case of the Higgs-portal DM scenario, we find that SM Higgs and heavy Higgs mixing $\sin\alpha$, heavy Higgs mass $m_{h_2}$, and quartic couplings $\lambda$ play a crucial role. On the other hand, for the $Z'$-portal DM scenario, the DM phenomenology is controlled by very few parameters, such as $g_1'$, $M_{Z'}$, $M_{\text{DM}}$, and $U(1)_X$ charge $x_H$. We also find that, to reproduce the observed DM relic density, it is required to fix the DM mass around $M_{Z'}/2$ for the case of $Z'$-portal DM. We further found that with the current collider constraint on $g_1'$ and $M_{Z'}$, $-2\leq x_H\leq 0$ is preferable to have the correct relic density.

%%%%%%%%%%%%%%%%%%%%%%%%%%%%%%%%%%%%%%%%%%%%%%%%%%%%%%%%%%%%%%%%%%%%%%%%%%%%%%%%%%
\chapter{A pseudo-scalar dark matter case in \texorpdfstring{$U(1)_X$}{U(1)X} extension of SM}
\label{chap_U1X2}
%%%%%%%%%%%%%%%%%%%%%%%%%%%%%%%%%%%%%%%%%%%%%%%%%%%%%%%%%%%%%%%%%%%%%%%%%%%%%%%%%%
%
In this chapter, we study a pseudo-scalar DM in a generic U(1)$_X$ extension of SM. The results are based on the work: Shivam Gola, "Pseudo scalar dark matter in a generic U(1)$_X$ model, Phys.Lett.B 842 (2023) 137982''.

\section{Introduction}
WIMP-type dark matter is heavily constrained by direct experimental searches~\cite{Lisanti:2016jxe,Schumann_2019}; therefore, to continue with WIMP, one has to find a way to evade the stringent direct detection bounds.

A pseudo-scalar DM can naturally evade the strong direct detection constraints as it has derivative couplings that imply a momentum suppression in the tree-level DM-nucleon scattering matrix, which vanishes in the non-relativistic limit~\cite{Gross_2017,Abe_2020,Abe_2021, Gola:2021abm,Okada_2021,Oda_2015,Das_2019,Das_2022}. Hence, it is interesting to seek a pseudo-scalar particle as a WIMP dark matter.

In this chapter, we consider a generic U(1)$_X$ model for pseudo-scalar DM and discuss its phenomenological implications in this study. The interesting aspect of the $U(1)_X$ models is that the three generations of right-handed neutrinos~(RHNs) are required to eliminate the gauge and mixed gauge-gravity anomalies \cite{Das:2016zue,Okada:2016tci,Bandyopadhyay:2017bgh,Das:2019pua}. The RHNs mix with active neutrinos of SM via type I seesaw mechanism~\cite{Minkowski:1977sc,Schechter:1980gr,Mohapatra:1979ia,Schechter:1981cv} to generate the required light neutrino masses and flavor mixing. A model that explains both the neutrino mass problem and the nature of dark matter would be a major step forward in high-energy physics~\cite{Ma:2006km,Hirsch:2013ola,Merle:2016scw,Avila:2019hhv,Mandal:2021yph,Mandal:2019oth}.

The pseudo-scalar is not protected by any symmetry; therefore, for it to be a DM candidate, it must have a lifetime much greater than the age of the universe. We study the feasible parameter space allowed by the lifetime constraint on our pseudo-scalar particle, then we scan for the allowed parameter space by relic and direct detection bounds while respecting several other theoretical and experimental constraints.

%%%%%%%%%%%%%%%%%%%%%%%%%%%%%%%%%%%%%%%%%%%%%%%%%%%%%%%%%%%%%

The chapter is organised as follows: In sec.~\ref{sec:model}, we introduce the model and discuss the details of the new fields and their interactions. In sec.~\ref{sec:constraints}, we discuss some of the relevant theoretical and experimental constraints. In sec.~\ref{sec:dark-matter}, we discuss relic density, direct detection, and other phenomenologically relevant studies. Finally, in sec.~\ref{sec:summary}, we summarize the chapter.

%%%%%%%%%%%%%%%%%%%%%%%%%%%%%%%%%%%%%%%%%%%%%%%%%%%%%%%%%%%%%
%\linebreak
\begin{table}[htbp]
\begin{center}
\begin{tabular}{||c|ccc||c||}
\hline
\hline
            & SU(3)$_c$ & SU(2)$_L$ & U(1)$_Y$ & U(1)$_X$ \\[2pt]
\hline
\hline
&&&&\\[-12pt]
$q_L^i$    & {\bf 3}   & {\bf 2}& $\frac{1}{6}$ & 		 $\frac{1}{6}x_H + \frac{1}{3}x_\Phi$   \\[2pt] 
$u_R^i$    & {\bf 3} & {\bf 1}& $\frac{2}{3}$ &  	  $\frac{2}{3}x_H + \frac{1}{3}x_\Phi$ \\[2pt] 
$d_R^i$    & {\bf 3} & {\bf 1}& $-\frac{1}{3}$ & 	 $-\frac{1}{3}x_H + \frac{1}{3}x_\Phi$  \\[2pt] 
\hline
\hline
&&&&\\[-12pt]
$\ell_L^i$    & {\bf 1} & {\bf 2}& $-\frac{1}{2}$ & 	 $- \frac{1}{2}x_H - x_\Phi$  \\[2pt] 
$e_R^i$   & {\bf 1} & {\bf 1}& $-1$   &		$- x_H - x_\Phi$  \\[2pt] 
\hline
\hline
$N_R$   & {\bf 1} & {\bf 1}& $0$   &	 $- x_\Phi$ \\[2pt] 
\hline
\hline
&&&&\\[-12pt]
$H$         & {\bf 1} & {\bf 2}& $\frac{1}{2}$  &  	 $\frac{x_H}{2}$  \\ 
$\Phi$      & {\bf 1} & {\bf 1}& $0$  & 	 $2 x_\Phi$   \\ 
$\chi$      & {\bf 1} & {\bf 1}& $0$  & 	 $-x_\Phi$   \\ 
\hline
\hline
\end{tabular}
\end{center}
\caption{
Particle content of  the minimal U$(1)_X$ model where $i(=1, 2, 3)$ represents the family index.}
\label{tab1}
\end{table}
%%%%%%%%%%%%%%%%%%%%%%%%%%%%%%%%%%%%%%%%%%%%%%%%%%%%%%%%%%%%%%%%%%%%

%%%%%%%%%%%%%%%%%%%%%%%%%%%%%%%%%%%%%%%%%%%%%%%%%%%%%%%%%%%
\section{Model}
\label{sec:model}
%%%%%%%%%%%%%%%%%%%%%%%%%%%%%%%%%%%%%%%%%%%%%%%%%%%%%%%%%%

We consider a BSM framework based on the $U(1)_X$ gauge group, which has been studied well in the literature~\cite{Okada:2016tci,Bandyopadhyay:2017bgh,Das:2019pua,Das:2022oyx}. The model has three RHNs~($N_R^i$) and two new scalars $\Phi,\chi$ additionally other than SM particles. The particle content and their respective charges are given in Table~\ref{tab1}, where the family index $i$ runs from 1 to 3. The $U(1)_X$ charges of the particles can be written in terms of only two charges, $x_H$ and $x_\Phi$, as shown in Table~\ref{tab1}.

%%%%%%%%%%%%%%%%%%%%%%%%%%%%%%%%%%%%%%%%%%%%%%%%%%%%%%%%%%%
\subsection{Scalar Sector}
The scalar part of the Lagrangian is given by,
\begin{align}
 \mathcal{L}_{s}=(D^{\mu}H)^{\dagger}(D_{\mu}H)+(D^{\mu}\Phi)^{\dagger}(D_{\mu}\Phi)+(D^{\mu}\chi)^{\dagger}(D_{\mu}\chi)-V(H,\Phi,\chi),
\end{align}
here the covariant derivative can be defined as
%%%%%%%%%%%%%%%%%%%%%%%%%%%%%%%%
\begin{align}
D_{\mu}=\partial_{\mu}-ig_{s}T^{a}G^{a}_{\mu}-igT^{a}W^{a}_{\mu}-ig_{1}Y B^1_{\mu}-ig_{X}Y_{X}B^2_{\mu},
\end{align}
%%%%%%%%%%%%%%%%%%%%%%%%%%%%%%
The scalar potential $V(H,\Phi,\chi)$ is given by,
\begin{align}
 V(H,\Phi,\chi) = &\mu_{H}^{2}H^{\dagger}H+\mu_{\Phi}^{2}\Phi^{\dagger}\Phi+m_{\chi}^{2}\chi^{\dagger}\chi-\lambda_{H}(H^{\dagger}H)^{2}-\lambda_{\Phi}(\Phi^{\dagger}\Phi)^{2}
-\lambda_{\chi}(\chi^{\dagger}\chi)^{2}\nonumber \\
 &-\lambda_{H\Phi}(H^{\dagger}H)(\Phi^{\dagger}\Phi)\nonumber 
 -\lambda_{\Phi\chi}(\Phi^{\dagger}\Phi)(\chi^{\dagger}\chi)-\lambda_{H\chi}(H^{\dagger}H)(\chi^{\dagger}\chi)+(\kappa_{T}\Phi^{\dagger}\chi^2 + \text{h.c.})
\end{align}
We parameterize the scalar fields as,
\begin{align}
 H=\frac{1}{\sqrt{2}}
 \begin{pmatrix}
  -i(\phi_{1}-i\phi_{2}) \\
  v_H+R_1+i\phi_{3}  \\
 \end{pmatrix},\hspace{0.2cm}
 \chi=\frac{1}{\sqrt{2}}(v_{\chi}+R_2+i\eta_{\chi}),
 \hspace{0.2cm}
 \Phi=\frac{1}{\sqrt{2}}(v_{\Phi}+R_3+i\eta_{\Phi})
\end{align}
$v_H, v_{\Phi}$ and $v_{\chi}$ are the VEVs of Higgs doublet and SM singlets $\Phi$ and $\chi$ respectively. $w^{\pm}=\phi_{1}\mp i\phi_{2}$ would be the Goldstone boson of $W^{\pm}$, while $\phi_{3},\eta_{\Phi}$ and $\eta_{\chi}$ will mix to give $G^0, G'$ and $A$, that would be the Goldstone bosons of the $Z$ and $Z^{'}$ bosons, and the physical pseudo-Nambu Goldstone boson respectively.
The real scalars $R_1, R_2$ and $R_3$ are not the mass eigenstates due to mixing which implies following mass matrix 
\begin{align}
 M^2_R=
\begin{pmatrix}
 2\lambda_{H}v^2_H  &   \lambda_{H\chi}v_H v_{\chi}  &  \lambda_{H\Phi}v_H v_{\phi}  \\
 \lambda_{H\chi}v_H v_{\chi} &  2\lambda_{\chi}v^2_{\chi}  &    (-\sqrt{2}\kappa_T+\lambda_{\Phi\chi}v_{\phi})v_{\chi}  \\
 \lambda_{H\Phi}v_H v_{\phi} &  (-\sqrt{2}\kappa_T+\lambda_{\Phi\chi}v_{\phi})v_{\chi}  &  2\lambda_{\Phi}v^2_{\phi}+\frac{\kappa_T v^2_{\chi}}{\sqrt{2}v_{\phi}}  \\
\end{pmatrix}
\end{align}
%%%%%%%%%%%%%%%%%%%%%%%%%%%%%%%%%%%%%%%%%%%%
The matrix $M^2_{R}$ can be diagonalized by an orthogonal matrix 
as follows~\cite{Darvishi_2021,Robens_2020}, 
\begin{align}
\mathcal{O}_{R} M_{R}^2 \mathcal{O}_R^T = 
\text{diag}(m_{h_1}^2,m_{h_2}^2,m_{h_3}^2),
\end{align}
 where
\begin{eqnarray}
 \label{rot1}
  \left( \begin{array}{c} h_1\\ h_2\\ h_3\\ \end{array} \right) = 
 \mathcal{O}_{R} \left( \begin{array}{c} R_1\\ R_2\\ R_3\\ 
 \end{array} \right). 
\label{eq:scalarmix}
 \end{eqnarray}
%%%%%%%%%%%%%%%%%%%%%%%%%%%%%%%%%%%%%%%%%%%%%%%%%%%%
We assume the mass eigenstates to be ordered by their masses $m_{h_1}\leq m_{h_2}\leq m_{h_3}$. We will use the standard parameterization $\mathcal{O}_{R} = R_{23} R_{13} R_{12}$ where
%%%%%%%%%%%%%%%%%%%%%%%%%%%%%%%%%%%%%%%%%%%%%%%%%%%%%
\begin{equation}
R_{12} = \left(
\begin{array}{ccc}
c_{12} & s_{12} & 0\\
-s_{12} & c_{12} & 0\\
0 & 0 & 1
\end{array} \right), 
\quad R_{13} = \left(
\begin{array}{ccc}
c_{13} & 0 & s_{13}\\
0 & 1 & 0\\
-s_{13} & 0 & c_{13}
\end{array} \right), 
\quad R_{23} = \left(
\begin{array}{ccc}
1 & 0 & 0\\
0 & c_{23} &  s_{23}\\
0 & -s_{23} & c_{23}
\end{array} \right)
\end{equation}
%%%%%%%%%%%%%%%%%%%%%%%%%%%%%%%%%%%%%%%%%%%%%%%%%%%%
$c_{ij} = \cos \theta_{ij}, s_{ij} = \sin \theta_{ij}$, where the angles $\theta_{ij}$ can be chosen to lie in the range $-\frac{\pi}{2}\leq\theta_{ij}\leq \frac{\pi}{2}$. The rotation matrix $\mathcal{O}_{R}$ is re-expressed in terms of the mixing angles in the following way: 
\begin{align}
\mathcal{O}_{R} = \left(
\begin{array}{ccc}
c_{12} c_{13}                          &  c_{13} s_{12}                        & s_{13}\\
-c_{23} s_{12} - c_{12} s_{13} s_{23}  & c_{23} c_{12} - s_{12} s_{13}  s_{23}  & c_{13} s_{23}\\
-c_{12} c_{23} s_{13} + s_{23} s_{12}  & -c_{23} s_{12} s_{13} - c_{12} s_{23} & c_{13} c_{23}
\end{array} \right).
\end{align}
%%%%%%%%%%%%%%%%%%%%%%%%%%%%%%%%%%%%%%%%%%%%%
It is possible to express the ten parameters of the scalar potential through the three physical Higgs masses, the three mixing angles, the three VEVs and $\kappa_T$~\cite{Robens_2020}. These relations are given by
%%%%%%%%%%%%%%%%%%%%%%%%%%%%%%%%%%%%%%%%%%%%%%
\begin{align}
        \lambda_H     & =\frac{1}{2 v_H^2} m_{h_i}^2 (\mathcal{O}_R^{i1})^2\,,        &
        \lambda_{\chi}        & =\frac{1}{2 v_\chi^2}  m_{h_i}^2 (\mathcal{O}_R^{i2})^2\,,      &
        \lambda_\Phi        & =\frac{1}{2 v_\Phi^2}  m_{h_i}^2 (\mathcal{O}_R^{i3})^2-\kappa_T\frac{v_\chi^2}{2\sqrt{2}v_\Phi^3}\,, \nonumber \\     
        \lambda_{H \chi} & =\frac{1}{v_H v_\chi}  m_{h_i}^2 \mathcal{O}_R^{i1} \mathcal{O}_R^{i2}\,,   &
        \lambda_{H \Phi} & =\frac{1}{v_H v_\Phi}  m_{h_i}^2 \mathcal{O}_R^{i1} \mathcal{O}_R^{i3}\,,   &
        \lambda_{\Phi \chi}    & =\frac{1}{v_\Phi v_\chi}  m_{h_i}^2 \mathcal{O}_R^{i2} \mathcal{O}_R^{i3}+\sqrt{2}\frac{\kappa_T}{v_\Phi}\,,
    \label{eq:quartic}
\end{align}
%%%%%%%%%%%%%%%%%%%%%%%%%%%%%%%%%%%%%%%%%
The squared mass matrix for the CP-odd scalars in the weak basis ($\eta_{\chi}$, $\eta_{\Phi}$) is given as
%%%%%%%%%%%%%%%%%%%%%%%%%%%%%%%%%%%%%%%%
\begin{align}
 M^2_{\eta}=\frac{\kappa_T}{\sqrt{2}}
\begin{pmatrix}
4v_{\Phi}  &    -2v_{\chi}  \\
-2v_{\chi}  & \frac{v^2_{\chi}}{v_{\Phi}} \\
\end{pmatrix}
\end{align}
%%%%%%%%%%%%%%%%%%%%%%%%%%%%%%%%%%%%%%%%%%%%%%%%%
This mass matrix can be diagonalized as $\mathcal{O}_\eta M^2_{\eta} \mathcal{O}_\eta^T=\text{diag}(0,m_A^2)$, where
\begin{align}
\mathcal{O}_\eta= 
\begin{pmatrix}
\cos\theta_{\eta} & \sin\theta_{\eta} \\
 -\sin\theta_{\eta} & \cos\theta_{\eta} \\
\end{pmatrix}, \,\, m^2_{A}=\frac{\kappa_T(4v^2_{\Phi}+v^2_{\chi})}{\sqrt{2}v_{\Phi}}
\label{psmass}
\end{align} 
and the null mass corresponds to the would-be Goldstone boson $G'$, which will be eaten by $Z'$.
%%%%%%%%%%%%%%%%%%%%%%%%%%%%%%%%%%%%%%%%%%%%%%%%%%%%%%
The gauge eigenstates $(\eta_\chi\,\, \eta_\Phi)$ are related with the mass eigenstates $(G'\,\, A)$ as
\begin{align}
 \begin{pmatrix}
  \eta_{\chi} \\
  \eta_{\Phi}  \\ 
 \end{pmatrix}
=
\begin{pmatrix}
 \cos\theta_{\eta} & -\sin\theta_{\eta} \\
 \sin\theta_{\eta} & \cos\theta_{\eta} \\
\end{pmatrix}
\begin{pmatrix}
  G' \\
  A \\ 
\end{pmatrix},\text{  where  }  \tan\theta_{\eta} = \frac{2v_{\Phi}}{v_{\chi}}
\end{align}
%%%%%%%%%%%%%%%%%%%%%%%%%%%%%%%%%
Note that fixing one of the Higgs masses to the mass of the observed Higgs boson, $m_{h_1}=125$ GeV, and fixing the Higgs doublet vev to its SM value, $v_H= 246$ GeV, leaves eight free input parameters from the scalar sector:
%%%%%%%%%%%%%%%%%%%%%%%%%%%%%%%%%%%%%%
\begin{align}
m_{h_2},\,\, m_{h_3},\,\, m_A,\,\, \theta_{12},\,\, \theta_{13},\,\, \theta_{23},\,\, v_\chi \text{ and } v_{\Phi}
\end{align}

%%%%%%%%%%%%%%%%%%%%%%%%%%%%%%%%%%%%%%
\subsection{Gauge Sector}
%%%%%%%%%%%%%%%%%%%%%%%%%%%%%%%%%%%%%%

We note that a kinetic mixing can occur provided there are two or more field strength tensors $B^1_{\mu\nu}$ and $B^2_{\mu\nu}$ which are neutral under some gauge symmetry. This only arises for the abelian $U(1)$ gauge group. Thus in our case with the gauge group $SU(3)_c\otimes SU(2)_L\otimes U(1)_Y\otimes U(1)_X$, there can be kinetic mixing between two abelian gauge group $U(1)_Y$ and $U(1)_X$. The kinetic terms can be expressed as follows
%%%%%%%%%%%%%%%%%%%%%%%%%%
\begin{align}
\mathcal{L}_{\rm Kinetic} = -\frac{1}{4}B^{1\mu\nu}B^1_{\mu\nu} - \frac{1}{4}B^{2\mu\nu}B^2_{\mu\nu} - \frac{\kappa}{2} B^{1\mu\nu}B^2_{\mu\nu},
\label{eq:kinetic-mixing}
\end{align}
%%%%%%%%%%%%%%%%%%%%%%%%%%%%
where $B^1_{\mu\nu}$ and $B^2_{\mu\nu}$ are the filed strength tensors of the gauge groups $U(1)_Y$ and $U(1)_X$, respectively. The requirement of positive kinetic energy implies that kinetic coefficient $|\kappa|<1$. One can diagonalize the kinetic mixing term as follows
%%%%%%%%%%%%%%%%%%%%%%%%%%%%%%%%%%%%%%%
\begin{align}
 \begin{pmatrix}
  \tilde{B}_{1} \\
  \tilde{B}_{2}  \\ 
 \end{pmatrix}
=
\begin{pmatrix}
 1 & \kappa \\
 0 & \sqrt{1-\kappa^2} \\
\end{pmatrix}
\begin{pmatrix}
 B_{1} \\
 B_{2} \\
\end{pmatrix}
\end{align}
%%%%%%%%%%%%%%%%%%%%%%%%%%%%%%%%%%%%%%%%
Let's first set kinetic mixing $\kappa=0$ to fix the notation. The diagonal component of the $SU(2)_L$ gauge filed $W_{3\mu}$ will mix with the $U(1)_Y$ and $U(1)_X$ gauge fields $B^1_{\mu}$ and $B^{2}_\mu$. To determine the gauge boson mass spectrum, we have to expand the scalar kinetic terms 
%%%%%%%%%%%%%%%%%%%%%%%%
\begin{align}
\mathcal{L}_{s}=(D^{\mu}H)^{\dagger}(D_{\mu}H)+(D^{\mu}\Phi)^{\dagger}(D_{\mu}\Phi)+(D^{\mu}\chi)^{\dagger}(D_{\mu}\chi)
\end{align}
%%%%%%%%%%%%%%%%%%%%
and have to replace the fields $H,\Phi$ and $\chi$ by the following expressions such as
%%%%%%%%%%%%%%%%%%%%%%%%%%%%%%%%%%
\begin{align}
H=\frac{1}{\sqrt{2}}\begin{pmatrix}
0\\
v_H+R_1\end{pmatrix},\,\, \braket{\Phi}=\frac{1}{\sqrt{2}} (v_\Phi+R_2) \text{ and } \braket{\chi}=\frac{1}{\sqrt{2}} (v_\chi+R_3).
\end{align}
%%%%%%%%%%%%%%%%%%%%%%%%%%%%%%%%%%%
With this above replacement, we can expand the scalar kinetic terms $(D^\mu H)^\dagger (D_\mu H)$, $(D^\mu\Phi)^\dagger (D_\mu\Phi)$ and $(D^\mu\chi)^\dagger (D_\mu\chi)$ as follows
%%%%%%%%%%%%%%%%%%%%%%%%%%%%%%
\begin{align}
& (D^{\mu}H)^{\dagger}(D_{\mu}H)\equiv\frac{1}{2}\partial^{\mu}R_1\partial_{\mu}R_1+\frac{1}{8}(R_1+v_H)^{2}\Big(g^{2}|W_{1}^{\mu}-iW_{2}^{\mu}|^{2}+(gW_{3}^{\mu}-g_{1}B^{1\mu}-\tilde{g}B^{2\mu})^{2}\Big) \\
& (D^{\mu}\Phi)^{\dagger}(D_{\mu}\Phi)\equiv\frac{1}{2}\partial^{\mu}R_2\partial_{\mu}R_2+\frac{1}{2}(R_2+v_\Phi)^{2}(2g_{1}^{'} B^{2\mu})^{2} \\
& (D^{\mu}\chi)^{\dagger}(D_{\mu}\chi)\equiv\frac{1}{2}\partial^{\mu}R_3\partial_{\mu}R_3+\frac{1}{2}(R_3+v_\chi)^{2}(g_{1}^{'} B^{2\mu})^{2}.
\end{align}
%%%%%%%%%%%%%%%%%%%%%%%%%%%%%%%%%%%%%%%%%%%
where we have defined $\tilde{g}=g_{X}x_H$ and $g_{1}^{'}=g_{X}x_\Phi$. SM charged gauge boson $W^{\pm}$ can be easily recognized with mass $M_{W}=\frac{g v_H}{2}$. On the other hand, the mass matrix of the neutral gauge bosons is given by
%%%%%%%%%%%%%%%%%%%%%%%%%%%%%
\begin{align}
\mathcal{L}_M=\frac{1}{2} V_0^T M_G^2 V_0
\end{align}
%%%%%%%%%%%%%%%%%%%%%%%%%%
where
\begin{align}
V_0^T= \begin{pmatrix}B^1_\mu & W_{3\mu} & B^2_\mu \end{pmatrix} \text{   and    }
M_G^2=
\begin{pmatrix}
\frac{1}{4}g^2_1 v_H^2 &  -\frac{1}{4} g g_1 v_H^2   &   \frac{1}{4} g_1 \tilde{g} v_H^2 \\
-\frac{1}{4} g g_1 v_H^2  &   \frac{1}{4} g^2 v_H^2  &  -\frac{1}{4} g  \tilde{g} v_H^2  \\
\frac{1}{4} g_1  \tilde{g} v_H^2  &  -\frac{1}{4} g  \tilde{g} v_H^2  &  \frac{1}{4}  \tilde{g}^2 v_H^2+ 4  g_1^{'2} v_T^2  
\end{pmatrix}
\end{align}
Following linear combination of $B_1^{\mu}$, $W_{3}^{\mu}$ and $B_2^{\mu}$ gives definite mass eigenstates $A_0^{\mu}$, $Z_0^{\mu}$ and $Z_0^{'\mu}$~(when $\kappa=0$),
%%%%%%%%%%%%%%%%%%%%%%%%%%%%%%%%%%
\begin{align}
 \begin{pmatrix}
  B_1^{\mu}\\
  W_{3}^{\mu}\\
  B_2^{\mu}
 \end{pmatrix}
= \begin{pmatrix}
 \cos\theta_{w}  & -\sin\theta_{w}~\cos\theta_0 &  \sin\theta_{w}~\sin\theta_0 \\
 \sin\theta_{w}  & \cos\theta_{w} \cos\theta_0  &  -\cos\theta_{w} ~\sin\theta_0 \\
 0                      & \sin\theta_0                        &  \cos\theta_0 \\
\end{pmatrix}
\begin{pmatrix}
 A_0^{\mu}\\
 Z_0^{\mu} \\
 Z_0^{'\mu} \\
\end{pmatrix}
\label{eq:gauge-tranformation}
\end{align}
where $\theta_{w}$ is the Weinberg mixing angle and,
\begin{align}
 \text{tan} 2\theta_0=\frac{2\tilde{g}\sqrt{g^{2}+g_{1}^{2}}}{\tilde{g}^{2}+16\left(\frac{v_T}{v_H}\right)^{2}g_{1}^{'2}-g^{2}-g_{1}^{2}} \text{ with } v_T=\frac{\sqrt{v_\chi^2+4 v_\Phi^2}}{2}.
\end{align}
%%%%%%%%%%%%%%%%%%%%%%%%%%%
When the kinetic mixing $\kappa$ is non-zero, the fields $A_0, Z_0$ and $Z'_0$ are not orthogonal. In the kinetic term diagonalized basis $\tilde{V}_0^T=(\tilde{B}_1\,\,W_3\,\,\tilde{B}_2)$, the mass matrix of the neutral gauge boson can be written as
%%%%%%%%%%%%%%%%%%%%%%%%%
\begin{align}
\mathcal{L}_M=\frac{1}{2} \tilde{V}_0^T S^T M_G^2  S \tilde{V}_0 = \frac{1}{2} \tilde{V}_0^T  \tilde{M}_G^2  \tilde{V}_0
\end{align}
%%%%%%%%%%%%%%%%%%
where
\begin{align}
S=\begin{pmatrix} 
1  &  0  & -\frac{\kappa}{\sqrt{1-\kappa^2}}  \\
0  & 1   & 0 \\
0  &  0  & \frac{1}{\sqrt{1-\kappa^2}}
\end{pmatrix},
\tilde{M}_G^2=S^T  M_G^2  S = \begin{pmatrix}
\frac{1}{4}g_1 v_H^2 &  -\frac{1}{4} g g_1 v_H^2   &   \frac{1}{4} g_1 \tilde{g}_t v_H^2 \\
-\frac{1}{4} g g_1 v_H^2  &   \frac{1}{4} g^2 v_H^2  &  -\frac{1}{4} g  \tilde{g}_t v_H^2  \\
\frac{1}{4} g_1  \tilde{g}_t v_H^2  &  -\frac{1}{4} g  \tilde{g}_t v_H^2  &  \frac{1}{4}  \tilde{g}_t^2 v_H^2+ 4  g_1^{''2} v_T^2  
\end{pmatrix}
\end{align}
%%%%%%%%%%%%
with $\tilde{g}_t=\frac{\tilde{g}-g_1\kappa}{\sqrt{1-\kappa^2}}$. Hence, comparing $M_G^2$ and $\tilde{M}_G^2$ we see that the overall effect of kinetic mixing introduction is just the modification of $\tilde{g}$ to $\tilde{g}_t$ and $g_1'$ to $g_1^{''}$. Hence, they can be related to orthogonal fields $A,Z$ and $Z'$ by the same transformation as in Eq.~\ref{eq:gauge-tranformation} but now $\theta_0$ is replaced by $\theta_{12}$
%%%%%%%%%%%%%%%%%%%%%%%%%
\begin{align}
 \begin{pmatrix}
  \tilde{B}_1^{\mu}\\
  W_{3}^{\mu}\\
  \tilde{B}_2^{\mu}
 \end{pmatrix}
= \begin{pmatrix}
 \cos\theta_{w}  & -\sin\theta_{w}~\cos\theta &  \sin\theta_{w}~\sin\theta \\
 \sin\theta_{w}  & \cos\theta_{w} \cos\theta  &  -\cos\theta_{w} ~\sin\theta \\
 0                      & \sin\theta                        &  \cos\theta \\
\end{pmatrix}
\begin{pmatrix}
 A^{\mu}\\
 Z^{\mu} \\
 Z^{'\mu} \\
\end{pmatrix}
\label{eq:final-gauge-tranformation}
\end{align}
%%%%%%%%%%%%%%%%%%%%%%%%
where
%%%%%%%%%%%%%%%%%%%%%%%
\begin{align}
 \text{tan} 2\theta=\frac{2\tilde{g}_t\sqrt{g^{2}+g_{1}^{2}}}{\tilde{g}_t^{2}+16\left(\frac{v_T}{v_H}\right)^{2}g_{1}^{''2}-g^{2}-g_{1}^{2}} \text{  with  } \tilde{g}_t=\frac{\tilde{g}-g_1\kappa}{\sqrt{1-\kappa^2}} \text{ and } g_1^{''}=\frac{g_1'}{\sqrt{1-\kappa^2}}.
\end{align}
Masses of physical gauge bosons $A$, $Z$ and $Z^{'}$ are given by,
%%%%%%%%%%%%%%%%%%%%%%%%%%%%%%
\begin{align}
 M_A=0,\,\,M_{Z,Z^{'}}^{2}=\frac{1}{8}\left(Cv_H^{2}\mp\sqrt{-D+v_H^{4}C^{2}}\right),
\end{align}
where,
\begin{align}
 C=g^{2}+g_{1}^{2}+\tilde{g}_t^{2}+16\left(\frac{v_T}{v_H}\right)^{2}g_{1}^{''2},
 \hspace{0.5cm}
 D=64v_H^{2}v_T^{2}(g^{2}+g_{1}^{2})g_{1}^{''2}.
\end{align}
%%%%%%%%%%%%%%%%%%%%%%%%%%%%%%%%%%%%%%%%%%
The covariant derivative also can be expressed in terms of the orthogonal fields $\tilde{B}_1$ and $\tilde{B}_2$ as 
%%%%%%%%%%%%%%%%%%%%%%%%%%%%%%%%%%%%%%%%
\begin{align}
D_{\mu}=\partial_{\mu}-ig_{s}T^{a}G^{a}_{\mu}-igT^{a}W^{a}_{\mu}-ig_{1}Y \tilde{B}^1_{\mu}-i\left(g_{X}Y_{X}\frac{1}{\sqrt{1-\kappa^2}}-g_1 Y \frac{\kappa}{\sqrt{1-\kappa^2}}\right)\tilde{B}^2_{\mu}
\end{align}
%%%%%%%%%%%%%%%%%%%%%%%%%%%%%%%%%%%%%%%%%%%%%%%%%%
\subsection{Yukawa Sector}
%%%%%%%%%%%%%%%%%%%%%%%%%%%%%%%%%%%%%%%%%%%%%%%%%%
The general form of the Yukawa interactions is given by, 
\begin{align}
 \mathcal{L}_{y} & =-y_{u}^{ij}\overline{q_{L}^{i}}\tilde{H}u^{j}_{R}-y_{d}^{ij}\overline{q_{L}^{i}}H d^{j}_{R}- y_{e}^{ij}\overline{\ell_{L}^{i}} H e^{j}_{R} - y_{\nu}^{ij}\overline{\ell_{L}^{i}}\tilde{H}\nu^{j}_{R}
 -\frac{1}{2} y_{M}^{ij}\Phi\overline{N_{R}^{ic}}N^{j}_{R} + \text{H.c.}
\label{Yukawa}
\end{align}
The last two terms are responsible for Dirac and Majorana masses of neutrinos. 
\subsection{Neutrino Mass}
Relevant light neutrino masses will come from the fourth and fifth terms of Eq.~\ref{Yukawa}. After the electroweak symmetry breaking, we can write the mass terms as,
%%%%%%%%%%%%%%%%%%%%%%%%%%%%%%%%%%%%%%%%%%%%%
 \begin{align}
  -\mathcal{L}_{M}=\sum_{i=1}^3\sum_{j=1}^3\overline{\nu_{jL}}m_{ij}^{D}N_{jR}+\frac{1}{2}\sum_{i,j=1}^3\overline{(N_{R})_{i}^{c}}M_{ij}^{R}N_{jR}+\text{H.c.},
 \end{align}
%%%%%%%%%%%%%%%%%%%%%%%%%%%%%%%%%%%%%%%%%%%%%%%
where $m_{ij}^{D}=\frac{y_{\nu}^{ij}v_H}{\sqrt{2}}$ and $M_{ij}^{R}=\frac{y_{M}^{ij}}{\sqrt{2}} v_{\Phi}$. Now we can write the $\mathcal{L}_{M}$ in the following matrix form,
\begin{align}
 -\mathcal{L}_{M}=\frac{1}{2}
 \begin{pmatrix}
  \overline{\nu_{L}} & \overline{(N_{R})^{c}} \\
 \end{pmatrix}
 \begin{pmatrix}
  {\bf 0}  &  m_{D} \\
  m_{D}^T  &  M_{R} \\
 \end{pmatrix}
\begin{pmatrix}
 (\nu_{L})^{c}\\
 N_{R} \\
\end{pmatrix} + \text{H.c.}
\label{eq:neutrino-mass}
\end{align}
%%%%%%%%%%%%%%%%%%%%%%%%%%%%%%%%%%%%%%%%%%%%%
In the seesaw approximation, this leads to the usual light neutrino mass matrix $m_{\nu}\approx -m_{D}M_R^{-1}m_D^T$. The $6\times 6$ matrix in Eq.~\ref{eq:neutrino-mass} is symmetric and can be diagonalized by the unitary matrix~(up to $\mathcal{O}(M_R^{-2})$)~\cite{Schechter:1981cv}
%%%%%%%%%%%%%%%%%%%%%%%%%%%%%%
\begin{align}
\begin{pmatrix}
U & S \\
-S^{\dagger}  &  V
\end{pmatrix}^{\dagger}
\begin{pmatrix}
  {\bf 0}  &  m_{D} \\
  m_{D}^T  &  M_{R} \\
 \end{pmatrix}
 \begin{pmatrix}
U & S \\
-S^\dagger  &  V
\end{pmatrix}^*=
\begin{pmatrix}
\widehat{M_{\nu}} & {\bf 0} \\
{\bf 0}  &  \widehat{M_{N}}
\end{pmatrix}
\end{align}
%%%%%%%%%%%%%%%%%%%%%%%%%%%%%%%%%%%%%%%

where, $\widehat{M_{\nu}}=\text{diag}\{m_1,m_2,m_3\}$, $\widehat{M_N}=\text{diag}\{M_{N_1},M_{N_2},M_{N_3}\}$, $U\approx U_{\text{PMNS}}$, $V\approx {\bf I}$ and $S\approx m_D M_R^{-1}$.

%%%%%%%%%%%%%%%%%%%%%%%%%%%%%%%%%%%%%%%%%%%%%%%%%%%%%%%%%
\section{Theoretical and experimental constraints}
\label{sec:constraints}
%%%%%%%%%%%%%%%%%%%%%%%%%%%%%%%%%%%%%%%%%%%%%%%%%%
We discuss the relevant theoretical and experimental constraints in this section.

%%%%%%%%%%%%%%%%%%%%%%%%%%%%%%%%%%%%%%%%
\subsection{Vacuum Stability condition}
%%%%%%%%%%%%%%%%%%%%%%%%%%%%%%%%%%%%%%%%%
The scalar potential $V(H,\Phi,\chi)$ must be bounded from below~\cite{Kannike:2012pe} and it can be determined from the following symmetric matrix which comes from the quadratic part of the potential,
%%%%%%%%%%%%%%%%%%%%%%%%%%%%%%%%%%%%%%%%%%%%%%%%%%%%
\begin{align}
 V^4_s=
\begin{pmatrix}
 \lambda_{H}    &   \frac{\lambda_{H\Phi}}{2}    &   \frac{\lambda_{H\chi}}{2}  \\
 \frac{\lambda_{H\Phi}}{2}  & \lambda_{\Phi}  &    \frac{\lambda_{\Phi\chi}}{2}  \\
 \frac{\lambda_{H\chi}}{2}  &  \frac{\lambda_{\Phi\chi}}{2}  & \lambda_{\chi}  \\
\end{pmatrix}
\end{align}
%%%%%%%%%%%%%%%%%%%%%%%%%%%%%%%%%%%%%%%%%%%%%%%%%%%%%%%
Above matrix will be positive-definite if following conditions are satisfied,
\begin{align}
 &\lambda_{H} > 0,
 \hspace{1cm}
 4\lambda_{H}\lambda_{\Phi}-\lambda_{H\Phi}^{2}>0 ,\nonumber \\
 &(-\lambda_{H}\lambda_{\Phi\chi}^{2}+\lambda_{H\Phi}\lambda_{\Phi\chi}\lambda_{H\chi}-\lambda_{\Phi}\lambda_{H\chi}^{2}+4\lambda_{H}\lambda_{\Phi}\lambda_{\chi}-\lambda_{H\Phi}^{2}\lambda_{\chi})>0.
\label{eq:stability}
\end{align}
%%%%%%%%%%%%%%%%%%%%%%%%%%%%%%%%%%%%%%%%%%%%
A stable vacuum can be achieved if conditions given in Eq.~\eqref{eq:stability} are satisfied. The perturbativity condition on quartic coupling $\lambda_i\leq 4\pi$ and gauge coupling $g_X \leq \sqrt{4\pi}$ is being ensured in the model.

%%%%%%%%%%%%%%%%%%%%%%%%%%%%%%%%%%%%%%%%%%%%%%%%%%
\subsection{Invisible Higgs width constraint}
\label{sec:inv-Higgs}
%%%%%%%%%%%%%%%%%%%%%%%%%%%%%%%%%%%%%%%%%%%%%%%%%%%%%%%

In our model, $h_1$ scalar is SM Higgs by choice, which is the mixed state of three real scalar fields from eq. \ref{eq:scalarmix}. 
%%%%%%%%%%%%%%%%%%%%%%%%%%%%%%%%%%%%%%%%%%
\begin{align}
h_{\text{SM}} = \sum_i \mathcal{O}_{R_{1i}} R_i
\label{eq:substitution}
\end{align}
%%%%%%%%%%%%%%%%%%%%%%%%%%%%%%%%%%%%%%%%%%%%%%%%%%%%%
If $m_{h_1} > m_{A}/2$, then the partial decay width to $AA$ is given by:
%%%%%%%%%%%%%%%%%%%%%%%%%%%%%%%%%%%%%%%%%%%%%%%%%%%%%%%%%%%%%
\begin{align}
\Gamma(h_1 \to A A) &= \frac{\lambda^2_{AAh_1}}{32\pi m_{h_1}}\sqrt{1-\frac{4m_{A}^2}{m_{h_1}^2}}
\end{align}
%%%%%%%%%%%%%%%%%%%%%%%%%%%%%%%%%%
here $\lambda_{AAh_1}$ is trilinear coupling for vertex $A-A-h_1$. 
  
\begin{align*}
\lambda_{AAh_1} =  2 \sqrt{2} \kappa_T \mathcal{O}_{\eta_{21}}^\dagger \mathcal{O}_{\eta_{22}}^\dagger \mathcal{O}_{R_{12}} - & \left( \mathcal{O}_{\eta_{22}}^\dagger \right)^2 (\lambda_{H\Phi} v_H \mathcal{O}_{R_{11}} + \lambda_{\Phi\chi} v_{\chi} \mathcal{O}_{R_{12}} + 2 \lambda_{\Phi} v_{\phi} \mathcal{O}_{R_{13}}) \ - \\& \left( \mathcal{O}_{\eta_{21}}^\dagger \right)^2 (\lambda_{\Phi\chi} v_H \mathcal{O}_{R_{11}} + 2 \lambda_{\chi} v_{\chi} \mathcal{O}_{R_{12}} + (\sqrt{2} \kappa_T + \lambda_{\Phi\chi} v_{\phi}) \mathcal{O}_{R_{13}})
\end{align*}     
      
The invisible decay width of SM Higgs boson $h_1$ is given by $\Gamma^{\text{inv}}(h_1)=\Gamma(h_1\to A A)$. Then, one can calculate the invisible Higgs branching ratio as follow~\cite{Djouadi_2008,Robens_2020} 
%%%%%%%%%%%%%%%%%%%%%%%%%%%%%%%%%%%%
\begin{align}
\text{BR}^{\text{inv}}(h_1)=
\frac{\Gamma^{\text{inv}}(h_1)}{\mathcal{O}^2_{R_{1i}}\Gamma^\text{SM}(h_1)+\Gamma^{\text{inv}}(h_1)},
\label{eq:inv-BR-Higgs}
\end{align}
%%%%%%%%%%%%%%%%%%%%%%%%%%%%%%%%%%%%%%%%%%%%%%%%%%%%%%%%%%
here $\Gamma^\text{SM}(h_1)=4.1$ MeV. The current upper limit on the invisible Higgs branching ratio is from ATLAS experiment~\cite{ATLAS:2020kdi,ATLAS:2019cid,CMS:2018yfx},
%%%%%%%%%%%%%%%%%%%%%%%%%%%%%%%%%%%%%%%

\begin{align}
\text{BR}^{\text{inv}}(h_1) \leq 0.11.
\label{eq:invisible}
\end{align}

%%%%%%%%%%%%%%%%%%%%%%%%%%%%%%%%%%%%%%%%%%%%%%%%%%
\subsection{Relic density constraint}
\label{sec:relic-constr}
%%%%%%%%%%%%%%%%%%%%%%%%%%%%%%%%%%%%%%%%%%%%%%%%%%%%%%%

The relic density bound is from Planck satellite data~\cite{Planck:2018vyg}.

\begin{align}
\Omega_{\text{DM}} h^2 = 0.12\pm 0.001
\label{eq:relic-density}
\end{align}

Any DM candidate must satisfy the relic bound given in equation~\ref{eq:relic-density}. 

%%%%%%%%%%%%%%%%%%%%%%%%%%%%%%%%%%%%%
\section{Dark Matter analysis}
\label{sec:dark-matter}
%%%%%%%%%%%%%%%%%%%%%%%%%%%%%%%%%%%%

The general charge assignment for $U(1)_X$ model can be described in terms of only two free charges $x_H, ~x_{\phi}$. The charges $x_H$, and $x_\Phi$ are the real parameters. For simplicity, one can fix $x_\Phi=1$ and vary $x_H$ only to characterize the $U(1)$ models. We can obtain the $B-L$ case by choosing $x_H=0$. When $x_H=-2$, the left-handed fermions have no interactions with the $Z^\prime$ which leads to an $U(1)_R$ model. The interactions of $e_R$ and $d_R$ with $Z^\prime$ are absent when $x_H=-1$ and $1$ respectively. Thus, one can study any $U(1)$ models by changing only one parameter $x_H$ in our generic model. We have utilized this generality of our model in doing dark matter analysis. 

\vspace{0.5mm}
\textbf{Benchmark :} We have fixed the following independent parameters throughout the paper for the study of our pseudo scalar particle candidate to DM.
%%%%%%%%%%%%%%%%%%%%%%%%%%%%%%%%%%%%

\begin{align*}
\textbf{BP :} \ \ \ \
x_{\phi}=1, \ \ \ \theta_{13,23}\approx 0, \ \ \ m_{h_3}=10^{13} ~ \text{GeV}, \ \ \ g_X=0.1, \ \ \ M_{Z'}=10^{14} ~ \text{GeV} 
\end{align*}
 
We have also set $m_{A}<<M_{N_i}$, therefore RHNs are not relevant for our DM analysis. Remaining free parameters such as $x_H$, $\theta_{12}$, $\kappa$, $m_{h_2}$, $v_{\phi}$, $v_{\chi}$ and $m_A$ are relevant for DM phenomenology.  

\begin{figure}[t]
\includegraphics[height=5cm, width=7cm]{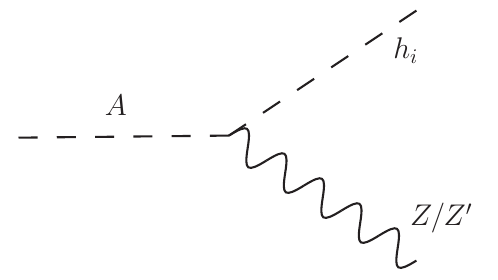} 
\hspace*{0.7cm}
\includegraphics[height=5cm, width=7cm]{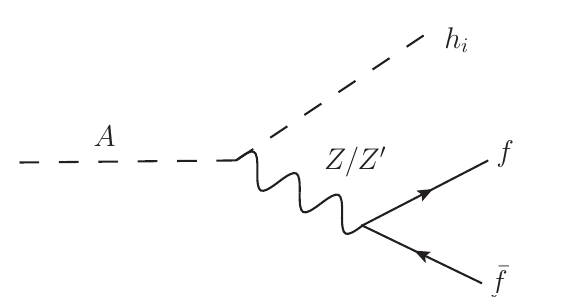}
\vspace*{0.5cm}
\begin{center}
\includegraphics[height=5cm, width=7cm]{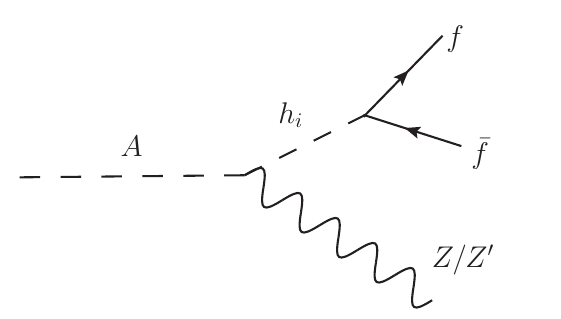}
\caption{The relevant two and three-body decay modes for pseudo scalar DM are shown here.} 
\end{center}
\label{fdiagrams}
\end{figure}

%%%%%%%%%%%%%%%%%%%%%%%%%%%%%%%%%%%%
\subsection{DM lifetime study}
\label{lifetime}
%%%%%%%%%%%%%%%%%%%%%%%%%%%%%%%%%%%%

A good candidate for DM must be stable or have a lifetime greater than the age of the universe. A conservative limit on the lifetime of dark matter is $\tau_{DM} \geq 10^{27}$ sec, or in terms of decay width, $\Gamma_{DM} \leq 6.6 \times 10^{-52}$ GeV, from the gamma rays observation of dwarf spheroidal galaxies~\cite{Baring_2016}. Our pseudo-scalar should also follow this criterion to be a good DM candidate.

There are two possible body decay modes, namely $A \rightarrow \nu\nu, ~h_i Z$. Particles $Z'$ and $N_i$ are chosen to be very heavy, such that $h_i Z'$ and $N N$ decay modes are not relevant to our study of DM phenomenology. The $A-\nu-\nu$ coupling arises due to scalar and active-sterile mixing. The $\nu\nu$ mode is proportional to neutrino mass and therefore strongly suppressed. $\Phi, \chi$ are SM singlets; therefore, the resultant pseudo-scalars do not couple to SM fermions and gauge bosons. The only two body decay channels are $h_i Z$; however, $h_3$ is also a massive particle. Hence, only $h_1 Z, ~h_2 Z$ modes are kinematically allowed in our interested mass range of DM.

Apart from the two body decays, there could be two possible; three body decays such as $A \rightarrow Z f \bar{f}, \ h_i f \bar{f}$ are also relevant and could be dominant or subdominant depending on the kinematics. The former process contributes to non-zero-gauge kinetic mixing. The latter process is mediated by $Z, ~Z'$, and a dominant three-body decay when kinematically allowed. We have shown the Feynman diagrams for these decay modes in figure~\ref{app:feynman diagram} and calculated the decay width expressions for all these modes as given in the following equations
\begin{align}
& \Gamma(A\to h_i X)=\frac{|\lambda_{AXh_i}|^2 m_A^3}{16\pi M_{X}^2}\lambda^{\frac{3}{2}}\Big(1,\frac{M_X^2}{m_A^2},\frac{m_{h_i}^2}{m_A^2}\Big), \\
& \Gamma(A\to N_i N_i)=\frac{\cos^2\theta_\eta}{8\pi v_\Phi^2}m_A M_{N_i}^2 \sqrt{1-4\frac{M_{N_i}^2}{m_A^2}}, \\
& \Gamma(A\to X f\bar{f})=\frac{|H_X(z)|^2 m^9 x_f^2}{128\pi^3 M_X^2 v_H^2} \big(z-4x_f^2\big)\lambda^{\frac{3}{2}}\big(1,\frac{M_X^2}{m_A^2},z\big)\frac{dz}{z} \lambda^{\frac{1}{2}}\big(1,x_{f}^2,x_f^2\big), \\
& d\Gamma(A\to h_i f\bar{f})=\frac{m_A^5}{384 \pi^3}\Big[\big(|A_L^{if}(z)|^2+|A_R^{if}(z)|^2\big) \big\{\lambda\big(1,x_{h_i}^2,z\big)+\frac{x_f^2}{z}\big(2(1-x_{h_i}^2)^2+2(1+x_{h_i}^2)z-z^2\big)\big\}\nonumber \\
& -6\text{Re}\big(A_L^{if*}(z) A_R^{if}(z)\big) x_f^2 (2(1+x_{h_i}^2)-z) \Big]\frac{dz}{z}\lambda^{\frac{1}{2}}\big(1,x_{h_i}^2,z\big) \lambda^{\frac{1}{2}}\big(1,x_{f}^2,x_f^2\big);\,\,\, 4 x_f^2 \leq z \leq (1-x_{h_i})^2,
\label{dwidth1}
\end{align}
where $X=Z,~Z'$ and $\lambda(a,b,c) = a^2 + b^2 + c^2 - 2 ab - 2 bc - 2 ca $ and couplings,
\begin{align*}
& \lambda_{AZh_i}=\frac{g_X x_\Phi \sin\theta}{\sqrt{1-\kappa^2}}(2\cos\theta_\eta O_{R_{i3}} -\sin\theta_\eta O_{R_{i2}}), \\
& \lambda_{AZ'h_i}=\frac{g_X x_\Phi \cos\theta}{\sqrt{1-\kappa^2}}(2\cos\theta_\eta O_{R_{i3}} -\sin\theta_\eta O_{R_{i2}}), \\
& A_{L(R)}^{if}(z)=\frac{\lambda_{AZh_i} g_{L(R)}^{Zf}}{z m_A^2 - M_Z^2 + i \Gamma_Z M_Z}+\frac{\lambda_{AZ'h_i} g_{L(R)}^{Z'f}}{z m_A^2 - M_Z^{'2} + i \Gamma_{Z'} M_{Z'}}, \\
& H_Z(z)=\sum_{i=1}^3 \frac{\lambda_{AZh_i} \mathcal{O}_{R_{i1}}}{z m_A^2-m_{h_i}^2+i\Gamma_{h_i}m_{h_i}},\,\, H_{Z'}(z)=\sum_{i=1}^3 \frac{\lambda_{AZ'h_i} \mathcal{O}_{R_{i1}}}{z m_A^2-m_{h_i}^2+i\Gamma_{h_i}m_{h_i}}, \\
& g^{Z\ell}_{R}=-\frac{1}{\sqrt{1-\kappa^2}}\Big(g_X \sin\theta \big(x_H+x_\Phi\big)-g_Z \big(\sin^2\theta_W\cos\theta \sqrt{1-\kappa^2}+\kappa \sin\theta \sin\theta_W\big)\Big),\nonumber\\
& g^{Z\ell}_{L}=-\frac{1}{2\sqrt{1-\kappa^2}}\Big(g_X \sin\theta \big(x_H+2x_\Phi\big)-g_Z \big(\kappa \sin\theta \sin\theta_W-(1-2\sin^2\theta_W)\cos\theta \sqrt{1-\kappa^2}\big)\Big),\nonumber \\
& g^{Zu}_{R}=\frac{1}{3\sqrt{1-\kappa^2}}\Big(g_X \sin\theta \big(2x_H+x_\Phi\big)-2g_Z\big(\cos\theta \sin^2\theta_W\sqrt{1-\kappa^2}+\kappa \sin\theta\sin\theta_W\big)\Big),\nonumber \\
& g^{Zu}_{L}=\frac{1}{6\sqrt{1-\kappa^2}}\Big(g_X\sin\theta \big(x_H+2x_\Phi\big)-g_Z\big(\cos\theta \sqrt{1-\kappa^2}(4\sin^2\theta_W-3)+\kappa\sin\theta\sin\theta_W\big)\Big),\nonumber 
\end{align*}
\begin{align*}
& g^{Zd}_{R}=\frac{1}{3\sqrt{1-\kappa^2}}\Big(g_X\sin\theta\big(x_\Phi-x_H\big)+g_Z\big(\cos\theta\sin^2\theta_W\sqrt{1-\kappa^2}+\kappa \sin\theta\sin\theta_W\big)\Big),\nonumber\\
& g^{Zd}_{L}=\frac{1}{6\sqrt{1-\kappa^2}}\Big(g_X\sin\theta \big(x_H+2x_\Phi\big)-g_Z\big(\cos\theta\sqrt{1-\kappa^2}(3-2\sin^2\theta_W)+\kappa\sin\theta\sin\theta_W\big)\Big), \nonumber
\end{align*}
where as $g^{Z'\ell/u/d}_{L/R}=g^{Z\ell/u/d}_{L/R}\Big(\sin\theta\to \cos\theta; \cos\theta\to -\sin\theta \Big)$.
%%%%%%%%%%%%%%%%%%%%%%%%%%%%%%%%%%%%%%%%%%%%%%

The relevant parameters for the study of DM lifetime constraints on our pseudo-scalar DM are U(1)$_X$ charge $x_H$, scalar mixing $\theta_{12}$ between Higgs h$_1$ and h$_2$, vev $v_{\phi}$ and $v_{\chi}$, gauge kinetic mixing $\kappa$ and masses $m_A$ and $m_{h_2}$. Other free parameters are the same as in the benchmark.

%%%%%%%%%%%%%%%%%%%%%%%%%%%%%%%%%%%%
\begin{figure}[htbp]
\includegraphics[height=5.5cm, width=8cm]{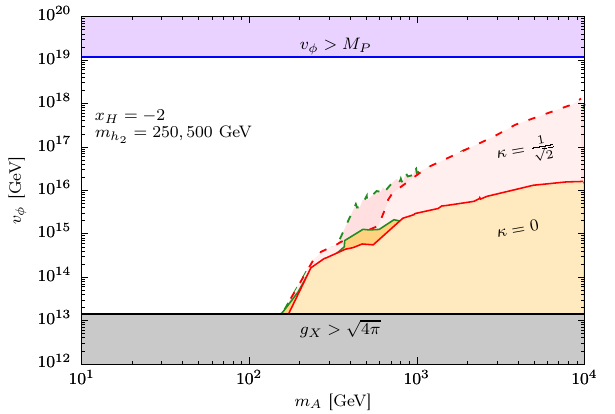}
\includegraphics[height=5.5cm, width=8cm]{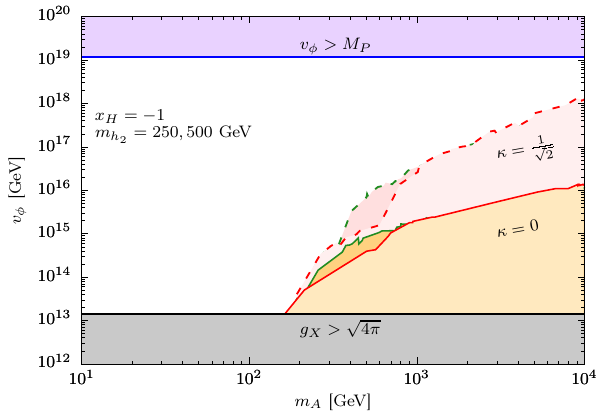}
\includegraphics[height=5.5cm, width=8cm]{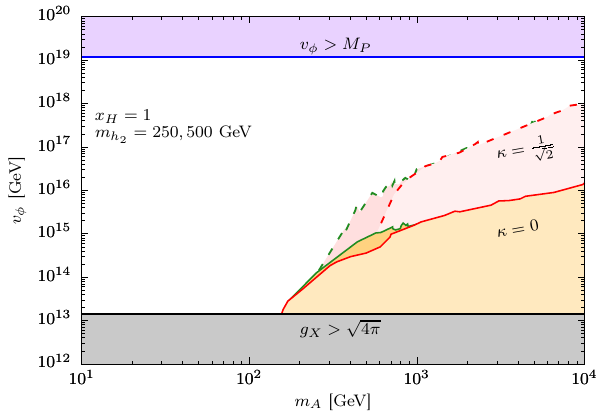}
\includegraphics[height=5.5cm, width=8cm]{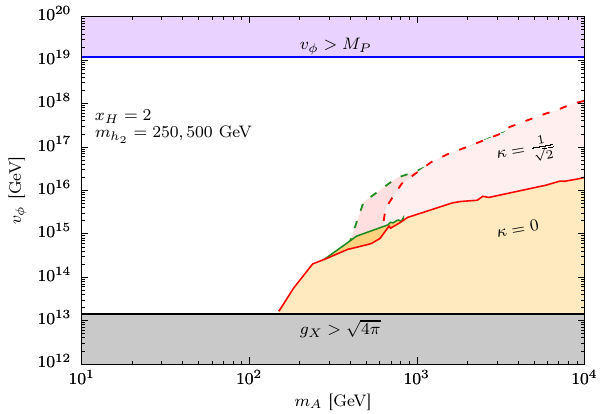}
\vspace{2mm}
\caption{Plots are displayed for $x_H = -2,-1,1,2$. In each plot, $\theta_{12}=0.1$ is fixed. The allowed parameter space for a pseudo-scalar is shown in the $(m_A, v_{\phi})$ plane. The coloured region is disfavored. The purple, pink, orange, and grey regions are not allowed by the Planck mass limit on vev, the DM lifetime constraint ($\tau > 10^{27}$ sec)~\cite{Baring_2016}, and the perturbative unitarity bound on gauge coupling $g_X$. The orange~(pink) region corresponds to kinetic mixing 0~($1/\sqrt{2}$). The dark~(light) orange and pink regions correspond to $m_{h_2}=250~(500)$ GeV.} 
\label{width1}
\end{figure}
%%%%%%%%%%%%%%%%%%%%%%%%%%%%%%%%%%%%

%%%%%%%%%%%%%%%%%%%%%%%%%%%%%%%%%%%%
\begin{figure}[t]
\includegraphics[height=5.5cm, width=8cm]{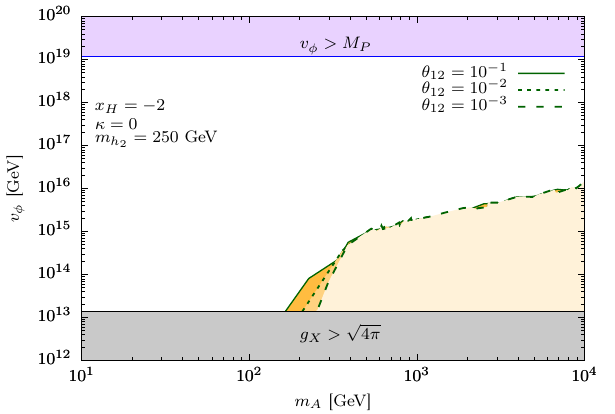}
\includegraphics[height=5.5cm, width=8cm]{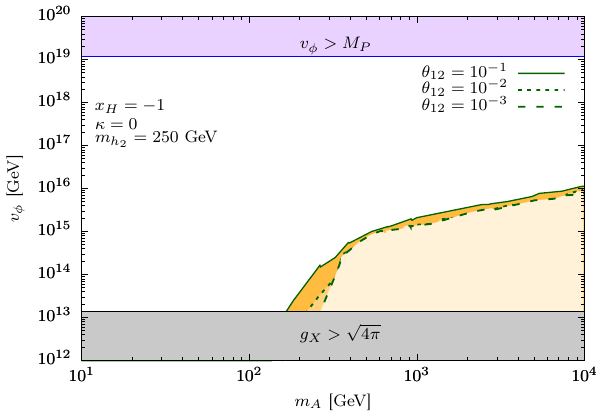}
\includegraphics[height=5.5cm, width=8cm]{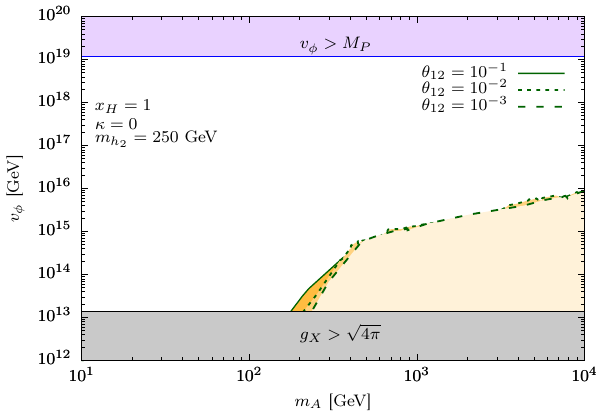}
\includegraphics[height=5.5cm, width=8cm]{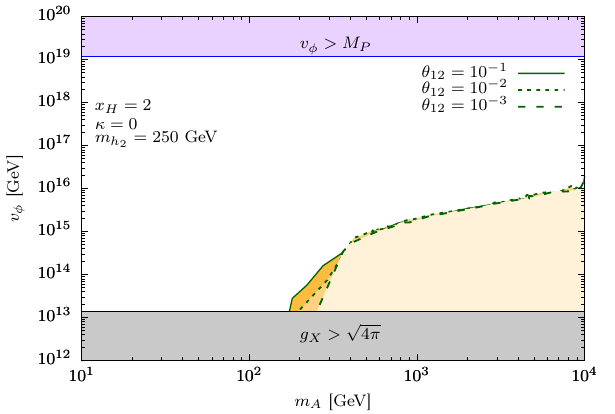}
\vspace{2mm}
\caption{Plots are displayed for $x_H = -2,-1,1,2$. The allowed parameter space for a pseudo-scalar is shown in the $(m_A, v_{\phi})$ plane. The disfavored region is coloured. The purple, orange, and grey regions are disfavored by the Planck mass limit on vev, the lifetime bound~\cite{Baring_2016}, and the perturbative unitarity limit on quartic coupling $\lambda_{\chi}$. The orange region~(from darker to lighter) corresponds to mixing $\theta_{12}=10^{-1,-2,-3}$.} 
\label{width2}
\end{figure}
%%%%%%%%%%%%%%%%%%%%%%%%%%%%%%%%%%%%

In fig~\ref{width1}, we show the allowed parameter space for DM in the $(m_A, v_{\phi})$ plane. The four plots correspond to $x_H=-2,-1,1,2$ from top left to bottom right. We have a fixed mixing angle $\theta_{12}=0.1$ in all the plots. Vev $v_{\chi}$ is a free parameter that is chosen from the range ($\sim 10 v_H - 10^4 v_H$), whereas $\kappa_T$ is determined using eq~\ref{psmass}, which is of the order of $\kappa_T \sim \mathcal{O}(\frac{m^2_A}{v_{\Phi}})$. The purple region on top of each plot is not allowed as vev $v_{\phi}$ gets larger than the Planck scale $M_P=1.2 \times 10^{19}$ GeV. The grey region on the bottom is disfavored by the perturbative unitarity bound of coupling $g_X > \sqrt{4\pi}$, which allows vev $v_{\phi}$ to be very high and small $\kappa_T$. The pink and orange regions in the middle of each plot correspond to kinetic mixing $\kappa=0,~1/\sqrt{2}$, respectively, and are disfavored by the DM lifetime constraint. The light~(dark) orange and red region is due to $m_{h_2}=250~(500)$ GeV. When allowed kinematically, the decay modes $h_i f \bar{f}, ~h_i Z$ are active in both pink and orange regions. However, in the pink region, where we turn on the kinetic mixing, which opens up an additional decay mode $Z f \bar{f}$, the lesser parameter space available by lifetime bound on the $(m_A, v_{\phi})$ plane,. One can also see that vev, $v_{\phi} > 10^{13}$ GeV, is favored for DM mass around $\sim$ 100 GeV.

In fig~\ref{width2}, we have done the analysis similar to above; however, we have fixed $m_{h_2}=250$ GeV, gauge kinetic mixing $\kappa=0$, and varied the scalar mixing $\theta_{12}$. The U(1)$_X$ charge $x_H$ is labelled in each plot. The orange region~(from darker to lighter) corresponds to mixing $\theta_{12}=10^{-1,-2,-3}$. Here, channels $h_i f \bar{f}$ and $h_i Z$ are the only relevant decay modes. The disallowed region increases as one increases the scalar mixing due to decay modes $h_1 Z, ~h_2 Z$ contributing significantly. The difference is significant only in a tiny region of parameter space, simply because of kinematics.
 
%%%%%%%%%%%%%%%%%%%%%%%%%%%%%%%%%%%%        
\subsection{Relic density analysis}
\label{relic}
%%%%%%%%%%%%%%%%%%%%%%%%%%%%%%%%%%%%

%%%%%%%%%%%%%%%%%%%%%%%%%%%%%%%%%%%%
\begin{figure}[t]
\includegraphics[height=5.5cm, width=8cm]{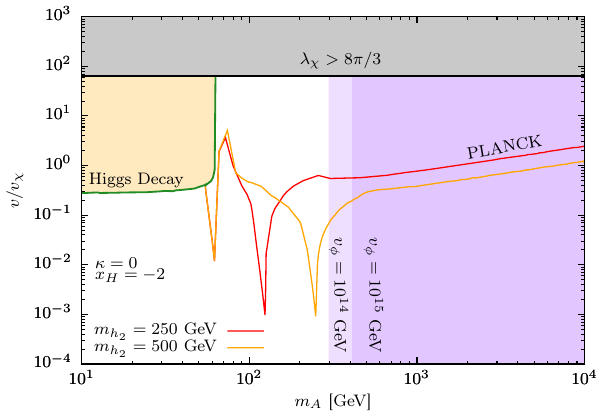}
\includegraphics[height=5.5cm, width=8cm]{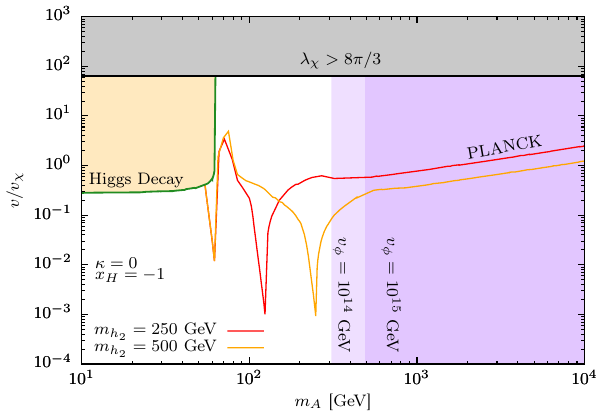}
\includegraphics[height=5.5cm, width=8cm]{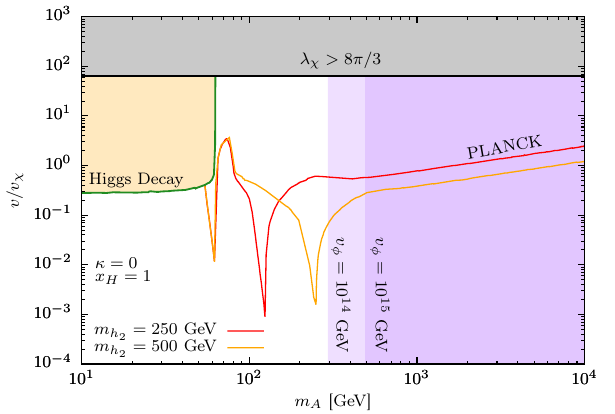}
\includegraphics[height=5.5cm, width=8cm]{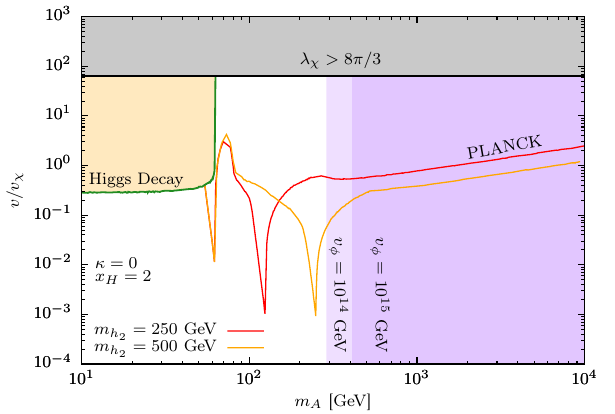}
\vspace{2mm}
\caption{Plots are displayed for $x_H = -2,-1,1,2$ when $\kappa=0$. Allowed parameter spaces are shown in the $(m_A, v/v_{\chi})$ plane. The disfavored region is coloured. The grey, orange, and purple regions are disfavored by the perturbative unitarity bound on quartic coupling~$\lambda_{\chi}$, the invisible Higgs width constraint~\ref{eq:invisible}, and the conservative lifetime bound~($\tau > 10^{27}$ sec)~\cite{Baring_2016} for $v_{\phi}=10^{14,15}$ GeV. Red and orange curves represent the correct thermal relic abundance satisfied by Planck data~\ref{eq:relic-density}. The two dips in the red~(orange) curve at 62.5, 125~(250) GeV are the resonances due to $h_1$ and $h_2$.} 
\label{relic21}
\end{figure}
%%%%%%%%%%%%%%%%%%%%%%%%%%%%%%%%%%%%

%%%%%%%%%%%%%%%%%%%%%%%%%%%%%%%%%%%%
\begin{figure}[t]
\includegraphics[height=5.5cm, width=8cm]{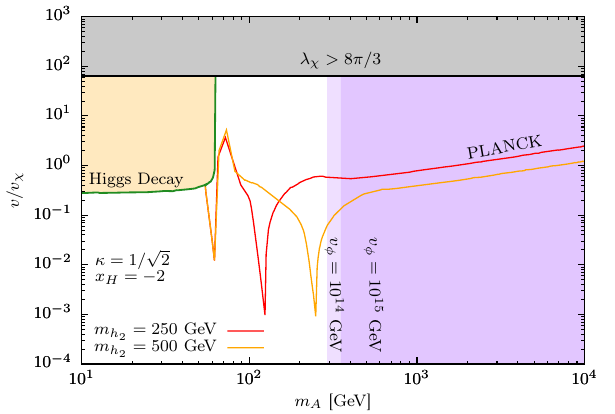}
\includegraphics[height=5.5cm, width=8cm]{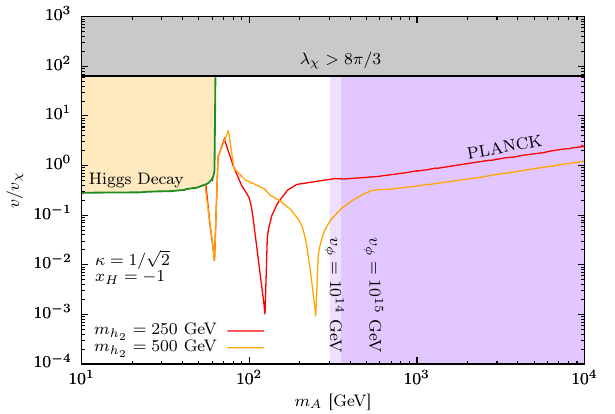}
\includegraphics[height=5.5cm, width=8cm]{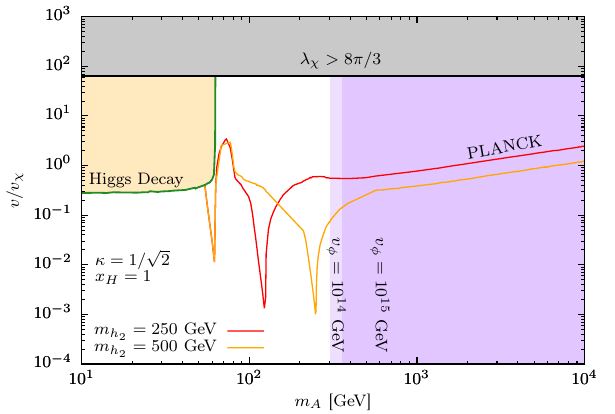}
\includegraphics[height=5.5cm, width=8cm]{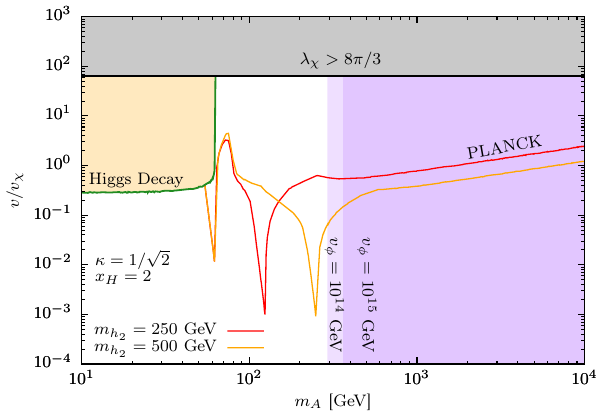}
\vspace{2mm}
\caption{Plots are displayed for $x_H = -2,-1,1,2$ when $\kappa=1/\sqrt{2}$.} 
\label{relic22}
\end{figure}
%%%%%%%%%%%%%%%%%%%%%%%%%%%%%%%%%%%%

We can now study the relic density constraint on our pseudo-scalar candidate to DM, which we did by micrOMEGAs package~\cite{Belanger:2018ccd}. In fig~\ref{relic21} and \ref{relic22}, we show the relic density analysis for our pseudo-scalar DM in the $(m_A, ~v/v_{\chi})$ plane. The plots in Fig. \ref{relic21}~(\ref{relic22}) correspond to gauge kinetic mixing $\kappa=0~(1/\sqrt{2})$. The scalar mixing $\theta_{12}=0.1$ is fixed in each plot. The U(1)$_X$ charge $x_H$ is also labelled in each plot. The grey region on top of each plot is disfavored by the perturbative unitarity of quartic coupling $\lambda_{\chi} < 8\pi/3$. The light and dark purple regions are disfavored by the DM lifetime bound, and they correspond to $v_{\phi}=10^{14,15}$ GeV, respectively. The light orange region in the middle is not allowed by the invisible Higgs width constraint~\ref{eq:invisible}. The red and orange curves correspond to $m_{h_2}=250,~500$ GeV, respectively, and they represent the correct relic abundance following the Planck data~\ref{eq:relic-density}. The main annihilation channels that contribute to relic abundances are: $W^+W^-, ~ZZ, ~h_1h_1, ~h_2h_2 $. The two dips in each plot are the two resonances due to two Higgs poles at $m_{h_1}/2$ and $m_{h_2}/2$. One can see the allowed parameter space in the $(m_A, ~v/v_{\chi})$ plane can be increased by increasing vev $v_{\phi}$ and decreasing gauge kinetic mixing $\kappa$.
      
%%%%%%%%%%%%%%%%%%%%%%%%%%%%%%%%%%%%
\subsection{Direct detection}
\label{direct}
%%%%%%%%%%%%%%%%%%%%%%%%%%%%%%%%%%%%

The main advantage of having a pseudo-scalar DM is that the DM-nucleon scattering cross section vanishes in the non-relativistic limit, and we need to check that this is the case in our model too. The relevant interaction vertices for scattering matrix is A$-$A$-h_i$, which simplify in the limit~$v_{\phi} >> v_H,~v_{\chi}$ and given as follows:

\begin{align*}
g_{A A h_1} \approx v_H \lambda_{H\chi} \cos\theta_{12} + 
  2. v_{\chi} \lambda_{\chi} \sin\theta_{12}, \hskip 0.5cm  
g_{A A h_2} \approx - v_H \lambda_{H\chi} \sin\theta_{12} + 
  2. v_{\chi} \lambda_{\chi} \cos\theta_{12}, \ \hskip 0.3cm
 g_{A A h_3} \approx 0
\end{align*} 

Hence, the scattering matrix is some of only two Feynman diagrams mediated by the two lighter Higgs $h_1,~h_2$.

\begin{align*}
\mathcal{M} \sim \frac{g_{A A h_1}g_{h_1 f \bar{f}}}{q-m^2_{h_1}} + \frac{g_{A A h_2}g_{h_2 f \bar{f}}}{q-m^2_{h_2}} 
\end{align*}

Here $g_{h_i f \bar{f}}$ is the coupling for Higgs-SM fermions interaction and q is the momentum transfer. One can see by using equation~\ref{eq:quartic} that the tree label amplitude vanishes in the non-relativistic limit, as it is shown for the simpler case in~\cite{Gross_2017}. However, one loop contribution could be finite, and it has been studied in ref~\cite{Ishiwata_2018}. There will be three types of Feynman diagrams, namely: 1. self-energy; 2. vertex corrections; and 3. box and triangle diagrams, as shown in ref~\cite{Ishiwata_2018}, which contribute to one loop. The generic expression for the scattering cross-section, which is the sum of all these Feynman diagrams, is given by
\begin{align}
\sigma^N_{\text{SI}}= \frac{1}{\pi}\left(\frac{m_N}{m_A+m_N}\right)^2 \vert f^N_{\text{scalar}}+f^N_{\text{twist2}}\vert^2
\label{sigmalo}
\end{align}
Here $m_N$ is the mass of the nucleon, and the definition of the functions $f^N_{\text{scalar}}, ~f^N_{\text{twist2}}$ is given in ref~\cite{Ishiwata_2018}.

%%%%%%%%%%%%%%%%%%%%%%%%%%%%%%%%%%%%
\begin{figure}[htbp]
\includegraphics[height=6cm, width=8cm]{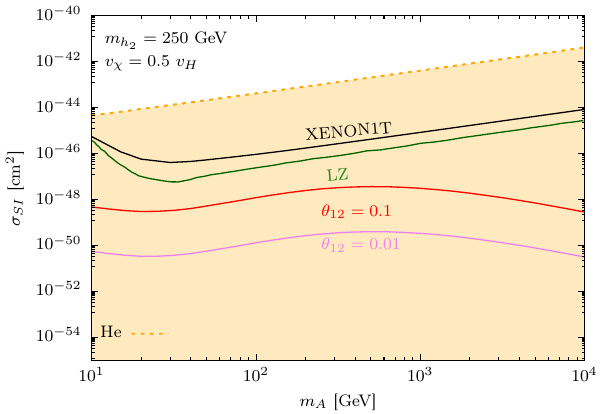}
\includegraphics[height=6cm, width=8cm]{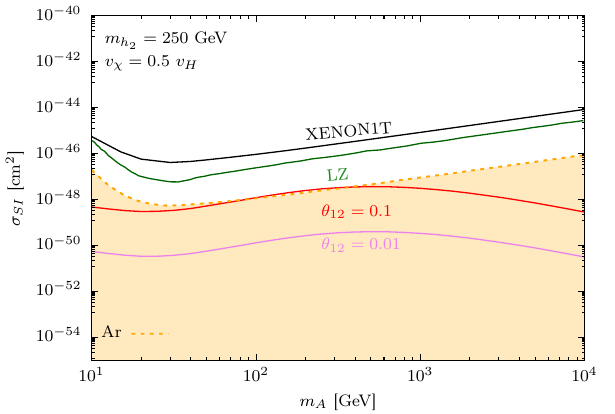}
\includegraphics[height=6cm, width=8cm]{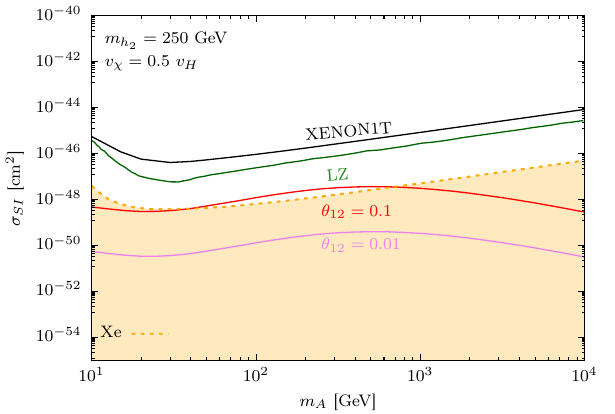}
\includegraphics[height=6cm, width=8cm]{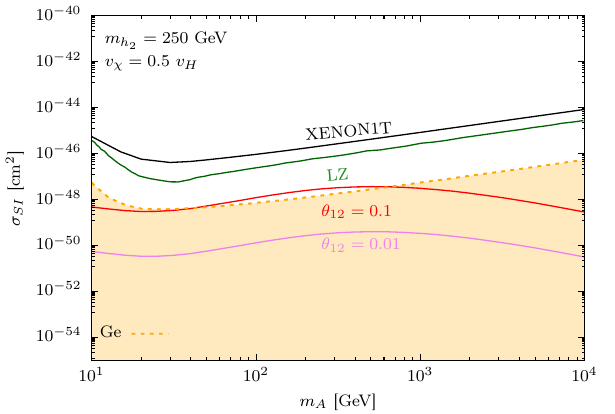}
\includegraphics[height=6cm, width=8cm]{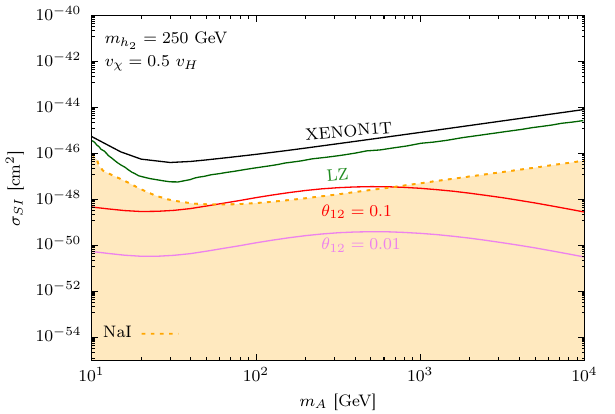}
\includegraphics[height=6cm, width=8cm]{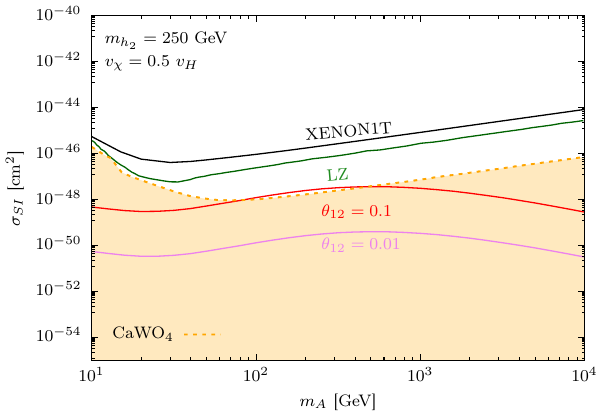}
\caption{We have fixed $m_{h_2}=250$ GeV and $v_{\chi}=0.5~v_H$ in each plot. The red and pink curves in each plot are the DM-nucleon spin-independent cross section~($\sigma_{SI}$) for scalar mixing $\theta_{12}=0.1,0.01$, respectively. The black and dark-green curves are the direct detection limits from the XENON1T~\cite{Aprile:2018dbl} and LZ~\cite{Akerib:2018lyp} experiments. The orange region is neutrino floor data~\cite{O_Hare_2021} for various targets such as Helium~(He), Argon~(Ar), Xenon~(Xe), Germanium~(Ge), NaI, and CaWO$_4$ as labeled in plots.}  
\label{dd1}
\end{figure}
%%%%%%%%%%%%%%%%%%%%%%%%%%%%%%%%%%%%

%%%%%%%%%%%%%%%%%%%%%%%%%%%%%%%%%%%%
\begin{figure}[htbp]
\includegraphics[height=6cm, width=8cm]{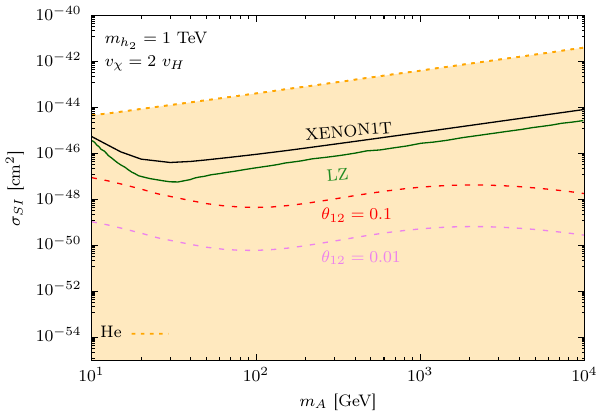}
\includegraphics[height=6cm, width=8cm]{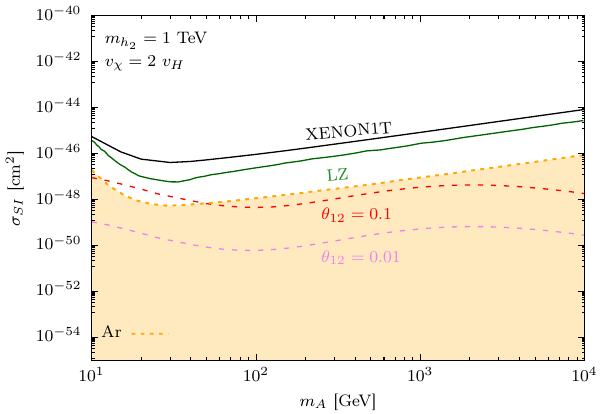}
\includegraphics[height=6cm, width=8cm]{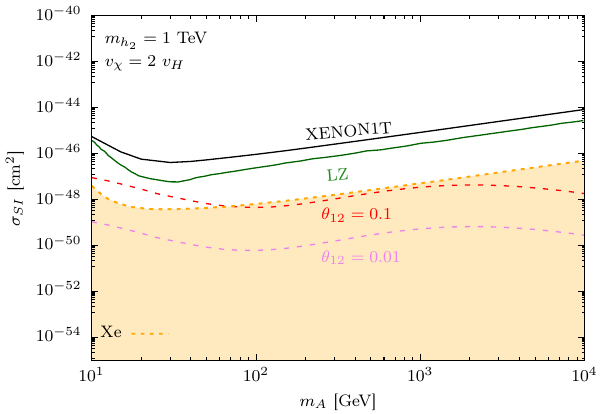}
\includegraphics[height=6cm, width=8cm]{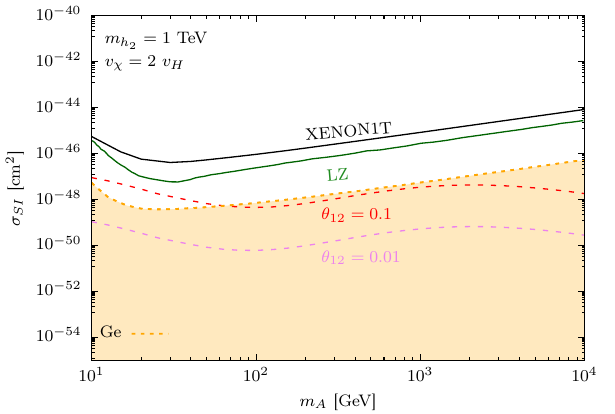}
\includegraphics[height=6cm, width=8cm]{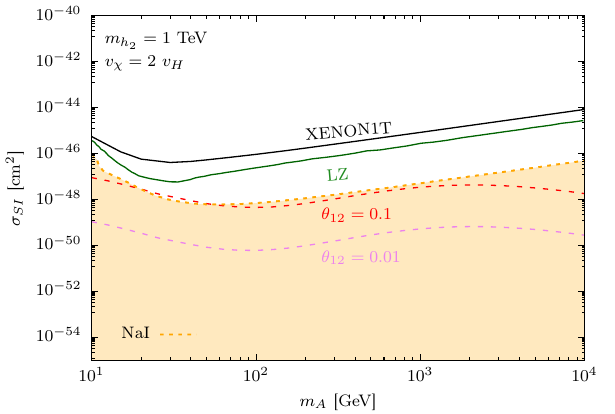}
\includegraphics[height=6cm, width=8cm]{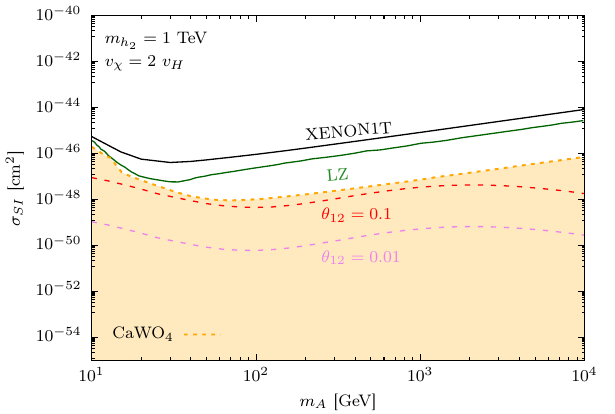}
\caption{We have fixed $m_{h_2}=1$ TeV and $v_{\chi}=2~v_H$ here in all the plots.}  
\label{dd2}
\end{figure}
%%%%%%%%%%%%%%%%%%%%%%%%%%%%%%%%%%%%

In our model, the third Higgs $h_3$ and Z$'$ are too heavy to play any significant role in the scattering matrix calculation; therefore, it is straightforward to follow the formalism developed in the reference to evaluate the scattering cross section at one loop level using equation \ref{sigmalo} in our model.

In fig~\ref{dd1}, \ref{dd2}, we check for the direct detection constraint on the $(\sigma_{\text{SI}}, m_A)$ plane. We have fixed $m_{h_2}=250$ GeV and $v_{\chi}=0.5~v_H$ in all plots of fig~\ref{dd1}, similarly $m_{h_2}=1$ TeV and $v_{\chi}=2~v_H$ are fixed in fig~\ref{dd2}. The red and blue curves in each plot are the predictions from the model for the $h_1-h_2$ mixing angle $\theta_{12}=0.1,~0.01$, respectively. The black and dark-green curve is the limit from XENON1T~\cite{Aprile:2018dbl} and LZ~\cite{Akerib:2018lyp}. The light orange region is the neutrino background in each plot due to six different targets: Helium~(He), Argon~(Ar), Xenon~(Xe), Germanium~(Ge), NaI, and CaWO$_4$. The neutrino floor data is taken from ~\cite{O_Hare_2021}. One can see the theoretical prediction, e.g., $\theta_{12}=0.1$, is above the neutrino floor around $m_A=200, 20$ GeV when $v_{\chi} = 0.5v_H, ~2v_H$, respectively, and the theoretical value is comparable to the LZ and XENON1T data for most of the $m_A$ range. The neutrino floor limit from six different target materials suggests an improved experiment setup to search for direct detection signals. The direct detection signals can be searched in the allowed parameter space of our model, as shown in fig.~\ref{dd1},~\ref{dd2}.

%%%%%%%%%%%%%%%%%%%%%%%%%%%%%%%%%%%%
\subsection{Constraining \texorpdfstring{$\sin{\theta_{12}}-m_{h_2}$}{sinmh2} plane}
\label{theta_mh2}
%%%%%%%%%%%%%%%%%%%%%%%%%%%%%%%%%%%%

\begin{figure}[htbp]
\includegraphics[height=6cm, width=8cm]{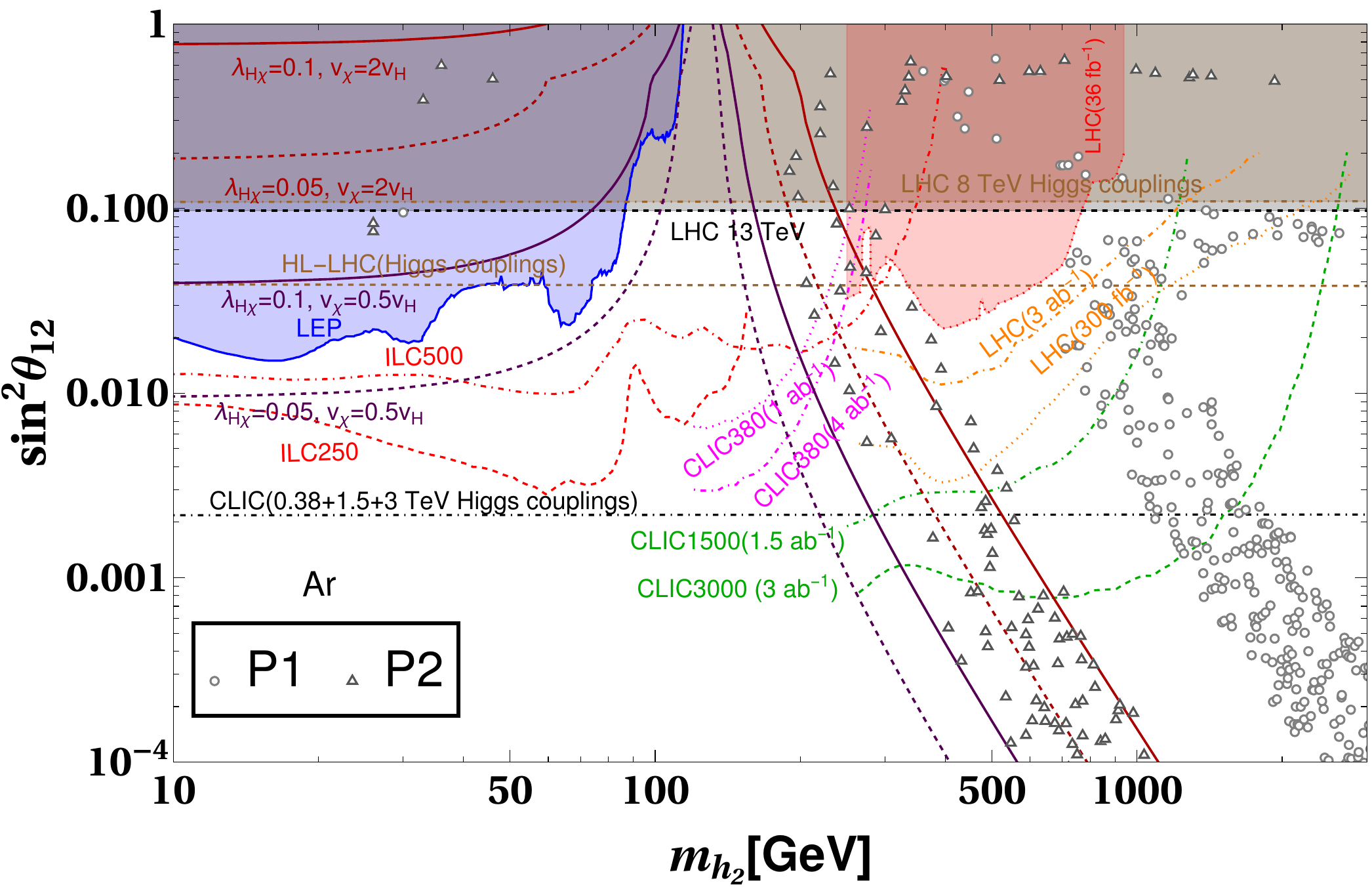}
\includegraphics[height=6cm, width=8cm]{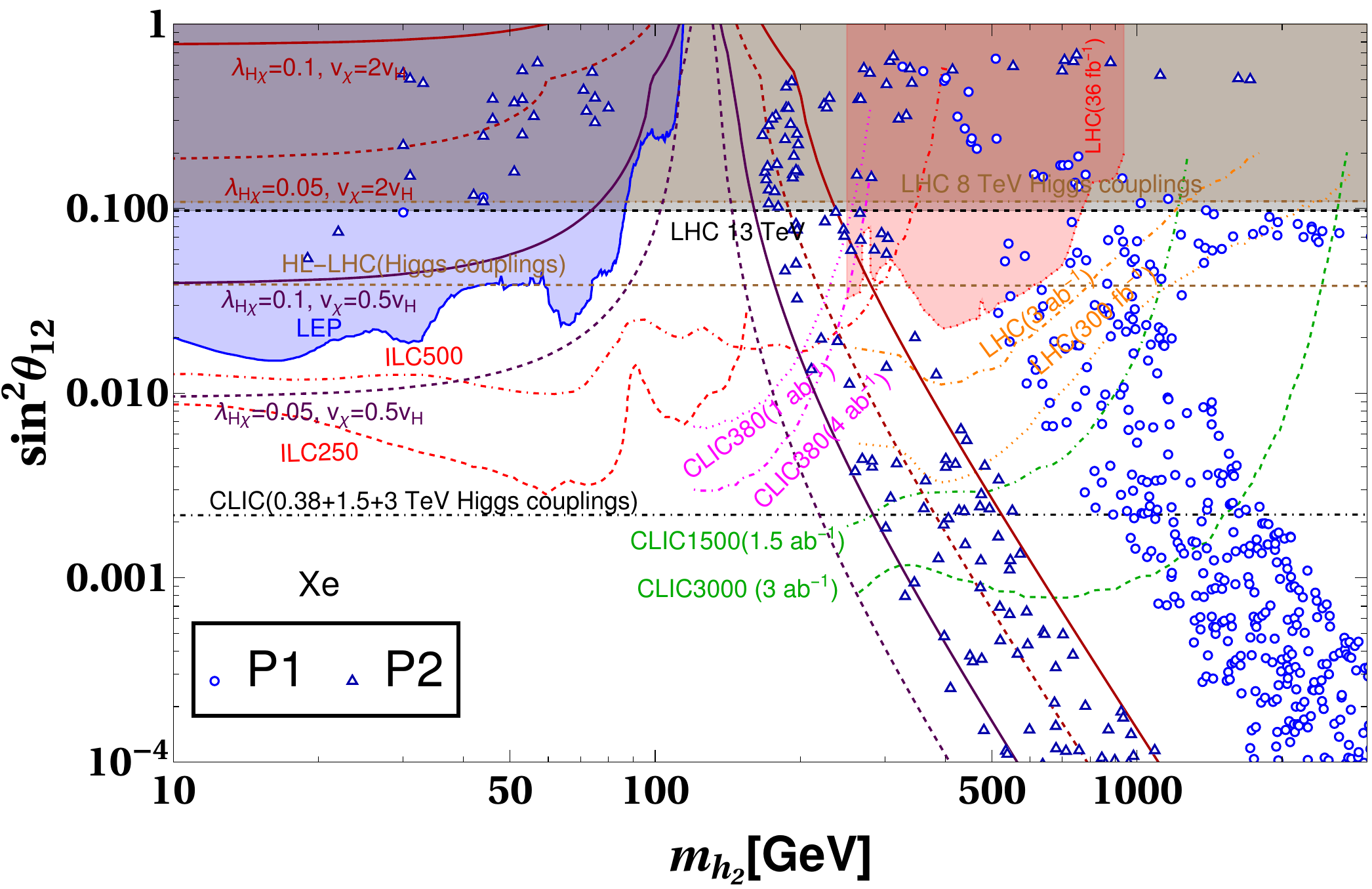}
\includegraphics[height=6cm, width=8cm]{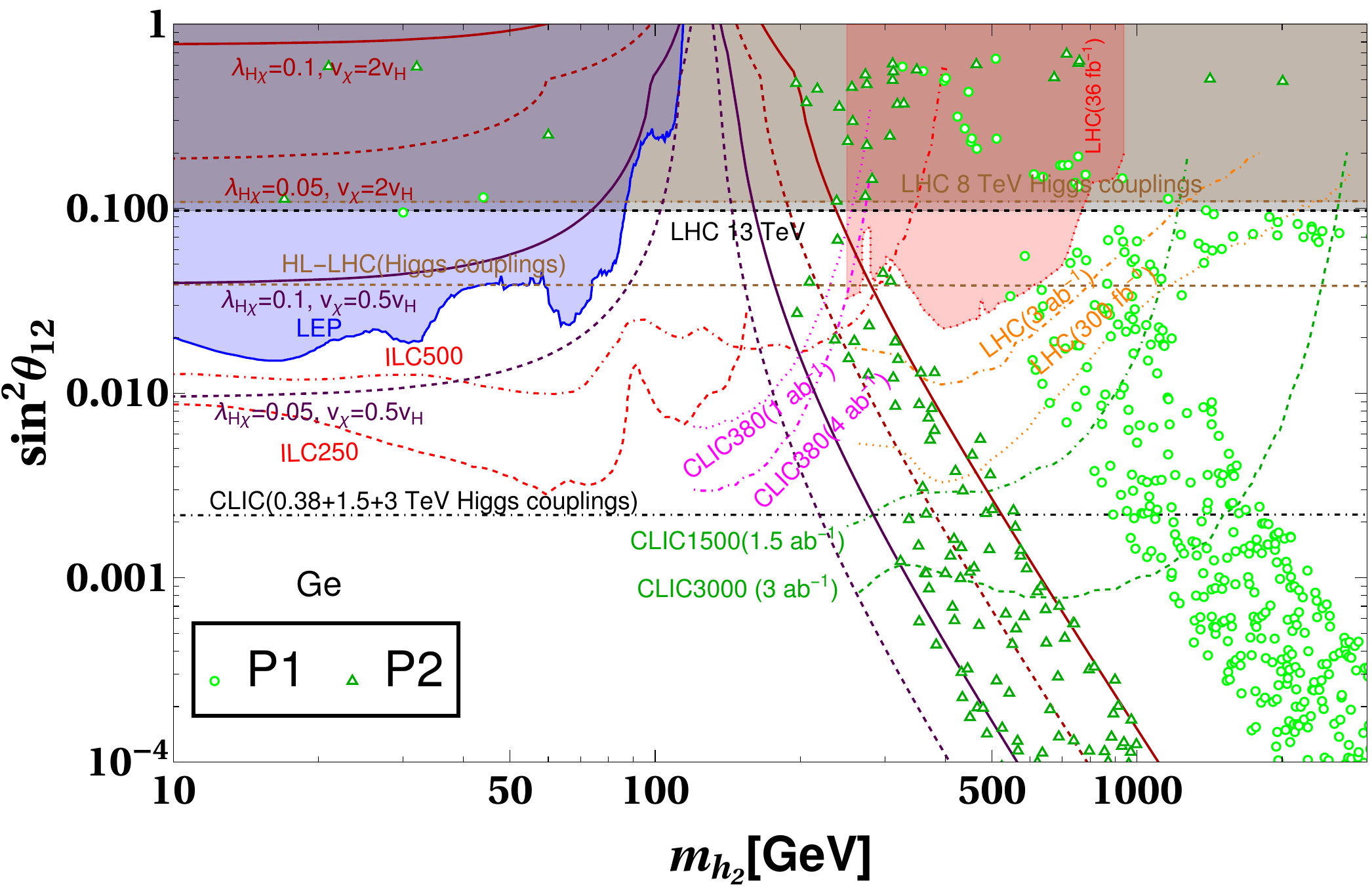}
\includegraphics[height=6cm, width=8cm]{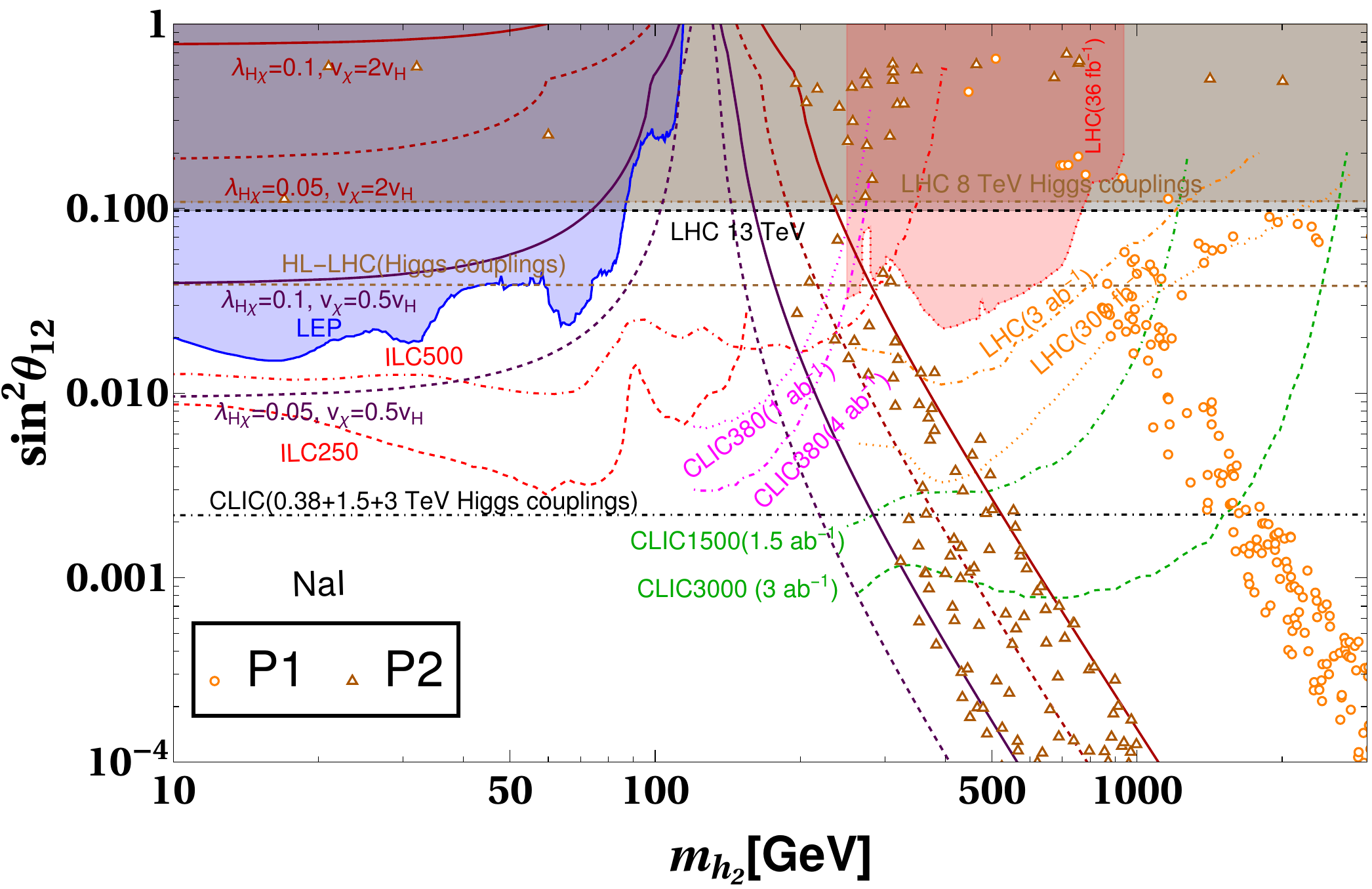}
\begin{center}
\includegraphics[height=6cm, width=8cm]{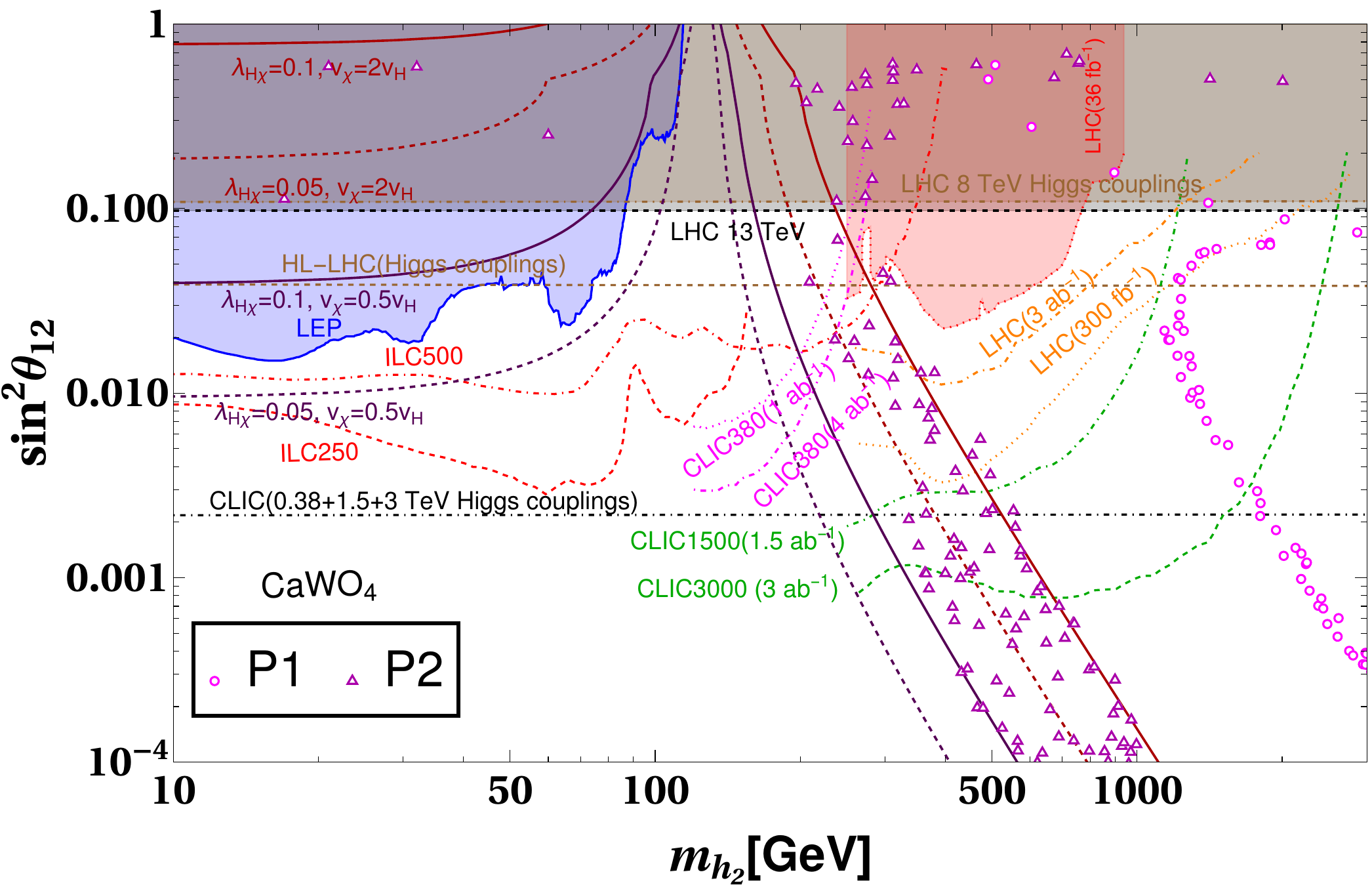}
\end{center}
\vspace{2mm}
\caption{The plot here is adapted from~\cite{Das:2022oyx}. Several bounds from colliders such as LHC \cite{deBlas:2018mhx}, LEP \cite{LEPWorkingGroupforHiggsbosonsearches:2003ing}, prospective colliders such as ILC \cite{Wang:2020lkq}, and CLIC \cite{deBlas:2018mhx} on the scalar mixing $\sin^2\theta_{12}$ as a function of $m_{h_2}$ are shown here. Shaded regions are ruled out by corresponding experiments. Limits on the scalar mixing from Eq.~\ref{eq:quartic} are also shown for different $\lambda_{H\chi}=$0.1, 0.05, respectively. Constraints from direct detection for $(m_A, v_{\chi}) =$ P1$\sim(20~\text{GeV}, ~2v_H)$,~ P2$\sim(200~\text{GeV}, ~0.5v_H)$ are shown when the neutrino floor is due to Ar, Xe, Ge, NaI, and CaWO$_4$. }  
\label{sinmh2}
\end{figure}

In this section, we constraint the ($\sin\theta_{12}$, $m_{h_2}$) plane using the direct detection study done in fig~\ref{dd1}, \ref{dd2}. To do this, we calculate the DM-nucleon cross section using equation~\ref{sigmalo} for $(m_A, v_{\chi}) = P1\sim(20~\text{GeV}, ~2 v_H),~ P2\sim(200~\text{GeV}, ~0.5 v_H)$, which is allowed points for $\theta_{12}=0.1$ by bounds as shown in direct detection plots. We then constrained the scattering cross section using limits from LZ and neutrino floor data for the corresponding points while taking $\theta_{12}$ and $m_{h_2}$ as free parameters. Since neutrino floor data is from five different targets such as Ar, Xe, Ge, NaI, and CaWO$_4$, we have five distinct plots for scalar mixing as shown in fig~\ref{sinmh2}. We do not consider neutrino floor data from the Helium target since it gives a much larger cross-section, as seen in figure~\ref{dd1}(\ref{dd2}); thus, `He' is not a good choice for direct searches, and hence, we do not consider it for further analysis.

%%%%%%%%%%%%%%%%%%%%%%%%%%%
%%%%%%%%%%%%%%%%%%%%%%%%%%%%%%%

There are several colliders bounds such as LHC \cite{deBlas:2018mhx}, LEP \cite{LEPWorkingGroupforHiggsbosonsearches:2003ing}, prospective colliders e.g., ILC \cite{Wang:2020lkq}, and CLIC \cite{deBlas:2018mhx} that are adapted here from the reference~\cite{Das:2022oyx}. The LEP bounds shown by the blue solid line consider both $Z\to \mu^+\mu^-$ and $e^+e^-$ modes. Bounds from LHC and High Luminosity LHC (HL-LHC) at 8 TeV are shown by the brown dot-dashed and dashed lines, respectively~\cite{deBlas:2018mhx}. LHC bounds at 13 TeV are shown by the black dashed line from the ATLAS results \cite{ATLAS:2020qdt}. The prospective bounds were estimated from $h_2 \to ZZ$ mode at the LHC at 300 (3000) fb$^{-1}$ luminosity using an orange dot-dashed (dotted) line from \cite{Buttazzo:2018qqp}. The prospective bounds from the combined CLIC analysis at 380 GeV, 1.5 TeV, and 3 TeV are shown by the black dot-dashed line.

We also superimpose the limits on the scalar mixing~$\theta_{12}$ using Eq.~\ref{eq:quartic} by considering two different choices of $\lambda_{H\chi}$ as $0.1$, $0.05$ for each point P1 and P2. The scalar mixing $\sin^2\theta_{12} < 0.001$ is respected by each constraint, as seen in fig~\ref{sinmh2}.

\section{Summary}
\label{sec:summary}

We have considered a generic $U(1)_X$ extension of SM that accommodates a pseudo-scalar DM candidate and neutrino mass generation mechanism. In the model, we have three RHNs and two complex scalars $\Phi$ and $\chi$ additionally, and all are charged under the $U(1)_X$ gauge group. $U(1)_X$ gauge is broken by the vev of $\Phi$. $\chi$ field also gets vev $v_{\chi}$, which results in a massive pseudo-scalar particle. The pseudo-scalar has coupling with Higgs and neutral gauge bosons, and its phenomenology is controlled by a few parameters such as $x_H$, $\kappa$, $m_{h_i}$, $g'_{1}$, $M_{Z'}$, and $m_A$. We have found that pseudo-scalars have sufficient parameter space, which is satisfied by the DM lifetime constraint while remaining consistent. We found the allowed parameter space using the relic density constraints, the invisible Higgs width constraint, and other bounds. A direct detection check is also done in our model. The tree-level DM-nucleon scattering amplitude vanishes in a non-relativistic limit, but one loop contribution is finite. We have shown the one-loop contribution to scattering cross-section against the experimental limit from XENON1T, LZ, and neutrino floors from five different targets. In our model, the pseudo-scalar DM has sufficient parameter space, which is consistent with several theoretical and experimental constraints and can be tested at future colliders and direct detection experiments.

%%%%%%%%%%%%%%%%%%%%%%%%%%%%%%%%%%%%%
\chapter{Summary}
\label{chap_conclusion}
%%%%%%%%%%%%%%%%%%%%%%%%%%%%%%%%%%%%%

This thesis examines several models of Weakly Interacting Massive Particle (WIMP) dark matter and their implications from a phenomenological standpoint. These models are particularly intriguing due to the stringent constraints of direct detection experiments. In chapters~\ref{chap_ALP_RHN} and \ref{chap_U1X2}, we explore models where the pseudo-scalar properties serve as both a mediator and a dark matter candidate, allowing them to evade the direct detection bounds. Additionally, in chapter~\ref{chap_U1X}, we delve into a two-component dark matter model and its relevant phenomenological features. Here, we provide a concise overview of each of these chapters.

Chapter~\ref{chap_ALP_RHN} focuses on an ALP portal fermion dark matter model. This model is minimal, consisting of three right-handed neutrinos (RHNs) and one pseudo-scalar mediator known as ALP. The two RHNs mix with the three active neutrinos, generating neutrino masses through Type I seesaw mechanism (detailed in Appendix-\ref{seasaw}). The stability of the third RHN is ensured by a Z2 symmetry, making it a viable dark matter candidate. We then proceed to analyze the constraints imposed by neutrino mixing and ALP couplings to Standard Model (SM) particles. By incorporating these experimental bounds, we investigate the parameter space that allows for fermion dark matter, taking into account relic density, direct detection, and indirect detection constraints. Remarkably, the model exhibits a significant parameter space that satisfies the relic density constraint, and it also possesses detectable signatures for indirect searches such as Fermi-LAT or HESS.

In chapter~\ref{chap_U1X}, we explore a model of dark matter consisting of a scalar and fermion as components. This model is based on a generic U(1)$_X$ extension of the Standard Model (SM). To ensure the consistency of the model, we introduce three right-handed neutrinos, which necessitate the presence of the U(1)$_X$ gauge symmetry. Additionally, we incorporate two new complex scalars into the model. These scalars are singlets under the SM but carry charges under the U(1)$_X$ gauge symmetry. One of the scalars breaks the U(1)$_X$ gauge symmetry, while the other scalar is responsible for the scalar dark matter (DM).

In this model, we also include two Z2 symmetries. One of these symmetries applies to the right-handed neutrinos, while the other symmetry applies to the new scalar. As a result, both the scalars and fermions can potentially act as DM candidates. Furthermore, the two remaining right-handed neutrinos mix with the three active neutrinos, leading to the generation of neutrino masses through a Type I seesaw mechanism.

We proceed to analyze various constraints that exist for this model. These constraints include the Higgs mixing angle ($\theta_{12}$) and the mass of the second Higgs boson ($m_{h_2}$), as well as the gauge coupling ($g'_1$) and the mass of the $Z'$ boson ($M_Z'$). By considering these experimental bounds, we investigate the parameter space that is consistent with relic density bounds for both scalar and fermion DM, taking into account scenarios where the two types of DM either interact or do not interact with each other. Additionally, we perform a scan to identify feasible parameter space that satisfies the direct detection bounds.

Finally, we explore a few more U(1) models that allow for parameter space accommodating the two-component DM. This model offers a rich phenomenology and provides opportunities for testing and collaboration with colleagues.

In chapter~\ref{chap_U1X2}, we explore a dark matter model with a pseudo-scalar in a generic U(1)$_X$ extension of the SM. This model shares similar particle and symmetry content with the two-component model discussed in chapter~\ref{chap_U1X}, but it does not possess Z2 symmetries. The U(1)$_X$ gauge symmetry is broken by one complex scalar, while the second scalar is responsible for the pseudo-scalar dark matter. The mixing of three right-handed neutrinos (RHNs) with three active neutrinos generates neutrino masses through a Type I seesaw mechanism. We then analyze various theoretical and experimental constraints, including the Higgs invisible width and the Planck bound, in the corresponding section. By imposing these constraints, we investigate the parameter space that satisfies the lifetime bound for the pseudo-scalar in different U(1) models. Additionally, we examine the relic constraint on the model by considering the lifetime bound in various U(1) models. Our findings reveal that the model possesses a sufficient parameter space that satisfies both the lifetime and relic bounds. Subsequently, we assess the feasible parameter space by considering the direct detection bounds when the neutrino floor is caused by multiple target materials such as helium (He), silver (Ag), xenon (Xe), germanium (Ge), sodium iodide (NaI), and calcium tungstate (CaWO$_4$). Our analysis demonstrates that the model exhibits a significant allowed space according to the direct bounds. Finally, we constrain the higgs mixing angle~{$(\theta_{12})$}-higgs mass~($m_{h_2}$) plane based on our direct detection study. The model successfully passes all the bounds, both those related to dark matter and non-dark matter, and it can be experimentally tested at colliders or through direct searches.

We have previously discussed the WIMP DM models within the framework of the ALP portal and U(1)$_X$ extension. These models have been thoroughly analyzed, taking into account various theoretical and experimental constraints. The model we have discussed addresses certain issues associated with WIMP models, particularly the lack of direct detection of WIMP DM. Additionally, our models incorporate neutrino physics, including oscillation and mass mechanisms. These bottom-up models offer a fascinating range of phenomenology that can be explored through collider-based experiments. However, it is worth noting that the standard DM mass in these WIMP DM models is limited to the range of a few GeV to TeV, which restricts the scope of our explored DM scenarios. There are other intriguing mass ranges for DM, such as ultralight dark matter, sub-GeV DM, and primordial black holes, among others. I intend to investigate these scenarios in future research. Furthermore, apart from considering different DM candidates, exploring the production mechanisms of DM is also an intriguing avenue for further exploration and study.
%%%%%%%%%%%%%%%%%%%%%%%%%%%%%%%%
% \let\cleardoublepage\clearpage
 \begin{appendices}
%%%%%%%%%%%%%%%%%%%%%%%%%%%%%%%

%%%%%%%%%%%%%%%%%%%%%%%%%%%%%%
\chapter{Standard model}\label{sm}
%%%%%%%%%%%%%%%%%%%%%%%%%%%%%%%%%%

The SM~\cite{Glashow:1961tr,Weinberg:1967tq} is a renormalizable, Lorentz-invariant quantum field theory based on the gauge group $SU(3)_{c}\otimes SU(2)_{L}\otimes U(1)_{Y}$, which contains the fundamental set of particles, the leptons, quarks, gauge bosons, and the Higgs.
\begin{table}[h!]
\begin{tabular}{c| c c c c}
\hline
\hline
      &  SU(3)$_C$  & SU(2)$_L$ & U(1)$_Y$  \\ 
\hline
$q^{i}_{L}$ & {\bf 3 }    &  {\bf 2}         & $\frac{1}{6}$ \\
$u^{i}_{R}$ & {\bf 3 }    &  {\bf 1}         & $\frac{2}{3}$   \\
$d^{i}_{R}$ & {\bf 3 }    &  {\bf 1}         & $-\frac{1}{3}$   \\
\hline
$\ell^{i}_{L}$ & {\bf 1 }    &  {\bf 2}         & $-\frac{1}{2}$  \\
$e^{i}_{R}$    & {\bf 1 }    &  {\bf 1}         & $-1$          \\
\hline
$H$            & {\bf 1 }    &  {\bf 2}         & $\frac{1}{2}$   \\
\hline
\hline
\end{tabular}
\hspace{1cm}
\begin{tabular}{c c}
\hline
\hline
     Gauge bosons  & Force  \\ 
\hline
$G_{\mu}^{a}$   &   Strong \\
$W_{\mu}^{\pm},\,Z_{\mu}^{0}$  &  Weak \\
$A_{\mu}$  &      Electromagnetic \\
\hline
\hline
\end{tabular}
\caption{The SM particle contents are shown in the table and $i=1, 2, 3$ is the generation index.}
\label{sm particle content}
\end{table}
The SM lagrangian can be written as
\begin{align}
\mathcal{L}_{\text{SM}}= \mathcal{L}_{\text{kinetic}} + \mathcal{L}_{\text{gauge}} + \mathcal{L}_{\text{Higgs}} + \mathcal{L}_{\text{Yukawa}},
\label{LSM}
\end{align}
 where,
\begin{align}
\mathcal{L}_{\text{kinetic}}= \mathcal{L}_{\text{kinetic}}^{\text{fermion}} + \mathcal{L}_{\text{kinetic}}^{\text{gauge}},
\label{Lkinetic}
\end{align} 
The kinetic term for the fermions~(leptons and quarks) is given by
\begin{align}
 \mathcal{L}_{\text{kinetic}}^{\text{fermion}} &= \ \bar{\ell_L}\gamma^\mu D^{\mu}\ell_{L} + \overline{e_R}\gamma^\mu D_{\mu}e_{R} + \overline{q_L}\gamma^\mu D_{\mu}q_{L} \nonumber \\
 \ & + \overline{u_R}\gamma^\mu D_{\mu}u_{R} + \overline{d_R}\gamma^\mu D_{\mu}d_{R} 
 \label{kinetic}
\end{align}
where $D_{\mu}$ is the covariant derivative.
The kinetic term for the gauge particles can be written as
\begin{align}
\mathcal{L}^{gauge}_{kinetic}= -\frac{1}{4} W_{\mu\nu}^{a} W^{a^{\mu\nu}} -\frac{1}{4} B_{\mu\nu}B^{\mu\nu} - \frac{1}{4}  G^{a}_{\mu\nu} G^{a\mu\nu},
\label{LG}
\end{align}

In SM, the masses of all the fermions and $SU(2)_{L}$ gauge bosons are generated through the spontaneous symmetry breaking~(SSB). The Lagrangian for the Higgs sector is given by

\begin{align}
 \mathcal{L}_{\text{Higgs}}=\left(D_{\mu}H\right)\left(D^{\mu}H\right)-\left(-\mu^{2}H^{\dagger}H+\lambda \left(H^{\dagger}H\right)^{2} \right)
\end{align}

The Yukawa interactions for the leptons and quarks are given by
\begin{align}
\mathcal{L}_{\text{Yukawa}} &= Y_{e} \overline{\ell_{L}} H e_{R} + Y_{u} \overline{q_{L}} \tilde{H} u_{R} + Y_{d} \overline{q_{L}} H d_{R} + H. c.
\label{LYuk}
\end{align}

%%%%%%%%%%%%%%%%%%%%%%%%%%%%%%
\chapter[Friedmann equations]{Friedmann equations}
\label{Friedman}
%%%%%%%%%%%%%%%%%%%%%%%%%%%%%%%%%%

Our universe at a large scale can be described well if we assume isotropy and homogeneity of space. The Friedmann-Lemaitre-Robertson-Walker~(FLRW) metric hold these assumptions, which is given by,
\begin{align}
ds^2=dt^2 - a^2(t)\left( \frac{dr^2}{1-\kappa r^2} + r^2 (d\theta^2 + \sin^2\theta d\phi^2) \right)
\end{align}
here r,$\theta$,$\phi$, t are comoving coordinates, a(t) is the scale factor. $\kappa$ is the curvature scalar and can take values +1,0,-1 correspond to positive, flat and negatively curved Universe respectively.
Friedmann equations can be derived using Einstein's equation,
\begin{align}
\mathcal{R}_{\mu\nu} - \frac{1}{2} \mathcal{R} g_{\mu\nu} = 8\pi G T_{\mu\nu} + \Lambda g_{\mu\nu}
\end{align}
 here $\mathcal{R}$, $\mathcal{R}_{\mu\nu}$, $T_{\mu\nu}$ and $\Lambda$ are Ricci scalar, Ricci tensor, energy-momentum tensor and cosmological constant respectively. Assuming the Universe is composed of fluid, the energy-momentum tensor
can be written as 
\begin{align}
T_{\mu\nu} = -p g_{\mu\nu} + (p+\rho)u_{\mu}u_{\nu}
\end{align} 
here p and $\rho$ are the pressure and energy density of the fluid respectively whereas $u_{\mu}$ is the velocity vector in comoving coordinates. One can now derive the Friedmann equations as
\begin{align}
\frac{\dot{a}^2}{a^2} = \frac{8\pi G\rho}{3} - \frac{\kappa}{a^2} + \frac{\Lambda}{3} 
\\
\frac{\ddot{a}}{a} = -\frac{4\pi G}{3} (\rho+3p) + \frac{\Lambda}{3}
\end{align}
Using the above two equations, one can derive the density evolution equation,
\begin{align}
\dot{\rho}=-3H(\rho+p)
\end{align}
where H is the Hubble parameter.

\let\cleardoublepage\clearpage
%%%%%%%%%%%%%%%%%%%%%%%%%%%%%%
\chapter{Type I seasaw mechanism}\label{seasaw}
%%%%%%%%%%%%%%%%%%%%%%%%%%%%%%%%%%  

In the type I seesaw mechanism, we have right-handed neutrinos $\nu_R$, which can have both Dirac and Majorana mass terms as follow~\cite{Minkowski:1977sc,Schechter:1980gr,Mohapatra:1979ia,Schechter:1981cv},
\begin{align}
\mathcal{L}^{\nu}_{mass} = m_D \bar{\nu}_R \nu_L + \frac{1}{2} m_L \bar{\nu}^c_L \nu_L + \frac{1}{2} M_R \bar{\nu}^c_R \nu_R + h.c.
\end{align}
here subscript c stands for charge conjugation. Now one can write the above expression as follows
\begin{align}
\mathcal{L}^{\nu}_{mass}=
 \begin{pmatrix}
  \overline{\nu^c_{L}} & \overline{\nu_{R}} \\
 \end{pmatrix}
 \begin{pmatrix}
   m_L   &  m_{D} \\
  m_{D}^T  &  M_{R} \\
 \end{pmatrix}
\begin{pmatrix}
 \nu_{L} \\
 \nu_{R} \\
\end{pmatrix}
\end{align}
One can diagonalize the above mass matrix and can get mass eigenstates corresponding to 
\begin{align}
m_{1,2}=\frac{1}{2} \vert (m_L + m_R) \pm \sqrt{(m_L-m_R)^2+4 m^2_D}\vert 
\end{align}
In the limit $m_D << M_R$, mass eigen values simplifies to
\begin{align}
m_1 \approx \frac{m^2_D}{M_R}
\\
m_2 \approx M_R
\end{align}   

This naturally explains the smallness of neutrino masses.

%%%%%%%%%%%%%%%%%%%%%%%%%%%%%%%%%%%%%%%%%%%%%%%%%%%
\chapter{Feynman diagrams in two-component DM model}
\label{app:feynman diagram}
\begin{figure}[htbp]
\begin{center}
\includegraphics[width=14.5cm,height=10.5cm]{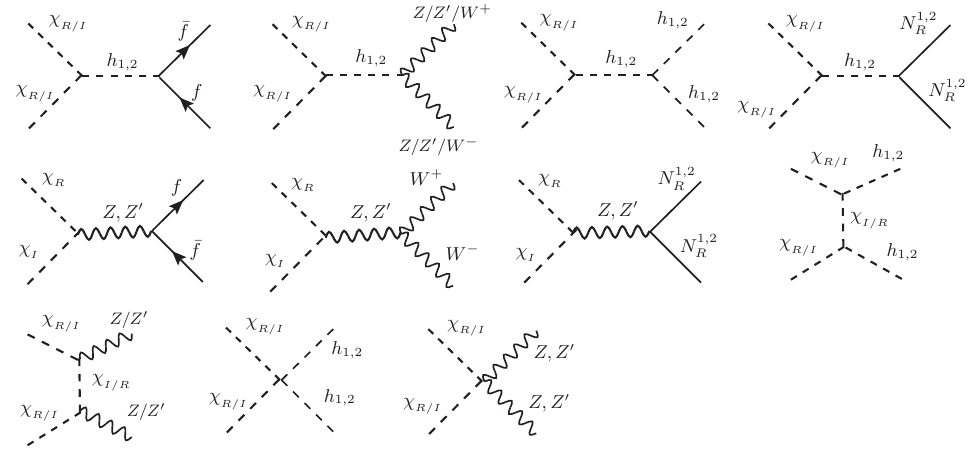}
\vspace{3mm}
\caption{Annihilation and coannihilation tree-level Feynman diagrams contributing to the relic abundance of $\chi_R$.}
\label{fig:annihilation-diagram-scalar}
\end{center}
\end{figure}

\begin{figure}[t!]
\begin{center}
\includegraphics[width=14.5cm,height=10.5cm]{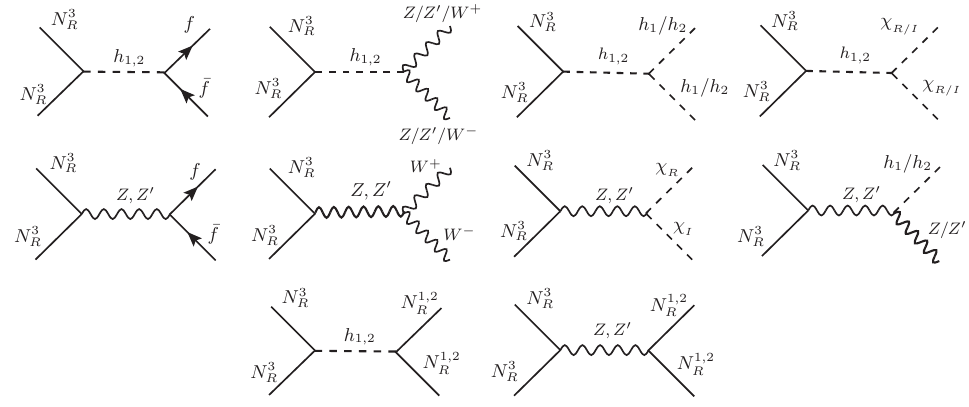}
\vspace{3mm}
\caption{Annihilation and coannihilation tree-level Feynman diagrams contributing to the relic abundance of $N_R^3$.}
\label{fig:annihilation-diagram-fermion}
\end{center}
\end{figure}

%%%%%%%%%%%%%%%%%%%%%%%%%%%%%%%%%%%%%%%%%%%%%%
The relevant Feynman diagram for relic density analysis of scalar and fermion DM are shown in Figs.~\ref{fig:annihilation-diagram-scalar} and \ref{fig:annihilation-diagram-fermion}, respectively. 
\vspace{1mm}
%%%%%%%%%%%%%%%%%%%%%%%%%%%%%%%%%%%%%%%%%%%

%%%%%%%%%%%%%%%%%%%%%%%%%%%%%%%%%%%%%%%%%%%%%%%
%%%%%%%%%%%%%%%%%%%%%%%%%%%%%%%%%%%%%%%%%%%%%%

\begin{comment}
\begin{figure}[htbp]
\begin{center}
\includegraphics[width=0.75\textwidth]{feyn_diag_U1XINR.pdf}
\includegraphics[width=0.75\textwidth]{feyn_diag_U1XnuI.pdf}
\caption{The Feynman diagrams that contribute to the process of DM conversion, $\chi\chi\leftrightarrow N_3 N_3$.}
\label{fig:annihilation-diagram-conversion}
\end{center}
\end{figure}
\end{comment}

\end{appendices}

\bibliographystyle{utphysM}
\bibliography{bibitem}
%utphysM

\end{document}